\documentclass[onecolumn, superscriptaddress, prmaterials, amsmath,amssymb,noeprint,12pt]{revtex4-2}

\usepackage{hyperref}
\hypersetup{backref,colorlinks=true,allcolors=MidnightBlue}
\usepackage{xr}
\usepackage{graphicx}
\graphicspath{{./figures/}}
\usepackage[dvipsnames]{xcolor}
\usepackage{tcolorbox}
\usepackage{amsmath,amssymb}
\usepackage{float}
\usepackage{multirow}
\usepackage{tabularx}
\usepackage{pifont}
\usepackage{wrapfig}
\usepackage[caption=false]{subfig}
\usepackage{miller}
\usepackage{todonotes}

\usepackage{tikz}
\usetikzlibrary{calc,arrows,shapes,positioning}
\usetikzlibrary{patterns,snakes}
\usetikzlibrary{positioning,decorations.pathreplacing}

\usepackage{epstopdf}
\AppendGraphicsExtensions{.tif}

\definecolor{sangria}{rgb}{0.57, 0.0, 0.04}
\definecolor{arsenic}{rgb}{0.23, 0.27, 0.29}
\definecolor{prussianblue}{rgb}{0.0, 0.19, 0.33}
\definecolor{phthalogreen}{rgb}{0.07, 0.21, 0.14}
\definecolor{dgreen}{rgb}{0.0, 0.4, 0.2}

\def\fig#1{Fig.~\ref{fig:#1}}
\def\eq#1{Eq.~\eqref{eq:#1}}
\def\tab#1{Table~\ref{tab:#1}}
\def\sec#1{Section~\ref{sec:#1}}
\def\sm{\textbf{SM}}

\begin{document}
\author{Musanna Galib}
\address{Department of Mechanical Engineering, University of British Columbia, 2054 - 6250 Applied Science Lane, Vancouver, BC, V6T 1Z4, Canada}
\author{Okan K. Orhan}
\address{Department of Mechanical Engineering, University of British Columbia, 2054 - 6250 Applied Science Lane, Vancouver, BC, V6T 1Z4, Canada}
\author{Jian Liu}
\address{School of Engineering, Faculty of Applied Science, The University of British Columbia, Okanagan Campus, Kelown, V1V 1V7, BC, Canada}
        
%
\author{Mauricio Ponga}
\address{Department of Mechanical Engineering, University of British Columbia, 2054 - 6250 Applied Science Lane, Vancouver, BC, V6T 1Z4, Canada}
%
\title{Residual Stress Development in Lattice Mismatched Epitaxial Thin Films via Atomic and Molecular Layer Depositions}
\begin{abstract}
Atomic and molecular layer deposition (ALD/MLD) coatings are promising solutions for preventing dendrite formation in aqueous and non-aqueous Li/Na/Zn metal batteries. Notably, alumina and alucone coatings have emerged as highly effective against dendrite formation in Zn anodes. Despite their demonstrated efficacy, a comprehensive understanding of their chemo-mechanical impact on anodes remains elusive. In this study, we take a bottom-up framework to these coatings on Zn foils, employing an approach that integrates \textit{ab initio} simulations with continuum theories to elucidate lattice misfit and chemical bonding. We use this insight to develop a macroscopic model to predict the epitaxial residual stresses generated during thin-film deposition. Our findings reveal a robust chemical bonding between the hydroxylated Zn surface and the thin film. This, in turn, generates large misfit strains that result in significant interfacial stresses during deposition. These results are then compared to experiments by measuring the curvature of the coated thin films, finding good agreement between experiments and theory. This novel understanding sheds light on the fundamental mechanisms underpinning the development of chemo-mechanical stresses in thin films, which impact dendrite suppression in anodes, offering valuable insights for the design of new coatings.
\end{abstract}

\maketitle

\section{\label{sec:Introduction}Introduction}
\par Zn metal (volumetric capacity of 5851 $\mathrm{mA \cdot h \cdot mL^{-1}}$) is one of the potential candidates for next-generation batteries in the field of scalable miniature energy storage devices~\cite{10.1021/acsnano.1c01389}. Aqueous zinc-ion batteries (AZIBs) can meet the prerequisites of flexible and wearable batteries, including high safety, affordability, reliable power, high volumetric capacity, and sustainability~\cite{10.1021/acsnano.1c01389}. However, the formation of sharp, needle-like metallic protrusions formed during electrodeposition (dendrites) and unwanted electrochemical reactions at the electrode/electrolyte interface (corrosion) restrain the widespread commercial use of AZIBs~\cite{10.1002/adma.202003021}. Recent studies have shown that corrosion in Zn anodes can be prevented using a thin layer of alumina ($\mathrm{Al_{2}O_{3}}$) or alucone~\cite{10.1039/D0TA00748J}. These thin layers are usually manufactured using atomic and molecular layer deposition (ALD/MLD) techniques, which are capable of generating a few nanometers thick coatings. Moreover, these coatings offer a certain level of chemo-mechanical strength, which can deter dendrite formation in Zn anodes ~\cite{10.1021/acsnano.0c07041}. As a result, the performance of AZIB can be greatly improved by applying ALD/MLD coatings. Thus, it is imperative to understand the chemo-mechanical integrity of these interfaces and their effects on the residual stresses to develop more robust, reliable devices ~\cite{10.1179/imr.1993.38.5.233, 10.1007/978-0-387-74365-3, 10.1016/j.tsf.2011.04.220}. 

\par Thin-film coatings have applications in many other fields, including electronic devices~\cite{10.1016/0304-3991(87)90045-3, 10.1111/j.1151-2916.1991.tb04086.x}, thermal barrier coatings~\cite{10.1179/imr.1993.38.5.233}, solid oxide fuel cells~\cite{10.1149/2.0591507jes}, lithium-ion batteries~\cite{10.1007/s41918-022-00146-6} and recently in AZIB~\cite{10.1039/D0TA07232J}. 
However, thin-film deposition techniques, such as ALD and MLD, usually experience residual stress regardless of the growth model or applied coating methods~\cite{10.1007/BF02666659}. Thus, it is paramount to comprehend the fundamental atomic-scale mechanisms and effects during film formation
to understand their impact on residual stress in thin-film coatings. Yet, determining the resulting residual stress due to thin-film coatings has proven difficult because of the variability in the processing and material characteristics (surface energy, diffusivity, growth rate, deposition flux, temperature, grain size, and morphology)~\cite{10.1063/1.4704683}. 
\par 
Several experimental techniques have been developed to measure the effect of residual stresses, including x-ray diffraction method~\cite{10.1116/1.573645}, Raman spectroscopy~\cite{10.1016/0038-1101(80)90164-1, 10.1016/j.jcrysgro.2005.03.056}, the wafer curvature method~\cite{10.1116/1.5011790}, and focused ion beam method~\cite{10.1116/1.5011790}. However, these techniques do not consider any stress relaxation brought on by lattice defects. They can also not differentiate between the separate contributions of thermal and lattice misfits to the developed stress and the need for a theoretical model to predict residual stresses. 
\par \sloppy Theoretical methods for estimating the residual stresses of coating-substrate structures have been developed by many researchers. Stoney proposed the first analytical equation for assessing thin coating stresses produced on a thick circular substrate~\cite{10.1098/rspa.1909.0021}. 
Though Stoney's equation~\cite{10.1098/rspa.1909.0021} and extended Stoney's equations~\cite{10.1017/CBO9780511754715, 10.1063/1.123722, 10.1016/S0022-5096(99)00070-8, 10.1063/1.1478137,10.1016/j.jmps.2018.06.021} can be easily applied to estimate the induced film stress using the radius of curvatures, these approaches are subjected to several assumptions such as uniform curvature and uniform stress distribution over the entire thin film system. As a result, different models have been developed to extend Stoney's original work. These extensions include, for instance, taking into account the impacts of film thickness and thin substrate~\cite{10.1017/CBO9780511754715, 10.1063/1.123722}, large deflection~\cite{10.1017/CBO9780511754715, 10.1063/1.123722, 10.1016/S0022-5096(99)00070-8}, non-uniform misfit strain~\cite{10.1016/j.tsf.2006.05.013, 10.1016/j.ijsolstr.2007.09.012}, plastic deformation~\cite{10.1063/1.1786339}, non-uniform thickness~\cite{10.1115/1.2745392}, non-isotropic stress~\cite{10.1063/1.1925328}, stress gradient~\cite{10.1088/0960-1317/16/2/024}, non-uniform temperature distributions~\cite{10.1016/j.jmps.2005.06.003}, structure dimensions~\cite{10.1063/1.2178400} and temperature mismatch~\cite{10.1080/01495739.2014.937249, 10.1016/j.engfailanal.2004.12.027, 10.1016/j.surfcoat.2008.06.178, 10.1016/S0040-6090(02)00699-5, 10.1016/j.matdes.2016.08.053}. However, all the aforementioned models are based on Stoney's formulation, which presumes a perfect interface. Because an ideal contact has no slip at the interface, the interfacial peeling and shear stresses are frequently negligible in an ideal interface~\cite{10.1115/1.2745387}.
\par For submicron film thickness, the lattice misfit might give rise to considerable stress, which is important to understand in multilayered hetero-epitaxial coatings~\cite{10.1016/j.surfin.2018.05.007}. Typically, when the lattice mismatch is small, films can develop pseudomorphically~\cite{10.1017/CBO9780511754715}. When the lattice misfit is smaller than $\sim$10\%, a coherent interface is predicted to form~\cite{10.1017/CBO9780511754715, 10.3390/met4030410, 10.1002/adfm.201000071} based on transmission electron microscopy (TEM) observations~\cite{10.1002/adfm.201000071, 10.1002/srin.201100086}. However, the true lattice misfit is different for the different film-substrate systems. The permitted lattice misfit for a given film-substrate system's pseudomorphic film development remains poorly understood. Above this threshold lattice misfit for pseudomorphic development, the elastic lattice strain cannot allow the lattice misfit, and dislocations form at the substrate-film interface to reduce the strain~\cite{10.1016/j.surfin.2018.05.007}. However, when ALD/MLD coatings are used, large lattice misfits exist between the film and substrate in a small length, generating large lattice strains and residual stresses in the interface. Unfortunately, the impact of lattice misfit on the development of stresses in coatings and the consequences for defect/dislocation formation at the interface is poorly understood.
\par To address the above-mentioned issues, we presented a bottom-up approach to calculate lattice misfit induced during epitaxial thin-film deposition and its effect on hetero-epitaxial residual stress. Owing to the potential applications of AZIB, we focus our study on Zn foils, with alumina and alucone coatings generated via ALD/MLD manufacturing techniques. We first use \textit{ab initio} simulations to understand the lattice misfit and chemo-mechanical stability of alumina/alucone coatings of several thicknesses in coherent and semi-coherent interfaces. These calculations are also used to predict the surface elastic constants of the thin films. An analytical model was developed to account for residual stresses in film and substrate due to lattice misfits in orthotropic hexagonal closed materials (hcp) using the \textit{ab initio} informed surface elastic values of the coating. The effect of plastic deformation and dislocation formation in the interface is also included when the lattice misfit of incoherent interfaces is large. The model is then benchmarked with finite element simulations and experimental results of Zn foils coated with alumina and alucone, finding reasonable agreements between experiments and our predictions. 

\section{\label{sec:Computational_and_Theoretical Method}Methods}
\subsection{\label{Density_Functional_Thoery_(DFT)_Modeling} Density Functional Thoery (DFT) modeling}
\par Density functional theory~\cite{Kohn_Walter} (DFT) simulations were conducted using the Vienna \textit{ab initio} simulation package (VASP)~\cite{10.1103/PhysRevB.47.558}. Electron wave functions were reported employing the projector augmented wave (PAW) method of Bl\"och~\cite{10.1103/PhysRevB.50.17953}, as implemented in VASP~\cite{10.1103/PhysRevB.59.1758}. The functional for the exchange-correlation energy ($\mathrm{E_{XC}}$) was formulated using the generalized gradient approximation (GGA) proposed by Perdew, Burke, and Ernzerhof (PBE)~\cite{10.1103/PhysRevLett.77.3865}. Due to the interest in understanding residual stresses in Zn cathodes used in AZIB, we modelled hexagonal closed-packed (hcp) Zn surfaces. A basal surface \hkl(0001) was built with three pairs of AB stacking Zn layers. The relaxed surface exhibited a surface energy of 42 $\mathrm{meV \cdot \text{\r{A}}^{-2}}$.
\par \sloppy Electronic configurations and energies were calculated using a self-consistent field iteration with a tolerance of $\mathrm{10^{-6}}$ eV. Six layers (two fully free layers and four layers with out-of-plane displacement fiexd) of Zn \hkl(0001) were used to model the surface with periodic boundary conditions applied on the supercell similar to a previous study~\cite{10.1021/acs.jpcc.2c06646}. Several simulations were carried out to verify that the surface energies were converged within 1 $\mathrm{meV \cdot \text{\r{A}}^{-2}}$ through a series of thorough convergence experiments with different vacuum gaps (e.g., 5, 10, 20, 30 \r{A}).  A vacuum gap of around 10 \r{A} was placed between the slabs to prevent inter-slab contact, compromising a good balance between accuracy and computational efficiency. All surface simulations were carried out with a 15 x 15 x 15 k-points mesh until the ground state energy difference was less than $\mathrm{1 ~meV}$ per atom. 
\begin{figure}[h]
\centering
\subfloat[]{
\label{}
\includegraphics[width=0.31\linewidth]{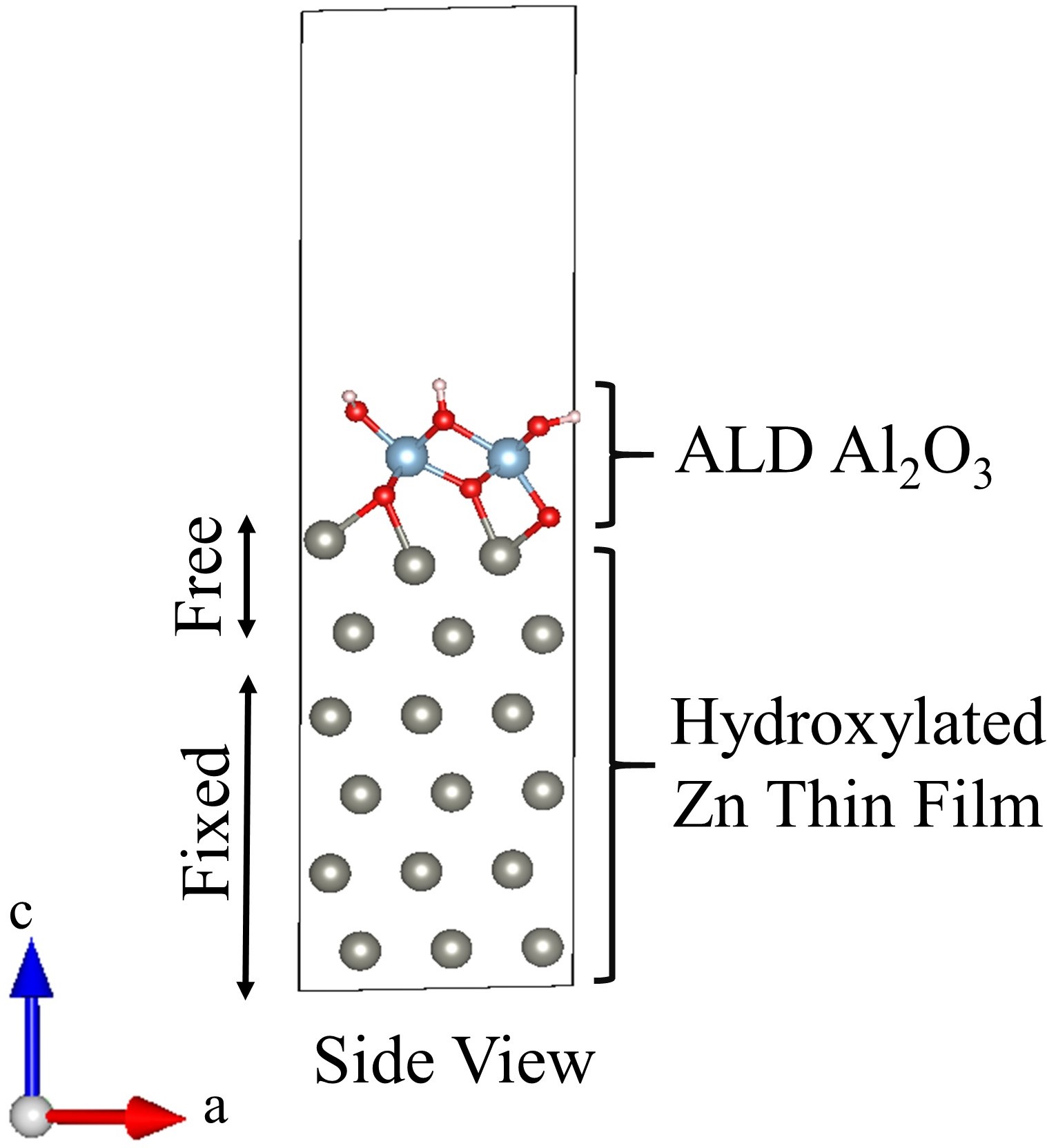}}
\qquad
\subfloat[]{
\label{}
\includegraphics[width=0.35\linewidth]{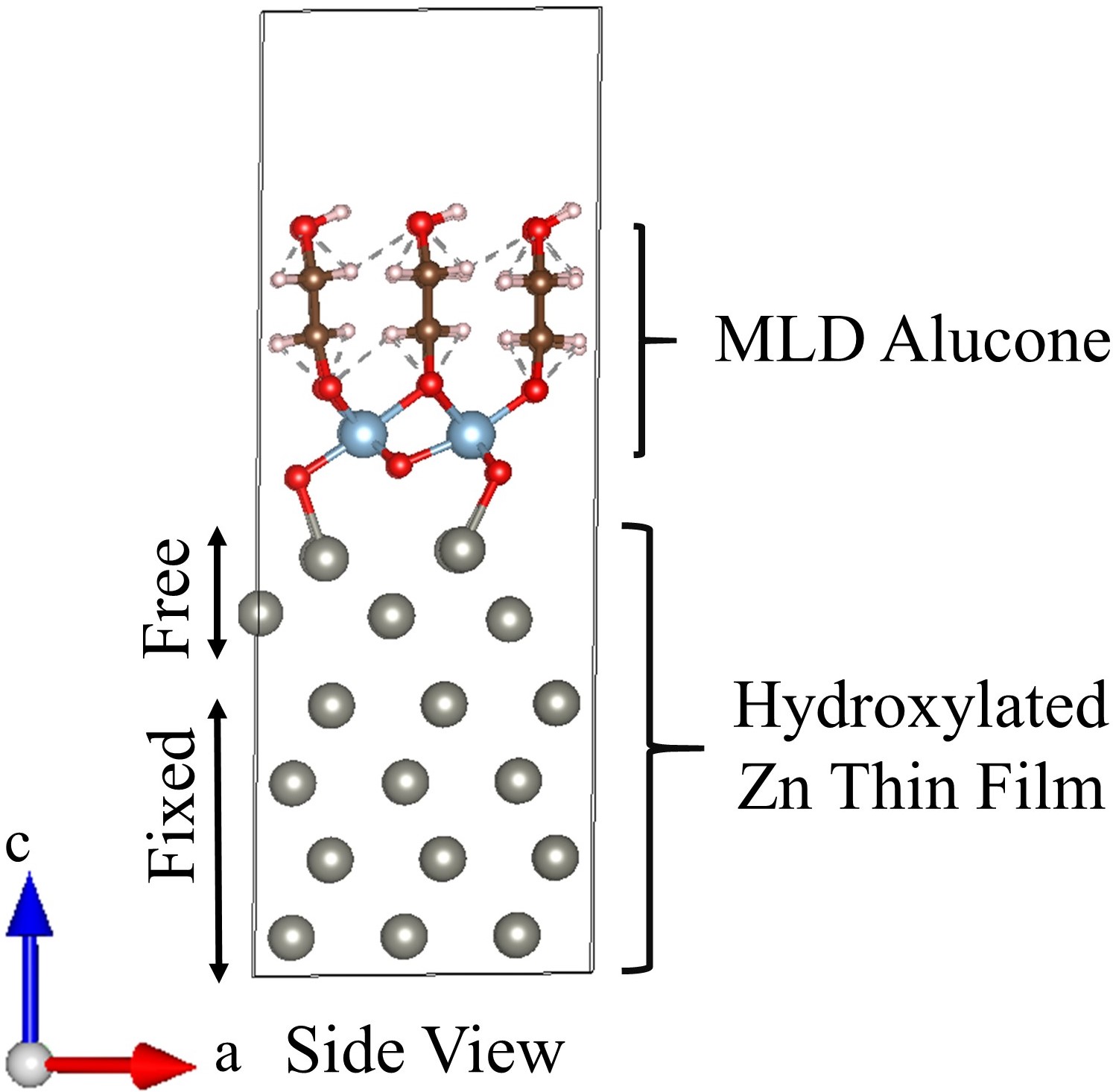}}
\qquad
\caption{Side view of (a) ALD $\mathrm{Al_{2}O_{3}}$ deposited hydroxylated Zn \hkl(0001) model and (b) MLD alucone deposited hydroxylated Zn \hkl(0001) model. Color code for the atoms: aluminum - blue, oxygen - red, hydrogen - pink, carbon - brown, and zinc - gray.}
\label{fig:ald_alumina_schematic}
\end{figure}
\subsection{\label{SurfaceElasticConstants} Surface properties}
To investigate the mechanical and chemical properties of (coated) surfaces, we proceed to compute the surface mechanical properties. The surface energy ($\gamma$)  is defined as the energy required to create a new surface in a bulk material. Conventionally, the surface energy is difficult to evaluate experimentally, but it can be easily computed computationally. Considering a computational cell containing $N_a$ atoms, the surface energy can be defined as 
\begin{equation} \label{eq:Gamma}
\gamma = \frac{1}{2A} \bigg( E_\textrm{surf} - N_a E_\textrm{cohesive} \bigg),
\end{equation}
where $E_\textrm{surf}$ is the total energy stored in the material with a new surface, and $E_\textrm{cohesive}$ is the cohesive energy per atom in the pristine material. Now, surfaces can be subjected to surface stresses, which can be evaluated using the \textit{surface elastic constants}. Shortly, the surface stresses can be evaluated for small deformations through the surface energy and strain as 
$
\boldsymbol \tau(\boldsymbol \epsilon) = \gamma \boldsymbol I+ \frac{\partial \gamma}{\partial \boldsymbol \epsilon},
$
~\cite{10.1103/PhysRevB.71.094104} where $\boldsymbol \epsilon$ is the infinitesimal strain tensor.  At zero strain, when the surface is relaxed, the surface energy and stress can be related through the strain as they are conjugate variables, i.e.,
 \begin{equation} \label{eq:ZeroStrain}
\boldsymbol \tau^0 = \bigg[ \frac{\partial \gamma}{\partial \boldsymbol \epsilon} \bigg]_{\boldsymbol \epsilon= \boldsymbol 0}.
\end{equation}
The term $\boldsymbol \tau^0$ is known as the \textit{surface stress} and can be related to the strain via a fourth-order surface elastic tensor ($C_{ijkl}^\textrm{s}$) that defines the properties of the surface for small deformations~\cite{10.1021/acs.jpcc.2c06646}
 \begin{equation} \label{eq:SurfaceElasticConstants}
C_{ijkl}^\textrm{s} = \frac{\partial \tau_{ij}^0}{\partial \epsilon_{kl}} = \bigg[ 2 \gamma \delta_{ik}  \delta_{jl} + \delta_{ij} \frac{\partial \gamma }{\partial \epsilon_{kl} } + \delta_{ij} \frac{\partial^2 \gamma }{\partial \epsilon_{ij} \epsilon_{kl} }  \bigg].
\end{equation}
The surface elastic constant tensor $C_{ijkl}^s$ will be used to compute the film biaxial modulus and residual stresses developed due to the coating. Notice that, the surface elastic constant tensor can also be collapsed into a two-dimensional array using Voight's notation. 
\subsection{\label{Analytical_Modeling_of_Residual_Stress_Evolution_in_Thin_Film_Systems}Analytical modeling of residual stress evolution in thin films}
\par The growth of films depends on interface arrangements between the lattices of the heteroepitaxial film and substrate. The epitaxial film's misfit strain can be computed by~\cite{10.1080/01418619708223740, 10.1016/j.surfin.2018.05.007} 
  \begin{equation} \label{eq:misfit_strain}
  \begin{split} 
   \varepsilon_{\textrm{m}} =
   \frac{a_{{\textrm{f}}}-a_{{\textrm{s}}}}{a_{{\textrm{s}}}},
   \end{split}
   \end{equation} 
where $a_{{\textrm{s}}}$ and $a_{{\textrm{f}}}$ are the lattice constants of the substrate and film, respectively. Here,  $\varepsilon_{\textrm{m}}$  presents the growth direction, which can be lateral/$a-$lattice constant mismatch ($\varepsilon_{\textrm{m}}^{a}$) or $c-$lattice mismatch for growth in \hkl[0001] direction ($\varepsilon_{\textrm{m}}^{c}$). 
A coating on a substrate will often be under compression if the film's lattice parameter is larger than the substrates ($\varepsilon_{\textrm{m}}>0$). Conversely, it will be under tension if the lattice parameter is less than the substrates ($\varepsilon_{\textrm{m}}<0$). Therefore, the lattice misfit and the strain resulting from the mismatch, referred to as mismatch strain, exhibit opposing directions, $\it{i.e.}$, mismatch strain is negative (compressive) when $\varepsilon_{\textrm{m}}>0$.
\subsubsection*{\label{}Effect of misfit dislocation}
\par When a semi-coherent hetero-epitaxial film grows in the substrate-film system, misfit dislocation will form. The residual homogenous strain in the partly relaxed film is~\cite{10.1007/BF02666659}
  \begin{equation} \label{eq:misfit_dislocation}
  \begin{split} 
   \varepsilon_\textrm{mf} = \varepsilon_\textrm{m}-\frac{b}{S},
   \end{split}
   \end{equation} 
where $b$ is the magnitude of the Burgers vector, and $S$ is the average distance between a square array of dislocations. The average distance between dislocations ($S$) is connected to the density of dislocation ($N$) by $N =\frac{1}{S^{2}}$~\cite{10.1002/1521-3951(199709)203:1<79::AID-PSSB79>3.0.CO;2-A,10.1016/0040-6090(92)90060-O}.
The corresponding strain energy is $E_\textrm{m}=Mh_\textrm{f}\left(\varepsilon_\textrm{m}-\frac{b}{S}\right)^{2}$ where $M$ is the effective modulus and ${h_\textrm{f}}$ is the film thickness.
\subsubsection*{Equilibrium spacing of dislocation}
\par The total energy per unit area ($E_\textrm{T}$) for a film over the critical thickness ($h>h_\textrm{c}$) may be approximated as the summation of three terms: the strain energy of the partly relaxed film ($E_\textrm{m}$), the energy connected to the misfit dislocations ($E_\textrm{d}$), and the dislocation core's energy ($E_{\textrm{core}}$), i.e., 
  \begin{equation} \label{eq:equilibrium_specing_of_dislocation}
  \begin{split} 
   E_\textrm{T}= E_\textrm{m}+E_\textrm{d}+E_\textrm{core}.
   \end{split}
   \end{equation} 
\par In Eq. (\ref{eq:equilibrium_specing_of_dislocation}), $E_\textrm{m}$ reflects the elastic strain energy per unit area owing to the induced strain coming from the layer's lattice mismatch, which is partly relieved by the growth of dislocations in accordance with~\eq{misfit_dislocation}. $E_\textrm{d}$ describes the impact of dislocation's self-energy. By minimizing $E_\textrm{T}$, equilibrium spacing of dislocations can be attained. According to \eq{misfit_dislocation}, and~\eq{equilibrium_specing_of_dislocation}, $E_\textrm{m}$ increases if misfit strain and film thickness increases. Thus, the self-energy of dislocations controls the minimization of $E_\textrm{T}$. Therefore, by minimizing $E_\textrm{T}$ with regard to the separation distance of dislocation ($S$), the equilibrium configuration of an array of dislocations is identified (for a particular film thickness). 
\par  For dislocation analysis of a hexagonal system, it is crucial to add the orientational dependence in the dislocation energy ($E_\textrm{d}$). Here, $E_\textrm{d}$ can be expressed as~\cite{10.1115/1.3167075}
  \begin{equation} \label{eq:Dislocation_energy}
  \begin{split} 
   E_\textrm{d} = \frac{1}{4\pi}b^{2}k(\phi) \ln \bigg(\frac{R}{r_{0}}\bigg),
   \end{split}
   \end{equation} 
where ${k}$ is the anisotropic energy factor, and  $r_\textrm{0}$ and $R$ are the inner and outer radii of the elastic field surrounding the dislocation. The angle between the Burgers vector and the dislocation's direction is denoted by the Greek letter $\phi$. 
\par Predominant slip systems in hcp crystals like Zn structures are \hkl<11-20> for \hkl{0001} basal (at room temperatures) and prismatic \hkl{10-10} (at high temperatures and less predominant) family of planes~\cite{10.1115/1.3167075, 10.1016/j.jmps.2017.12.009}. Burgers vector of stable dislocations in Zn crystal are typically $\frac{1}{3}$\hkl<11-20>  where $|b| = a $. It is worth noting that we consider the dislocation core's radius as ${r_\textrm{0}=b}$ where linear elasticity is not valid, which is also supported through large-scale simulations of dislocations in hcp materials \cite{10.1016/j.jcp.2020.109249}. $R$ is usually associated with the average distance between dislocations. The logarithmic term includes the cut-off radius for the dislocation's elastic field. In particular, because the film's surface is traction-free, the outer cut-off radius increases with film thickness (${h_\textrm{f}}$).
For misfit dislocations with a square array, the misfit line length per unit area is ${2/S}$; hence, the dislocation energy per unit area is
  \begin{equation} \label{eq:Dislocation_energy_per_unit_area}
  \begin{split} 
   E_\textrm{d} = \frac{1}{4\pi}b^{2}k(\phi)\frac{2}{S}\ln\bigg(\frac{\beta h_{\textrm{f}}}{b}\bigg).
   \end{split}
   \end{equation} 
Here, $\beta$ is a constant. Therefore, the total energy per unit film area:
  \begin{equation} \label{eq:total_energy}
  \begin{split} 
   E = Mh_\textrm{f}\left(\varepsilon_{\textrm{m}}-\frac{b}{S}\right)^2 + \frac{1}{4\pi}b^{2}k(\phi)\frac{2}{S} \ln \bigg(\frac{\beta h_{\textrm{f}}}{b} \bigg) + E_\textrm{core}.
   \end{split}
   \end{equation} 
\begin{figure}[h]
\centering
\includegraphics[width=0.33\linewidth]{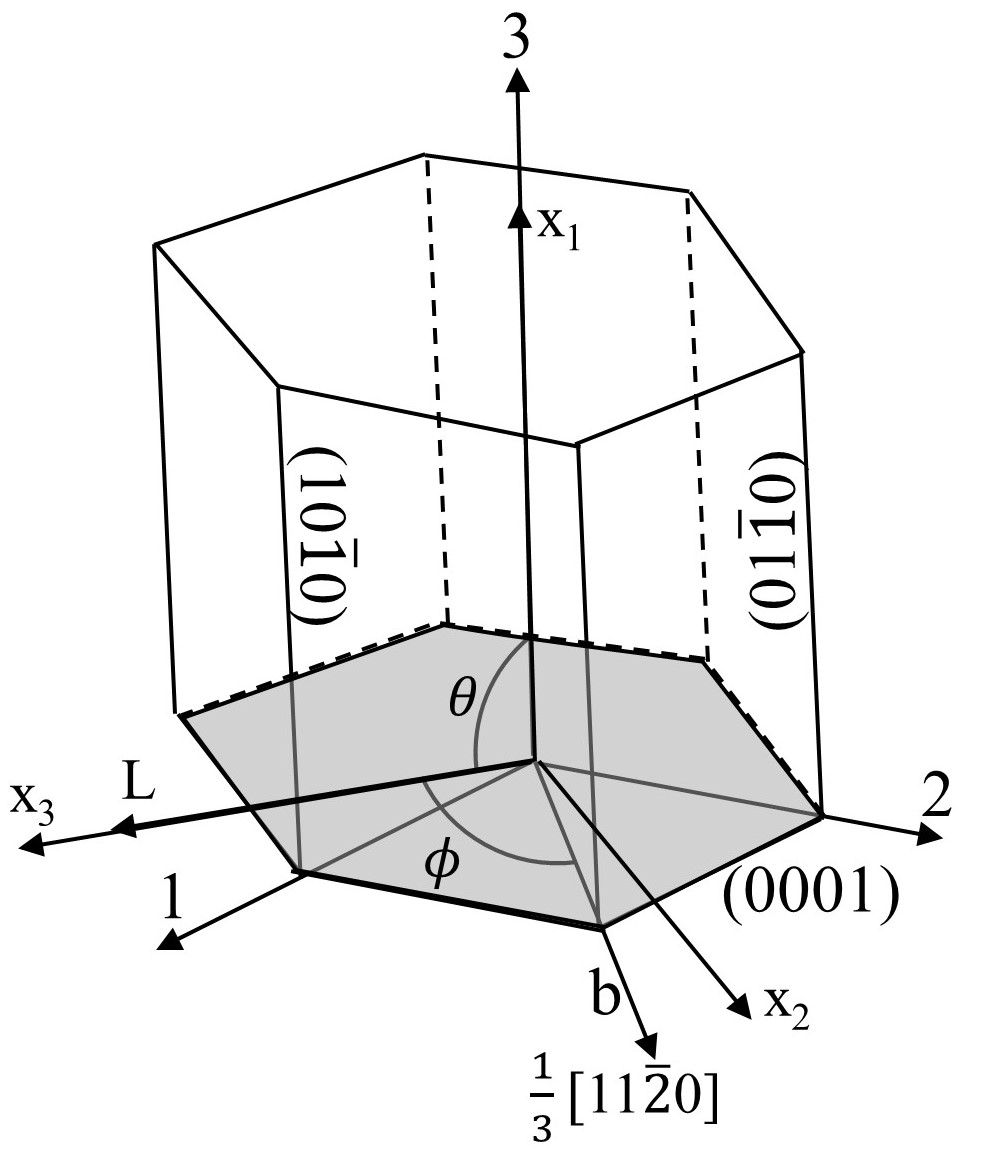}
\caption{Schematic of dislocation in the basal plane  \hkl(0001) in a hcp system. Here, L is the dislocation line, and b is the Burgers vector. 1-2-3 defines the crystallographic coordinate system and $x_1-x_2-x_3$ is the dislocation coordinate system.}
\label{fig:dislocation_schematic}
\end{figure}
\par In general, the energy factor ($k$) is an anisotropic quadratic combination of the components of the Burgers vector ($b_i$):~\cite{10.1002/pssa.2210350240} 
  \begin{equation} \label{eq:energy_factor_general form}
  \begin{split} 
    kb^2 = k_{ij}b_{i}b_{j}\:\: (i,j = 1,2,3).
   \end{split}
   \end{equation} 
When the $x_2$ axis is in the basal plane, and the $x_3$ axis is parallel to the dislocation line in the dislocation coordinate system as shown in~\fig{dislocation_schematic}, the expression for energy factor ($k$) for a dislocation in an hcp Zn crystal is 
\cite{10.1016/0025-5416(70)90075-3}  
  \begin{equation} \label{eq:energy_factor_for_hexagonal}
  \begin{split} 
   kb^2 = k_{11}b_{1}^2+k_{22}b_{2}^2+k_{33}b_{3}^2+2k_{13}b_{1}b_{3}.
   \end{split}
   \end{equation} 
Here, $b_1, b_2$, and $b_3$ are the components of Burgers vector in the dislocation coordinate system. Axes 1 and 2 of the crystallographic coordinate system are in the basal plane, and axis 3 is oriented along the c-axis of a primitive cell as shown in~\fig{dislocation_schematic}. Therefore, only elastic constants ($C_{ij}$) of the hcp Zn crystal and the angle ($\theta$) between the dislocation line and the c-axis are the variables that affect the coefficients $k_{ij}$. 
\par Since the basal plane of hexagonal crystals is isotropic~\cite{10.1115/1.3167075}, the energy factor ($k$) should have the same dependence on $\phi$ as in isotropic system ($k = \frac{\mu}{1-\nu}\sin^2{\phi}+\mu \cos^2{\phi}$), as seen below:~\cite{10.1002/pssa.2210350240}
  \begin{equation} \label{eq:energy_factor_for_hexagonal_in_terms_of_elastic_constant}
  \begin{split} 
   k = \left[\frac{(C_{11}-C_{12})C_{44}}{2}\right]^{1/2} \cos^{2}{\phi} + (\lambda^2C_{33}+C_{13})\times\\ \left[\frac{C_{44}(\lambda^2C_{33}-C_{13})}{C_{33}(\lambda^2C_{33}+C_{13}+2C_{44})}\right]^{1/2} \sin^{2}{\phi}.
   \end{split}
   \end{equation} 
Here, $\lambda^2 = \big(\frac{C_{11}}{C_{33}} \big)^{1/2}, ~\theta = \pi/2, b_{1} =0, ~b_{2}=b \sin(\phi), b_{3}=b \cos(\phi)$. 
\subsubsection*{Critical thickness}
\par An epitaxial layer can develop pseudomorphically on a substrate below a specific thickness, known as the critical thickness ($h_\textrm{c}$), but thicker epilayers $h>h_\textrm{c}$ result in a relaxation of misfit strain through plastic deformation. 
At the critical film thickness ($h_\textrm{c}$), initial dislocation introduces no change in the film's energy. Misfit dislocation will develop if the energy is reduced in this way~\cite{10.1007/BF02666659}: 
  \begin{equation} \label{eq:Critical_thickness_condition}
  \begin{split} 
   \frac{dE_\textrm{T}}{d(1/S)}\Bigg|_{b/S=0} = 0.
   \end{split}
   \end{equation} 
Consequently, the critical thickness at which it is favorable to develop misfit dislocations is as follows: 
  \begin{equation} \label{eq:Critical_thickness}
  \begin{split} 
   \frac{h_\textrm{c}}{ \ln\left(\frac{\beta h_\textrm{c}}{b}\right)} = \frac{kb}{4\pi M\varepsilon_m}
   \end{split}
   \end{equation} 
If $h<h_{c}$, misfit dislocations cannot grow and are unstable.
On the other hand, $h>h_{c}$, Misfit dislocations are stable, although they do not always grow (metastable state).  ~\eq{Critical_thickness} provides an implicit equation for the critical thickness ($h_{c}$), since it is also a function in the argument of the logarithm term. This is analogous to Matthews and Blakeslee's equation where $h_{c} = -AW\left(-\frac{1}{AB}\right)$, where $A = \frac{kb}{4\pi M\varepsilon_m}$ and $B = \frac{\beta}{b}$ are two material constants, and $W$ denotes the Lambert $W-$function~\cite{10.1016/S0022-0248(02)00941-7, 10.1007/s10853-010-5026-y}. Though the iterative numerical approach for solving~\eq{Critical_thickness} is straightforward, the numerical solution heavily depends on the initial guess, which can lead to local minima. Lambert's $W-$function $W(x)$ is a multi-valued complex function with unlimited branches, but just two of them have real roots for $-1/e\leq x\leq 0, ~x \in  \mathbb{R}$, namely $W_{0}(x) (-1\leq W(x))$ and $W_{-1}(x) ( W(x) \leq -1)$. The prefactor and the $W$-function argument always have the same sign, which might be either positive or negative. However, the critical thickness $h_{c}$ always has to be positive.
\subsubsection*{Residual stress in deposited films}
\par
\begin{figure}[h]
\centering
\includegraphics[width=0.50\linewidth]{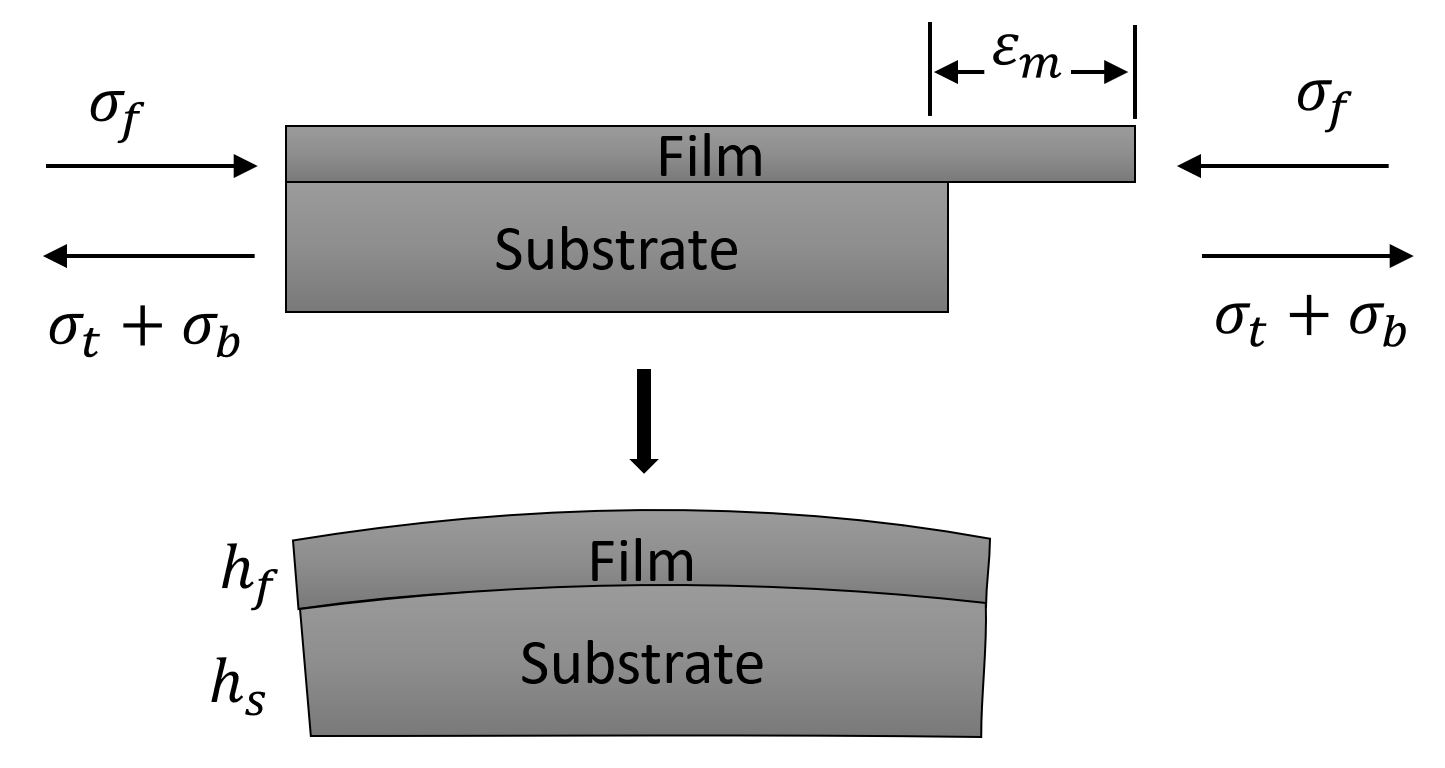}
\caption{Schematic of a substrate-film system with induced misfit strain and its effects.}
\label{fig:continuum_schematic}
\end{figure}

 The substrate-film interaction is illustrated schematically in~\fig{continuum_schematic}. Elastic stresses will develop in both the substrate and the film to fulfill the need for strain compatibility at the substrate-film interface since they are mutually constricted.
The compatibility equation for total misfit strain can be presented as
  \begin{equation} \label{eq:compatibity_eqn_for_total_misfit_strain}
  \begin{split} 
   \varepsilon_\textrm{T} &= (\varepsilon_\textrm{sub}^\textrm{epi}-\varepsilon_\textrm{film}^\textrm{epi})+(\varepsilon_\textrm{sub}^\textrm{Th}-\varepsilon_\textrm{film}^\textrm{Th})\\
   &= \varepsilon_\textrm{m}+\Delta \alpha \Delta T
   \end{split}
   \end{equation} 
Where $\varepsilon_\textrm{sub}^\textrm{epi}$ and $\varepsilon_\textrm{film}^\textrm{epi}$ are epitaxial misfit strains and $\varepsilon_\textrm{sub}^\textrm{Th}$ and $\varepsilon_\textrm{film}^\textrm{Th}$ are thermal strains in substrate and film, respectively. Lattice misfit strains is predicted from~\eq{misfit_strain} and~\eq{misfit_dislocation} whereas thermal strains can be estimated from $ \varepsilon_\textrm{sub}^\textrm{Th} = \alpha_\textrm{sub}\Delta T $ and $ \varepsilon_\textrm{film}^\textrm{Th} = \alpha_\textrm{film}\Delta T $. Here, $\alpha_\textrm{film}$ and $\alpha_\textrm{sub}$ are coefficients of thermal expansion (CTE) of the thin film and substrate, respectively.
\par Film anisotropy and film stress are strongly connected for a generic mismatch strain. Here, the definition of mismatch strain only considers the components of $\varepsilon_{11}^\textrm{m}, ~\varepsilon_{22}^\textrm{m}, ~\varepsilon_{12}^\textrm{m}$ with respect to the global frame since this only includes straining in directions parallel to the interface~\cite{10.1017/CBO9780511754715}. 
 Assuming the film surface at $x_{3} = h_\textrm{f}$ is traction-free for any mismatch strain, $\sigma_{33}^\textrm{m}=\sigma_{23}^\textrm{m}=\sigma_{13}^\textrm{m}=0$.
\par For an hcp structure, assuming material coordinates coincide with global coordinate systems (rotation matrices are identity matrices, ~\eq{eqns_for_coordiate_transformation} for details), the effective biaxial modulus (M) are 
  \begin{equation} \label{eq:effective_biaxial_modulus}
  \begin{split} 
  M_{11} = C_{11}+C_{12}-\frac{C_{13}^{2}+C_{23}C_{13}}{C_{33}}\\
  M_{22} = C_{12}+C_{22}-\frac{C_{23}^{2}+C_{23}C_{13}}{C_{33}}
   \end{split}
   \end{equation} 
Using surface elastic constants ($C_{ij}^\textrm{S}$), the effective biaxial modulus of~\eq{effective_biaxial_modulus} for thin film can be expressed as $M_{ij}^\textrm{T}=M_{ij}+(h_\textrm{f}/2)C_{ij}^\textrm{S}$. From the schematic diagram in~\fig{continuum_schematic}, the film will be in compression, and the substrate will be in tension and bending deformation. For the substrate and film to fit together perfectly by allowing the misfit $\varepsilon_\textrm{f}^\textrm{elastic}=\varepsilon_\textrm{mf}+\varepsilon_\textrm{s}^\textrm{elastic}$.
Therefore, the film stresses can be expressed as 
  \begin{equation} \label{eq:film_stress}
  \begin{split} 
  \sigma_{ij}^\textrm{film} = \frac{\delta_{ij}\varepsilon_\textrm{m}}{\left(\frac{1}{M_{ij}^\textrm{T}}+\frac{4h_\textrm{f}}{h_{s}M_{ij}}\right)}\:\: (i,j=1,2)
   \end{split}
   \end{equation} 
Here, $\delta_{ij}$ is the Kronecker delta. The total in-plane forces must equal zero in order to meet the equilibrium condition, i.e., $\sigma^\textrm{sub}h_{s}+\sigma^\textrm{film}h_\textrm{f}=0$. Therefore, the substrate's stresses can be expressed as
  \begin{equation} \label{eq:substrate_stress}
  \begin{split} 
  \sigma_{ij}^\textrm{sub} = -\frac{\delta_{ij}\varepsilon_{m}\frac{h_\textrm{f}}{h_\textrm{s}}}{\left(\frac{1}{M_{ij}^\textrm{T}}+\frac{4h_\textrm{f}}{h_\textrm{s}M_{ij}}\right)}
   \end{split}
   \end{equation} 
\par Here, both ~\eq{film_stress}  and \eq{substrate_stress} include the effect of surface elastic constants (surface stress) for orthotropic systems in the biaxial stresses.
\subsection{\label{sec:Finite_Element_Modeling}Finite Element Modeling}
\par In the aforementioned continuum modeling, the stress states are considered homogeneous over the whole substrate-film system. To evaluate the validity of this hypothesis comprehensively, we used finite element (FE) modeling to understand the evolution of the film and substrate stresses and the boundary effects. The force induced on the substrate bound to the film changes shape according to the kind of mismatch stress: for tensile mismatch stress, it evolves as concave down, and for compressive mismatch stress, it evolves as convex up~\cite{10.1016/j.tsf.2011.04.220}. As this deformation takes place, additional shear stresses at the interface need to develop to make the deformation compatible. Thus, shearing stresses at the substrate-film interface are estimated using FE. 
A time-independent linear elastic and anisotropic material model simulated the film and substrate. The elastic constants for alucone film and substrate were obtained from our previous work ~\cite{10.1021/acs.jpcc.2c06646}. The elastic constants for alumina are obtained from our DFT simulations and presented in Section \ref{sec:thin_film_properties}.  The film and substrate were in contact using tied conditions, ensuring the no-slip condition between the two interfaces. The misfit was generated by setting up a lattice mismatch for alumina of $\epsilon = - 0.032$, obtained from our \textit{ab initio} simulations.  To generate this misfit, a thermal expansion was used, setting a coefficient of thermal expansion of $\alpha_\textrm{T} = $ -0.032, and a change of temperature of $\Delta T = 1^\circ$ C, which adds the necessary force vector to the equilibrium equations ~\cite{10.1016/B978-0-08-098356-1.00009-6}. While our FE analysis neglects intrinsic stresses, these are included in the elastic properties used for thin films, including surface effects. All finite element analyses (FEA) were performed using the ABAQUS 6.24 software. Different film thicknesses were used for the FE model in this investigation (film thickness varied from 10 to 100 times thinner than the substrate). The epitaxial film thickness varied from  5 to 50 nm, including 10, 20, 30, and 40 nm cases. The thickness of the substrate measured from the substate-film contact is 500 nm, and the width is 1500 nm. 
\subsection{\label{sec:ALD_MLD_coating} Experimental assesement of stresses in ALD/MLD coated surfaces}
\begin{figure}[h]
\centering
\subfloat[]{
\label{}
\includegraphics[width=0.40\linewidth]{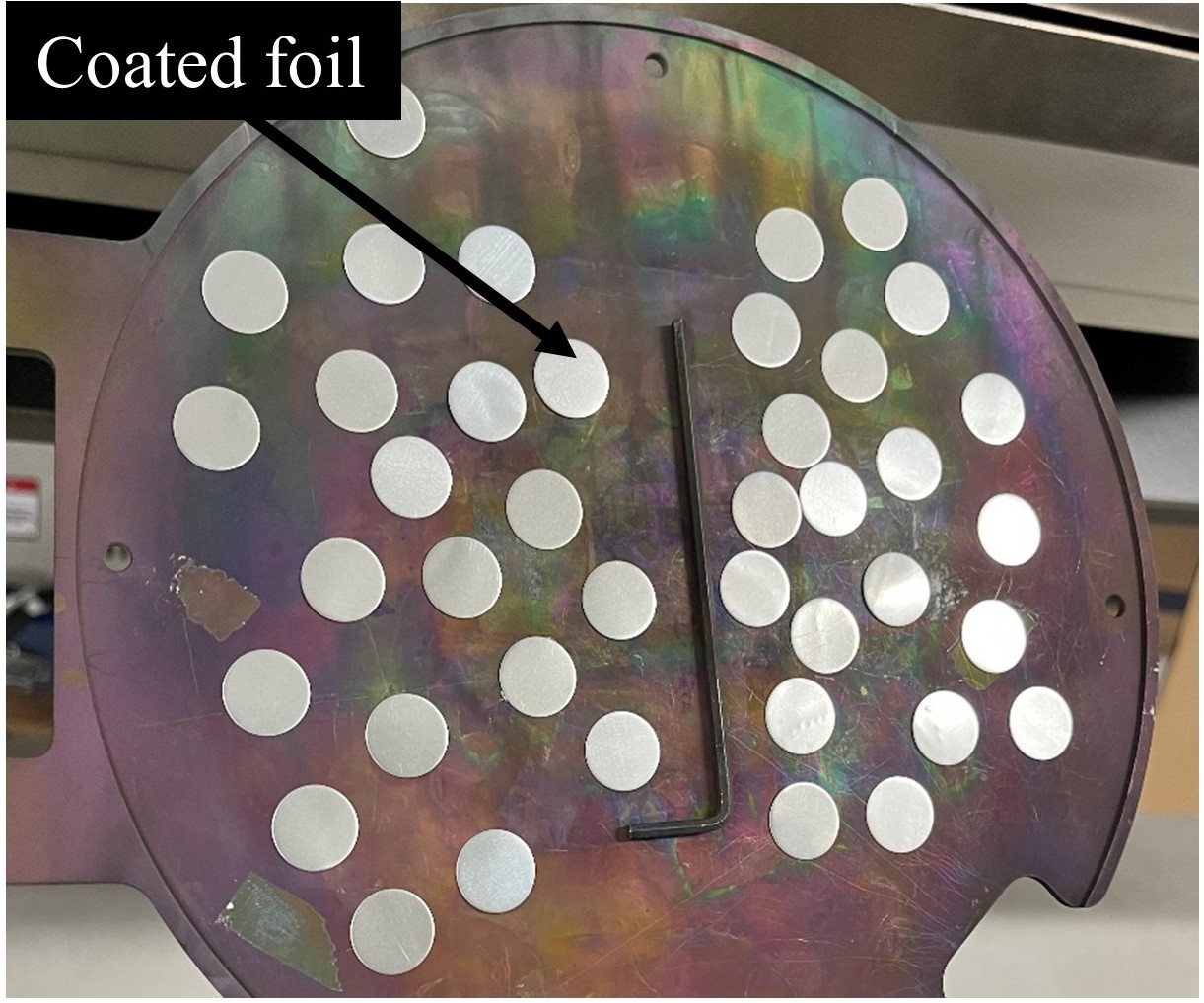}}
\qquad
\subfloat[]{
\label{}
\includegraphics[width=0.38\linewidth]{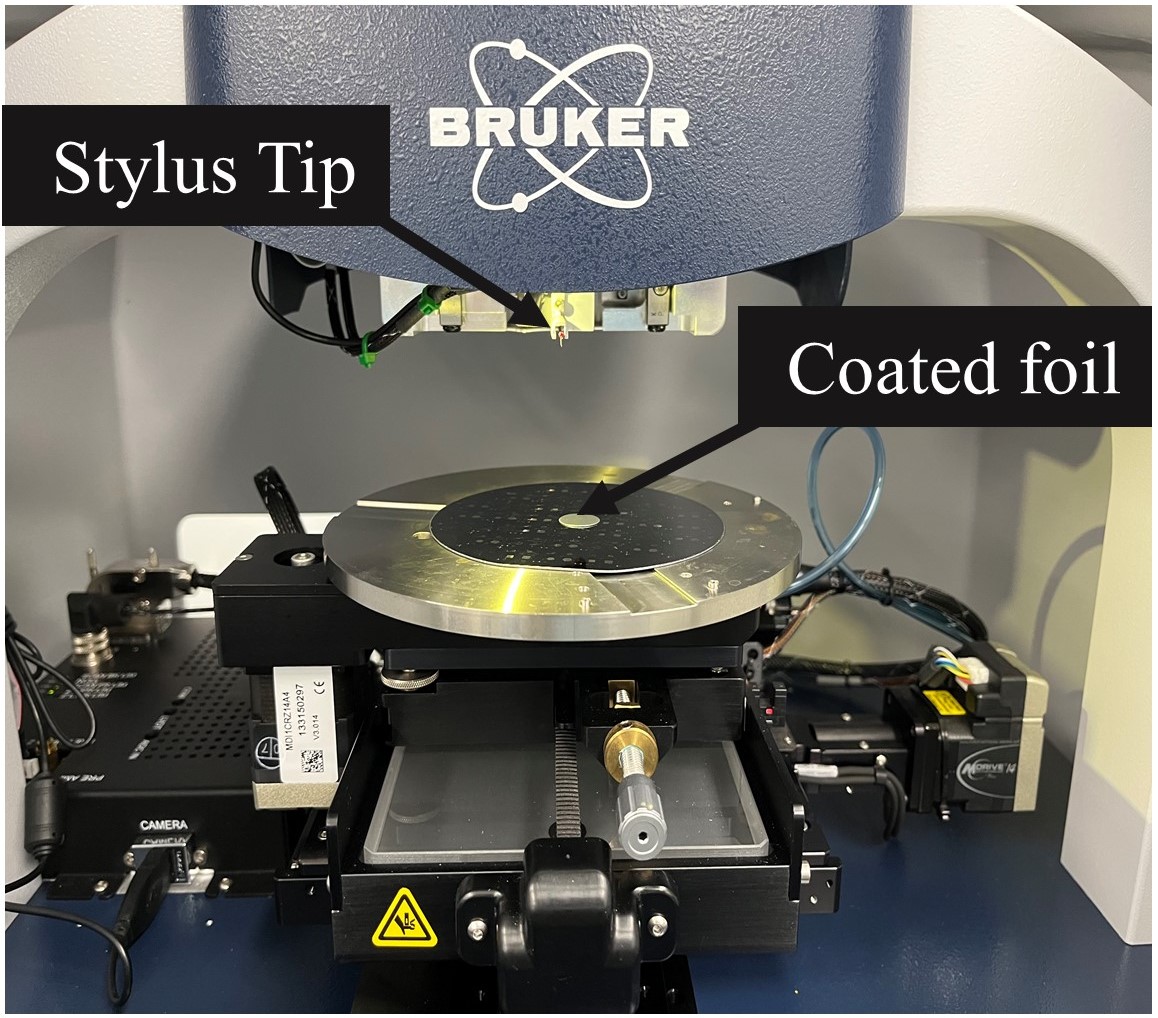}}
\qquad
\subfloat[]{
\label{}
\includegraphics[width=0.40\linewidth]{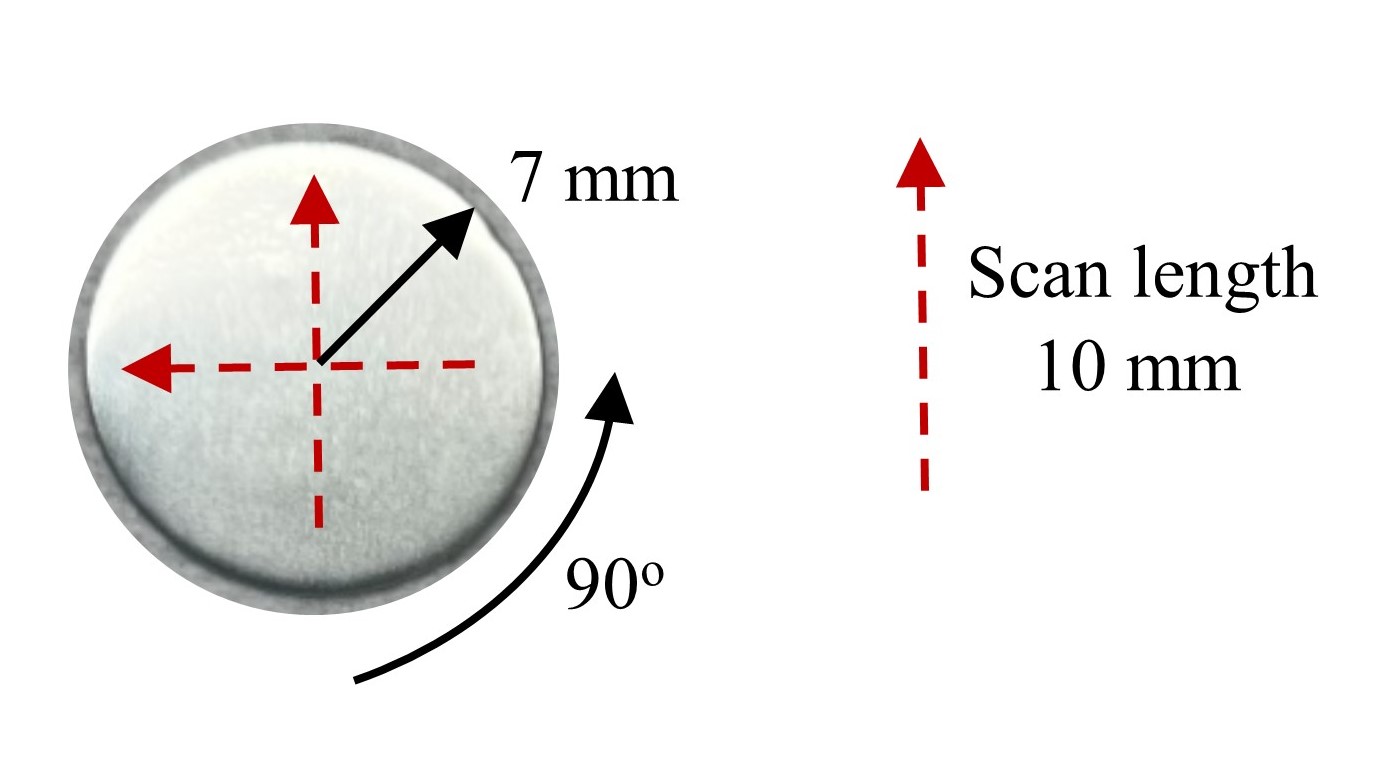}}
\caption {
\label{fig:experiment_images} (a) ALD/MLD coated zinc foils, (b) scanning profilometer setup for surface roughness and curvature measurement, and (c) scan length and direction on coated zinc foils. }
\end{figure}

We performed a series of experimental tests to evaluate the residual stresses developed by ALD/MLD manufacturing processes. Alumina and alucone coatings were applied to Zn foils with different coating layers, including 1, 5, and 10 ALD/MLD cycles (layers), to assess the dependence of the coating thickness. Zn thin foils of 0.1 mm in thickness and 14 mm in diameter were used. These thin foils were cleaned in deionized water and isopropanol, followed by natural air drying. Subsequently,  alumina coating on the Zn foils was carried out at 100$^\circ$C by alternatively providing trimethylaluminum (TMA) and $\mathrm{H_2O}$, whereas alucone coating was performed by providing TMA and ethylene glycol (EG: $\mathrm{HO-CH_2-CH_2-OH}$) alternatively into a commercial ALD system (GEMStar XT Atomic Layer Deposition System). During the ALD process, the chamber's pressure remains constant at around 500-700 mTorr, when a 200 ms pulsing of the TMA occurs. The chamber is then unperturbed for 5 seconds. After this, it is open for 15 seconds to release the extra byproducts from the chamber. The same procedure is used for pulsing EG, but the pulse duration is 21 ms. The resulting coated foils are shown in~\fig{experiment_images}(a). The coated foils were scanned with a DektakXT stylus profiler in a cleanroom as shown in~\fig{experiment_images}(b) and (c) with a scan length of 10 mm, scan duration of 60 seconds, and stylus force of 6 mg. The coated samples were rotated 90 degrees to capture the average radius of curvature of that particular sample, and at least two different samples have been chosen here as shown in~\fig{experiment_images} (c).
\section{\label{sec:Results_and_Discussion}Results and Discussion}
Although it is recognized that thin films by ALD are efficacious in enhancing the performances of different systems, the chemo-mechanical stability of the ALD alumina thin film method on the Zn surface remains unclear. Chemo-mechanical stability analysis of ALD alumina has been presented here using charge density, and Bader charges analysis, whereas the chemo-mechanical stability of MLD alucone has been studied previously~\cite{10.1021/acs.jpcc.2c06646}.
\subsection{\label{sec:thin_film_properties}Thin Film Properties}
\begin{figure}[h]  
\centering
\subfloat[]{
\label{}
\includegraphics[width=0.38\linewidth]{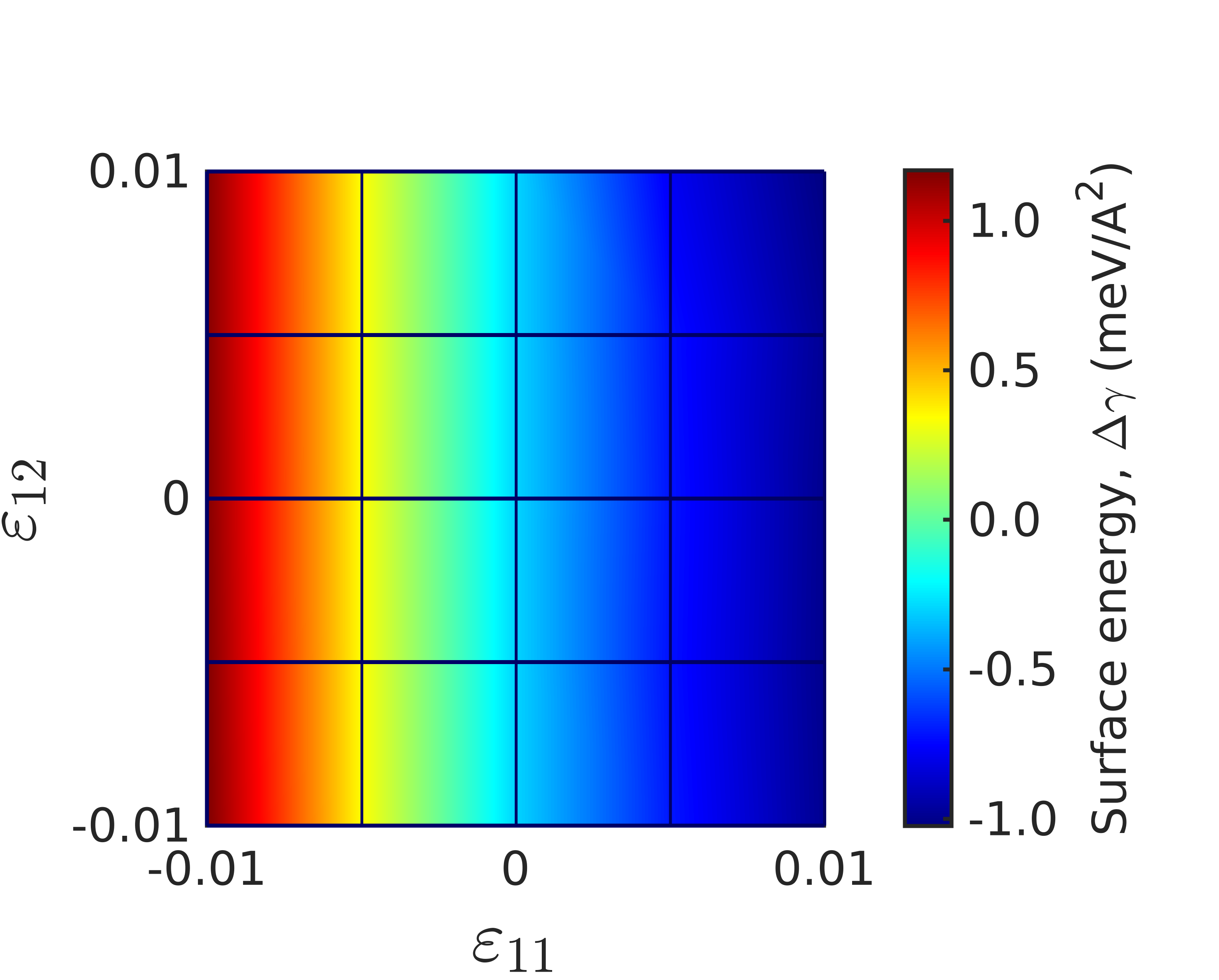}}
\qquad
\subfloat[]{
\label{}
\includegraphics[width=0.38\linewidth]{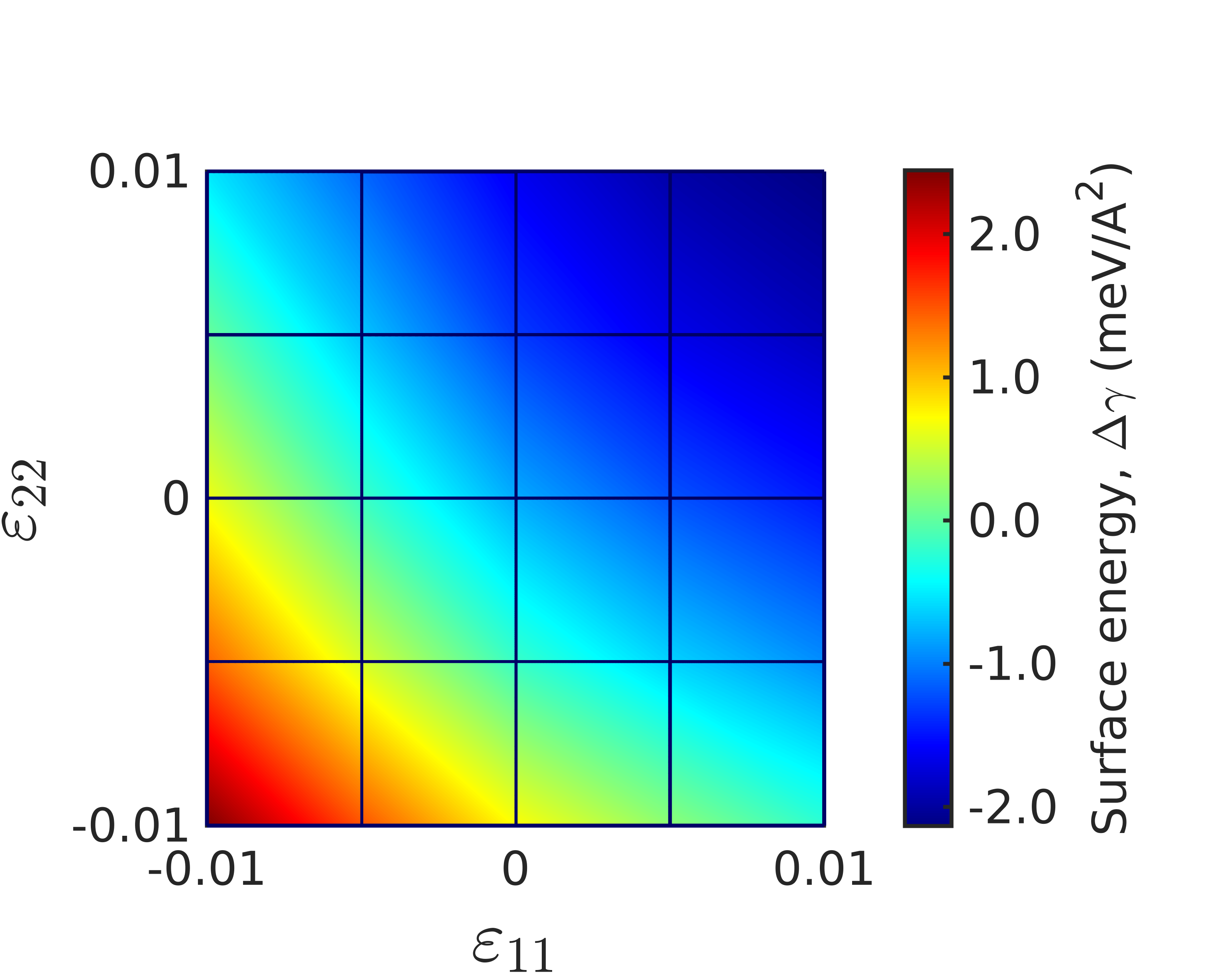}}
\qquad
\subfloat[]{
\label{}
\includegraphics[width=0.38\linewidth]{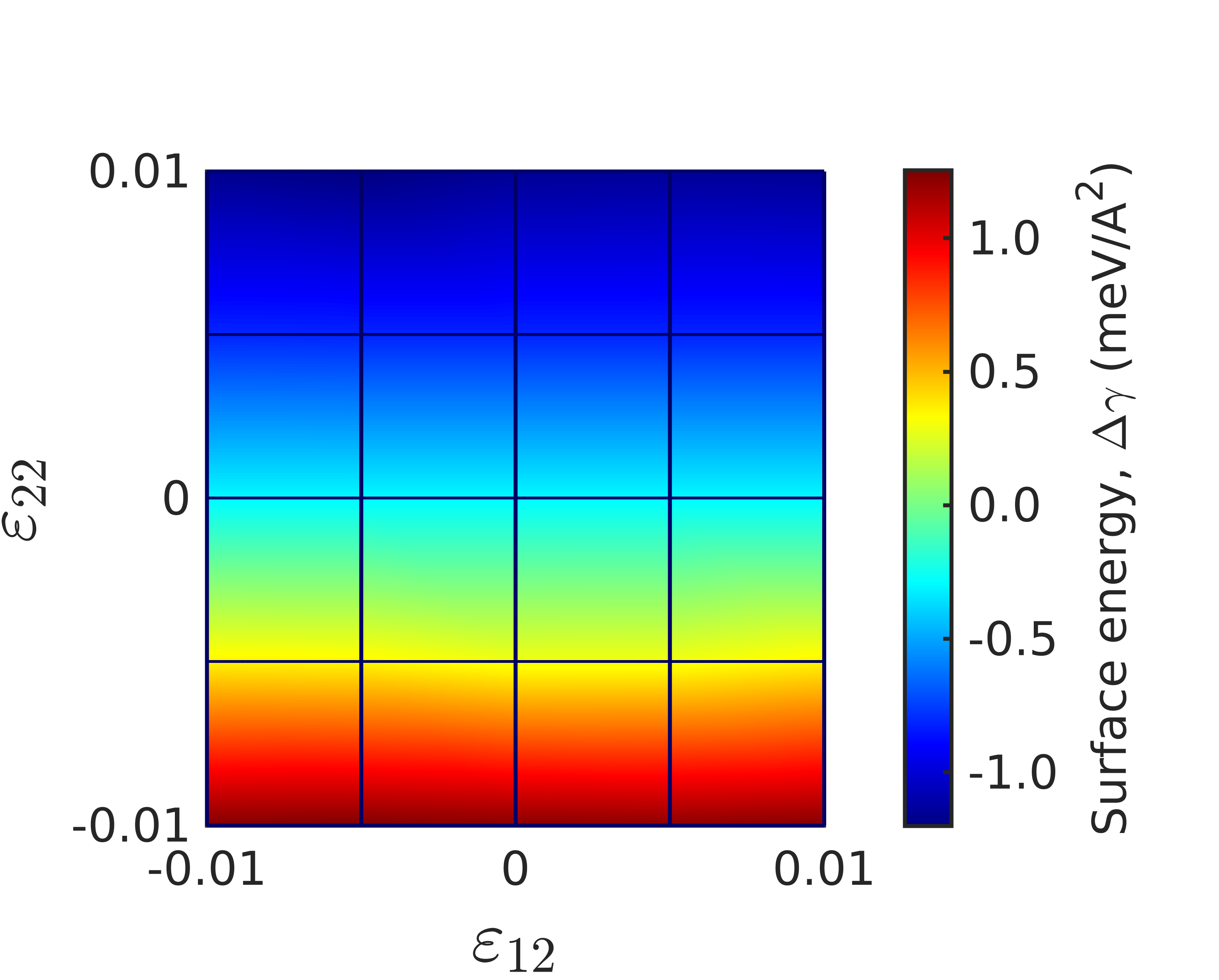}}
\qquad
\caption{\label{fig:surface_energy_with_strain} Variation of the surface energy in contour plots with different biaxial strain for ALD alumina coated hydroxylated Zn \hkl(0001). Reference surface energy at zero strain is $\mathrm{\gamma_{0} = 0.623}$ $\mathrm{eV/\text{\r{A}}^{2}}$. (a) Change of the surface energy as a function of $\varepsilon_{11}$ \textit{vs.} $\varepsilon_{12}$, (b) as a function of $\varepsilon_{11}$ \textit{vs.} $\varepsilon_{22}$, and (c) as a function of $\varepsilon_{12}$ \textit{vs.} $\varepsilon_{22}$.} 
\end{figure}
The surface energy (\(\gamma\)), which is a measurement of the surface atoms' extra energy owing to a number of reasons, including the broken bonds, describes the stability of a surface. Understanding surface reconstruction, roughening, and the equilibrium form of the crystal all depend on this fundamental number~\cite{10.1038/sdata.2016.80}. Despite its significance, it is difficult and infrequent to experimentally determine surface energy, particularly for certain aspects~\cite{10.1063/1.1735524}. To obtain surface energy as shown in~\tab{ALD_alumina_surface_energy}, the ALD alumina coatings were allowed to relax in their unstrained form. The energy of the strained coating was calculated for the 25 combinations of infinitesimal strains ($\mathrm{\varepsilon_{11}}$, $\mathrm{\varepsilon_{12}}$; $\mathrm{\varepsilon_{11}}$, $\mathrm{\varepsilon_{22}}$; and $\mathrm{\varepsilon_{12}}$, $\mathrm{\varepsilon_{22}}$) shown in~\fig{schematic_surface_energy_with_strain}. The negative surface energies in~\fig{surface_energy_with_strain} (a,b,c) are due to the change in Gibbs free energy due to chemical effects and exothermic adsorption energies (see~\sm~for further details on adsorption energies) of alumina on solid Zn surface. To obtain surface elastic constants shown in~\tab{ALD_alumina_surface_elastic_constants}, the surface energy was interpolated using the different shape functions using the Finite Difference Method (FDM)~\cite{10.1016/S0377-0427(00)00507-0}, Local Maximum Entropy (LME)~\cite{10.1002/nme.1534} and the Higher Order LME Scheme (HOLMES) ~\cite{10.1007/978-3-642-32979-1_7}, and appropriate first and second derivatives were taken following the process to calculate fourth-order surface elastic tensor ($C_{ijkl}^\textrm{s}$) shown in~\eq{SurfaceElasticConstants}. The interpolation of surface energy with different methods is shown in~\tab{FDM_HOLMES_LME}. The bulk elastic constants ($C_{ij}$ in Voigt's notation) then was converted to $C_{ij}^{\textrm{plane}}$ using plane stress conditions, and then film elastic properties (shown in~\tab{ALD_alumina_elastic_constants}) were calculated using $C_{ij}^{\textrm{film}} = C_{ij}^{\textrm{plane}}+ \frac{2}{h_{f}}C^{\textrm{S}}_{ij}$~\cite{10.1016/j.tsf.2004.03.034}. 
\begin{table}[h]  \caption{\label{tab:ALD_alumina_surface_energy} List of surface energy $(\gamma_0)$  ($\mathrm{meV\cdot A^{-2}}$), surface stress $(\tau^0)$ ($\mathrm{meV \cdot A^{-2}}$) of bare and coated  (MLD alucone and ALD alumina) zinc for unstrained cases.}
    \centering
    \begin{tabular}{c c c c c }
\hline Surface & $\gamma_{0}$ & $\tau^{0}_{11}$  & $\tau^{0}_{22}$ & $\tau^{0}_{12}$ \\ 
 \hline \hline 
Zinc~\cite{10.1021/acs.jpcc.2c06646} & 42 & -96 & -95 & 0 \\ 
MLD alucone deposited zinc~\cite{10.1021/acs.jpcc.2c06646} & -133 & -373 & -374 & -3 \\ 
ALD alumina deposited zinc & -623 & -729 & -740 & 0 \\ \hline
    \end{tabular}
    \end{table}
\begin{table}[H]
    \caption{\label{tab:ALD_alumina_surface_elastic_constants} List of fourth-order surface elastic-stiffness coefficients $(\mathrm{{\it{C}}_{ijkl}^{\textrm{s}}})$ ($\mathrm{eV \cdot A^{-2}}$) of bare and coated  (MLD alucone and ALD alumina) zinc.}
    \centering
    \begin{tabular}{c c c c c c c c c c}
\hline  & $C_{1111}^{\textrm{s}}$  & $C_{1122}^{\textrm{s}}$ & $C_{1212}^{\textrm{s}}$ & $C_{2222}^{\textrm{s}}$ & $C_{2211}^{\textrm{s}}$ \\ 
 \hline \hline
Zinc~\cite{10.1021/acs.jpcc.2c06646}  & -1.49 & 11.39 & -6.44  & - & - \\
MLD alucone deposited zinc~\cite{10.1021/acs.jpcc.2c06646} & 2.1 & 2.1 & 0.3 & 8.5 & 2.1 \\ 
ALD alumina deposited zinc & 7.45 & 1.53 & 0 & 6.13 & 1.54 \\
 \hline
& $C_{1112}^{\textrm{s}}$ & $C_{1222}^{\textrm{s}}$ & $C_{2212}^{\textrm{s}}$ &  $C_{1211}^{\textrm{s}}$\\ 
 \hline
Zinc & - & - & -& - \\ 
MLD alucone deposited zinc &-1.2 & 1.4 & 1.4 & -1.2 \\ 
ALD alumina deposited zinc & 0.10 & 0.99 & 0.99 & 0.10 \\ \hline
    \end{tabular}
    \end{table}
\begin{table}[h]
    \caption{\label{tab:ALD_alumina_elastic_constants}  List of second-order effective elastic-stiffness coefficients $(C_{ij}^{\textrm{film}})$ (GPa),  different modulus (Bulk (B), Shear(G), Young (E) (GPa)), Poisson's ratio \((\nu)\) and Pugh ratio \((p)\) of bare and coated  (MLD alucone and ALD alumina) zinc.}
    \centering
    \begin{tabular}{c c c c c c c c c c}
\hline  & $C_{11}^{\textrm{film}}$  & $C_{22}^{\textrm{film}}$ & $C_{12}^{\textrm{film}}$ & $C_{21}^{\textrm{film}}$ & $C_{13}^{\textrm{film}}$ & $C_{23}^{\textrm{film}}$ & $C_{31}^{\textrm{film}}$\\ 
 \hline \hline
Zinc~\cite{10.1021/acs.jpcc.2c06646} & 141.26 & 149.52 & 1.43  & - & -  & -  & - \\
MLD alucone \\ deposited zinc~\cite{10.1021/acs.jpcc.2c06646} & 143.6 & 147.7 & 6.9 & 6.9 & 0.0 & 0.0 & 0.0 \\
ALD alumina \\ deposited zinc & 146.99 & 146.15 & 6.54 & 6.54 & 0.06 & 0.63 & 0.06 \\
 \hline 
  & $C_{32}^{\textrm{film}}$ &  $C_{33}^{\textrm{film}}$ & B & G & E & $\nu$ & $p$\\ 
 \hline \hline
Zinc & - & -&  16.68 & 13.21 & 31.35 & 0.19 & 1.26\\ 
MLD alucone \\ deposited zinc & 0.0 & 67.5 & 18.6 & 11.4 & 28.5 & 0.24 & 1.6\\
ALD alumina \\ deposited zinc & 0.63 & 67.48 & 18.58 & 12.55 & 30.73 & 0.22 & 1.48 \\ \hline
    \end{tabular}
    \end{table}
\subsection{\label{sec:charge_density_difference}Charge density difference and Bader charge analysis}
\begin{figure}[h]
\centering
{\subfloat[]{
\label{}
\includegraphics[width=0.35\linewidth]{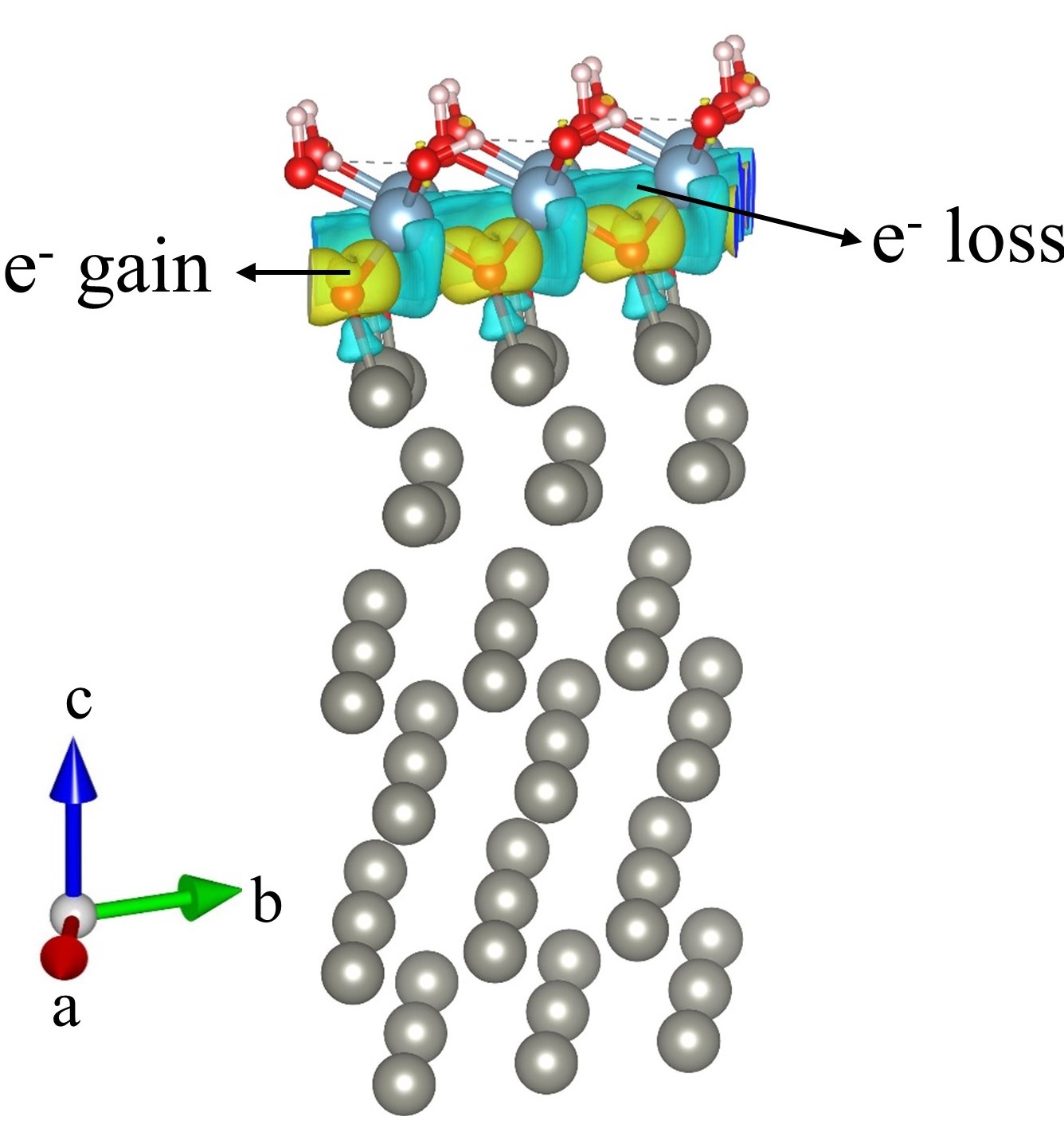}}}
\qquad
{\subfloat[]{
\label{}
\includegraphics[width=0.25\linewidth]{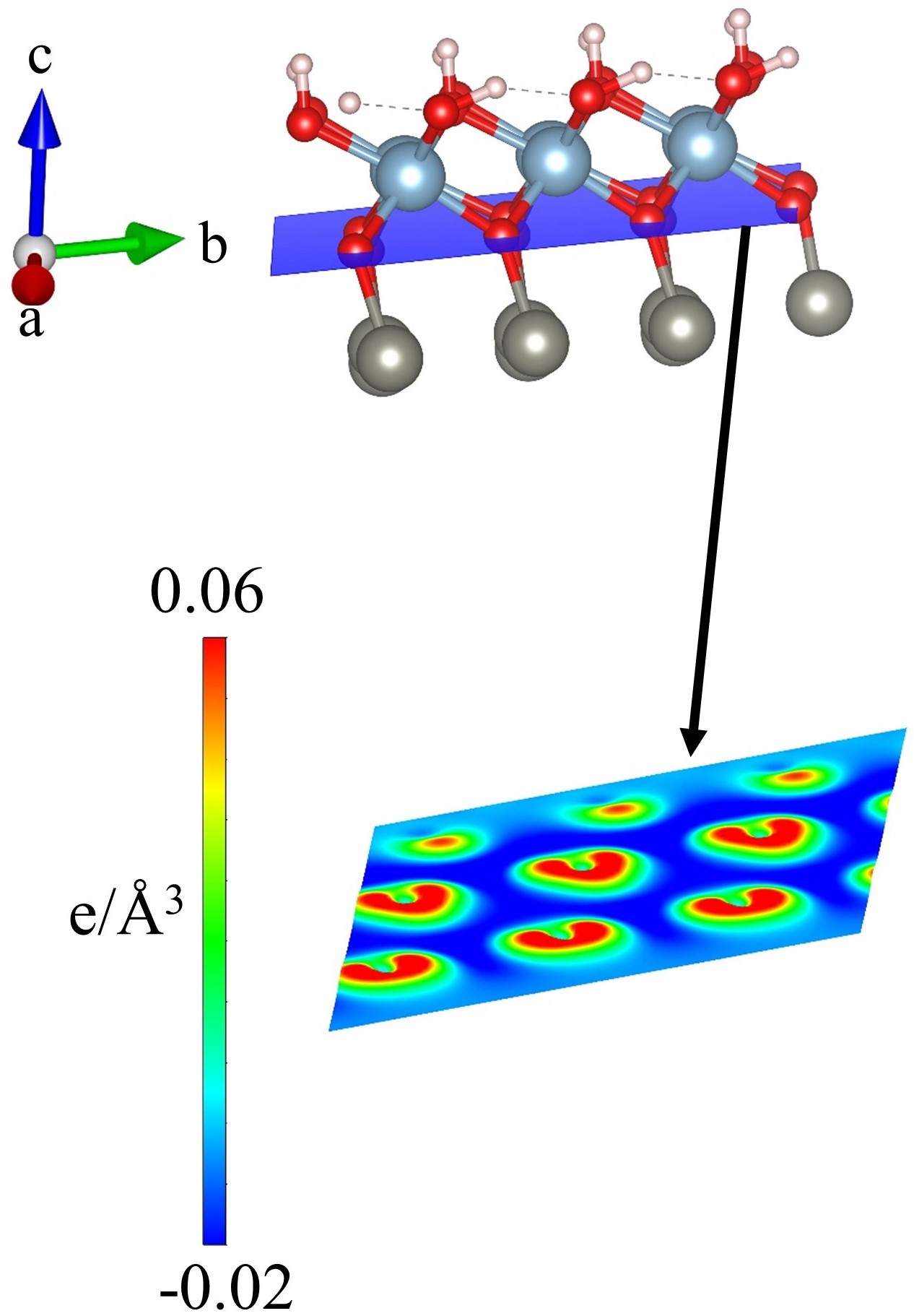}}}
\qquad
\caption {
Differential charge density for the ALD alumina adsorbed hydroxylated Zn. (a) The isosurface displays an electron accumulation in the yellow area and a reduction in electrons in the blue area, (b) a section parallel to the ab plane, where the red areas show the oxygen's electron gains on the hydroxylated Zn. 
} \label{fig:charge_density_difference}
\end{figure}

To understand the charge redistribution of ALD $\mathrm{Al_{2}O_{3}}$ adsorbed on a hydroxylated Zn surface, the differential charge density due to the adsorption has been computed and analyzed. The differential charge density ($\Delta\rho_\textrm{c}$) of the system is defined as the difference in charge between the deposited hydroxylated Zn surface and the reference $\mathrm{Al_{2}O_{3}}$, and hydroxylated Zn surface ~\cite{10.1016/j.apsusc.2020.148461}.

\fig{charge_density_difference} maps the rearranged electron densities for the ALD $\mathrm{Al_{2}O_{3}}$ adsorption on the hydroxylated Zn. As depicted in ~\fig{charge_density_difference}(a), an increment of $\rho_\textrm{c}$ at the Zn-O bond (yellow isosurfaces) indicates absorption of electrons at the expense of a reduction of $\rho_\textrm{c}$ at the Al-O bonds (light blue isosurfaces). As shown in~\fig{charge_density_difference}(b), there is a sharp rise in $\rho_\textrm{c}$ around the O atoms of the hydroxyl group (red areas). The significant increase in $\rho_\textrm{c}$ suggests the presence of covalent bonding (chemisorption) due to the deposition.

To quantitatively analyze the chemisorption of the ALD coating process, we turn our attention to the Bader charge analysis. This analysis allows us to quantitatively show the degree of the chemical reactivity, e.g., the decomposition of precursors in the hydroxyl groups ~\cite{10.1088/0953-8984/21/8/084204}. Table~\ref{tab:bader_charge} shows the computed average Bader charge and electron gain/loss for different atoms near the coating zone. It is evident from Table~\ref{tab:bader_charge} that the Al atoms lose electrons, giving these away to the O atoms in the hydroxyl group. Remarkably, H in the hydroxyl group had a significant electron loss, making an unequal charge between the atoms in that group. As a result, the Al-hydroxylated Zn surface bonds get stronger, indicating strong chemisorption of the ALD coating.
\begin{table}[h]
    \caption{\label{tab:bader_charge}Average Bader atomic charges and electron gain and loss.}
    \centering
    \begin{tabular}{ccc}
    	\hline Atom & Charge & Electron \\ \hline \hline
        $\textrm{Al}$ & $~~2.47\text{ e}$ & Loss\\
        $\textrm{O} \text{ (hydroxylated Zn)}$ & $-1.42\text{ e}$ & Gain\\
        $\textrm{O} \text{ (Alumina)}$ & $-0.75\text{ e}$ & Gain\\
        $\textrm{Zn} \text{ (free layers)}$ & $-0.01\text{ e}$ & Gain\\
        $\textrm{H}$  & $~~0.68\text{ e}$ & Loss\\ \hline
    \end{tabular}
    \end{table}   
\subsection{\label{sec:Multi_layer_misfit}Multi-layer misfit}
\begin{figure}[h]
\centering
\subfloat[]{
\label{}
\includegraphics[width=0.44\linewidth]{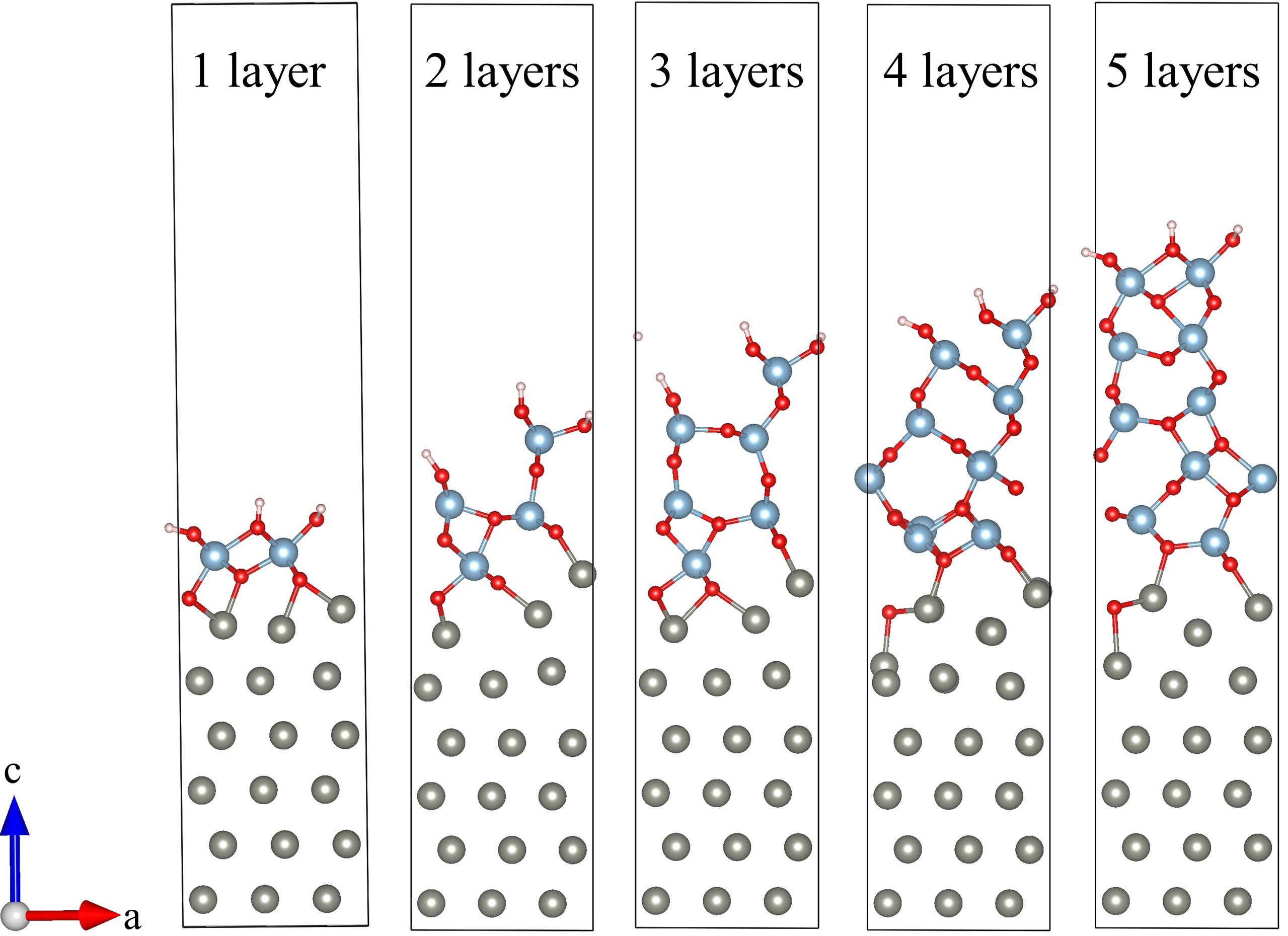}}
\qquad
\subfloat[]{
\label{}
\includegraphics[width=0.45\linewidth]{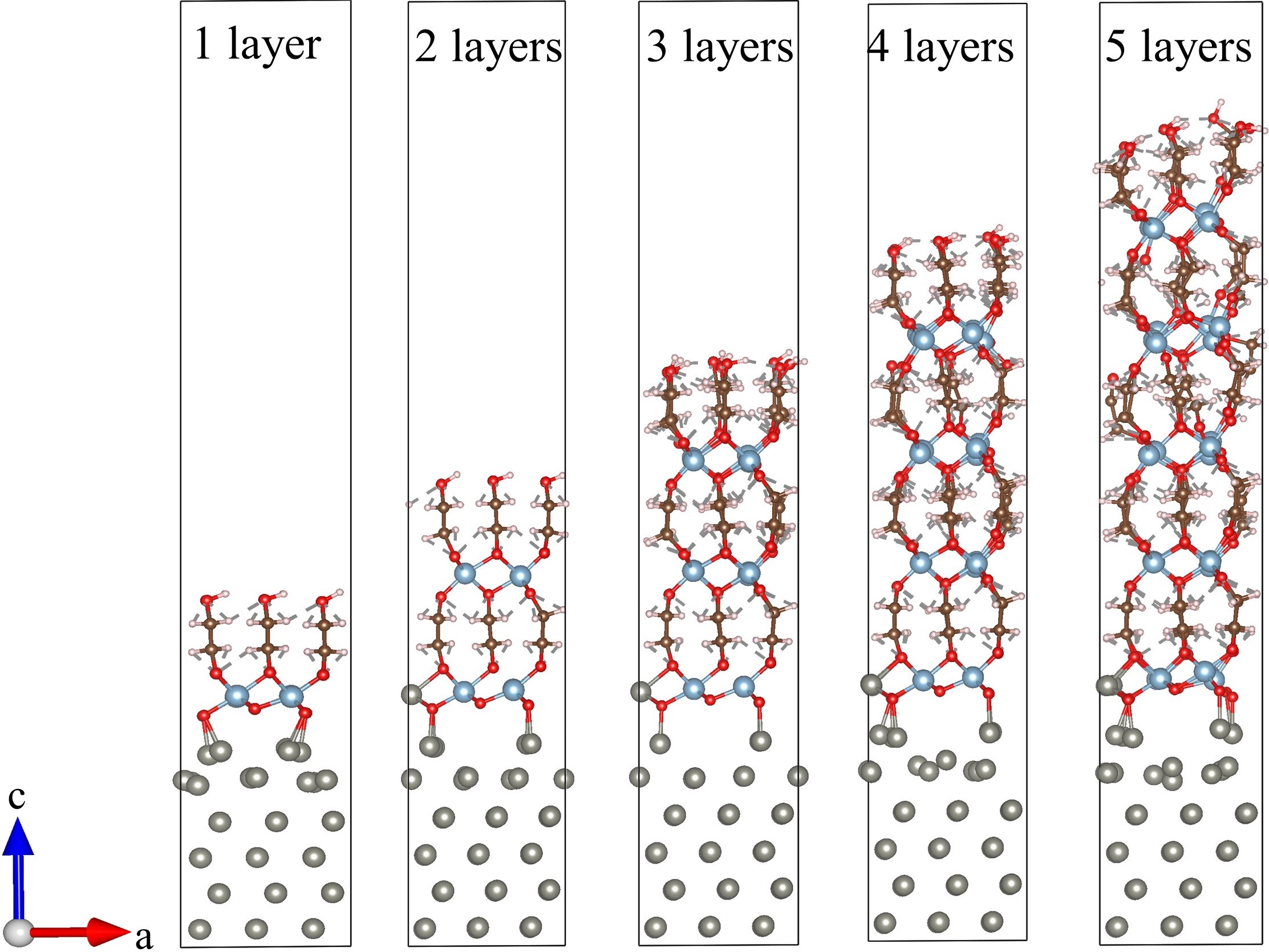}}
\qquad
\caption{\label{fig:multilayers_ALD_MLD_optimized_cell} DFT optimized multi-layer thin film on Zn surface (a) ALD alumina and (b) MLD alucone. }
\end{figure}

The misfit strain calculated in this study is based on the DFT calculation of five ALD/MLD coating layers deposited on the Zn substrate. The effect of multi-layers on simulation cell optimization will change the cell volume and atomic positions, but cell shape should not be changed from bulk zinc's hcp shape (see~\sm~\sec{Implementation_of_multi_layer_misfit} for further details). The optimized simulation cells with multi-layer ALD alumina and MLD alucone coating are shown in~\fig{multilayers_ALD_MLD_optimized_cell}.
\begin{figure}[h]
\centering
\label{}\includegraphics[width=0.40\linewidth]{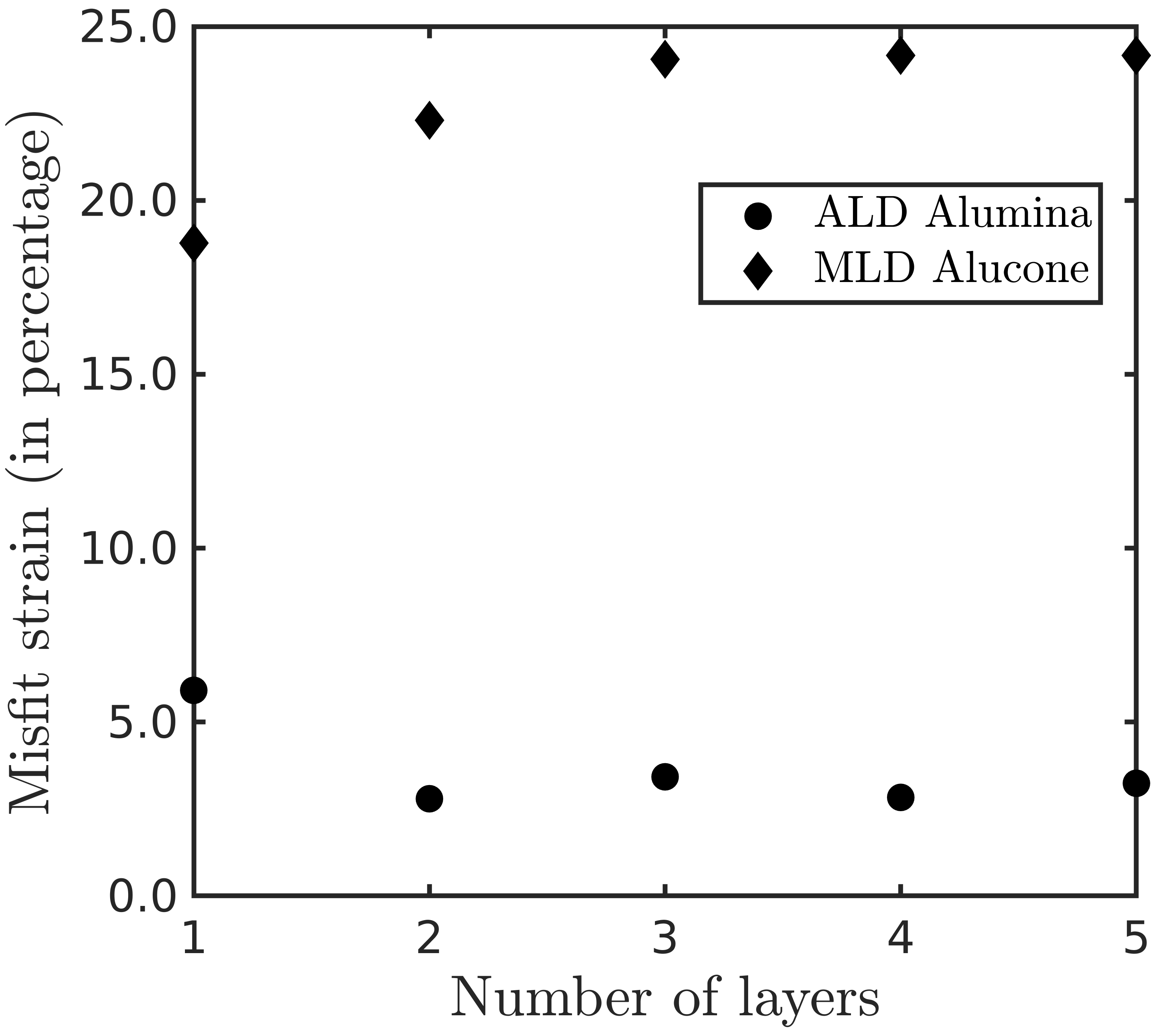}
\caption{In-plane misfit strain for multilayer coating on Zn substrate.}
\label{fig:multilayer_misfit}
\end{figure}
\par The in-plane lattice parameter of MLD alucone films grown on Zn substrates rises with increasing film thickness before saturation at 24\%, as shown in~\fig{multilayer_misfit}.  It is worth mentioning that this large misfit is because the simulation was carried out using a coherent hetero-epitaxial coating, meaning that the interface does not generate misfit dislocations. This is because to explicitly simulate dislocations in the thin films, we need to have large simulation cells out of the reach of state-of-the-art \textit{ab initio} simulations. In turn, misfit dislocations and critical film thickness will be analyzed next. 
%
For ALD alumina, we observed that the misfit strain decreases with coating thickness, suggesting a coherent interface between Zn and coating. While part of the misfit strain could be compensated by plastic deformation for thicker coatings, we found that the ALD alumina film thickness for 5 layers was $ \sim$ 1.5 nm, whereas, for 3.24\% of misfit strain, the critical thickness for stable dislocation was 3.76 nm. Therefore, plastic deformation did not reduce misfit strain, as the film thickness did not exceed the critical value.
\subsection{\label{sec:Effect_of_Lattice_Misfit_on_Residaul_stresses}Effect of Lattice Misfit on Residual Stresses}
\begin{figure}[h]
\centering
\subfloat[]{
\label{}
\includegraphics[width=0.40\linewidth]{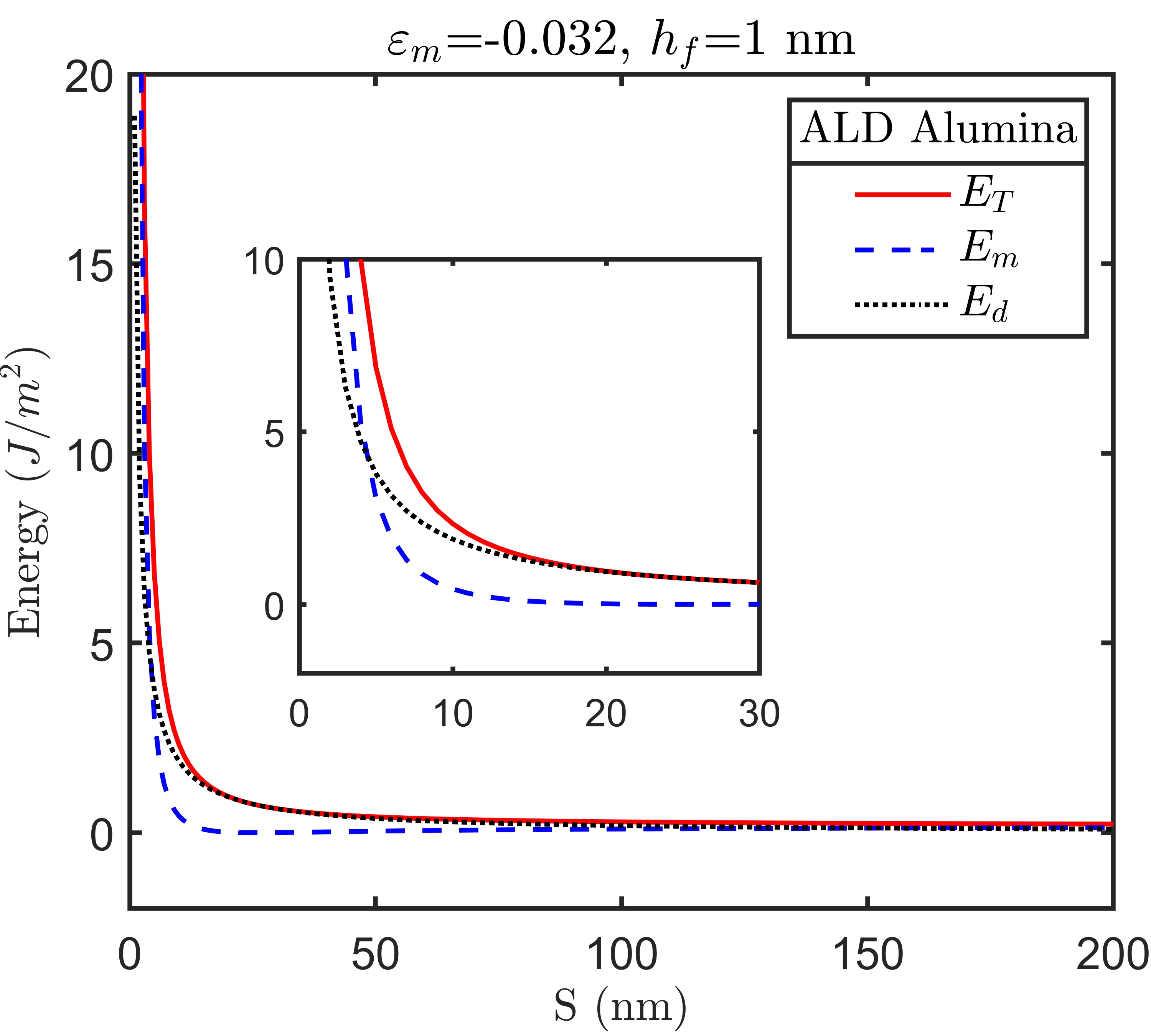}}
\qquad
\subfloat[]{
\label{}
\includegraphics[width=0.40\linewidth]{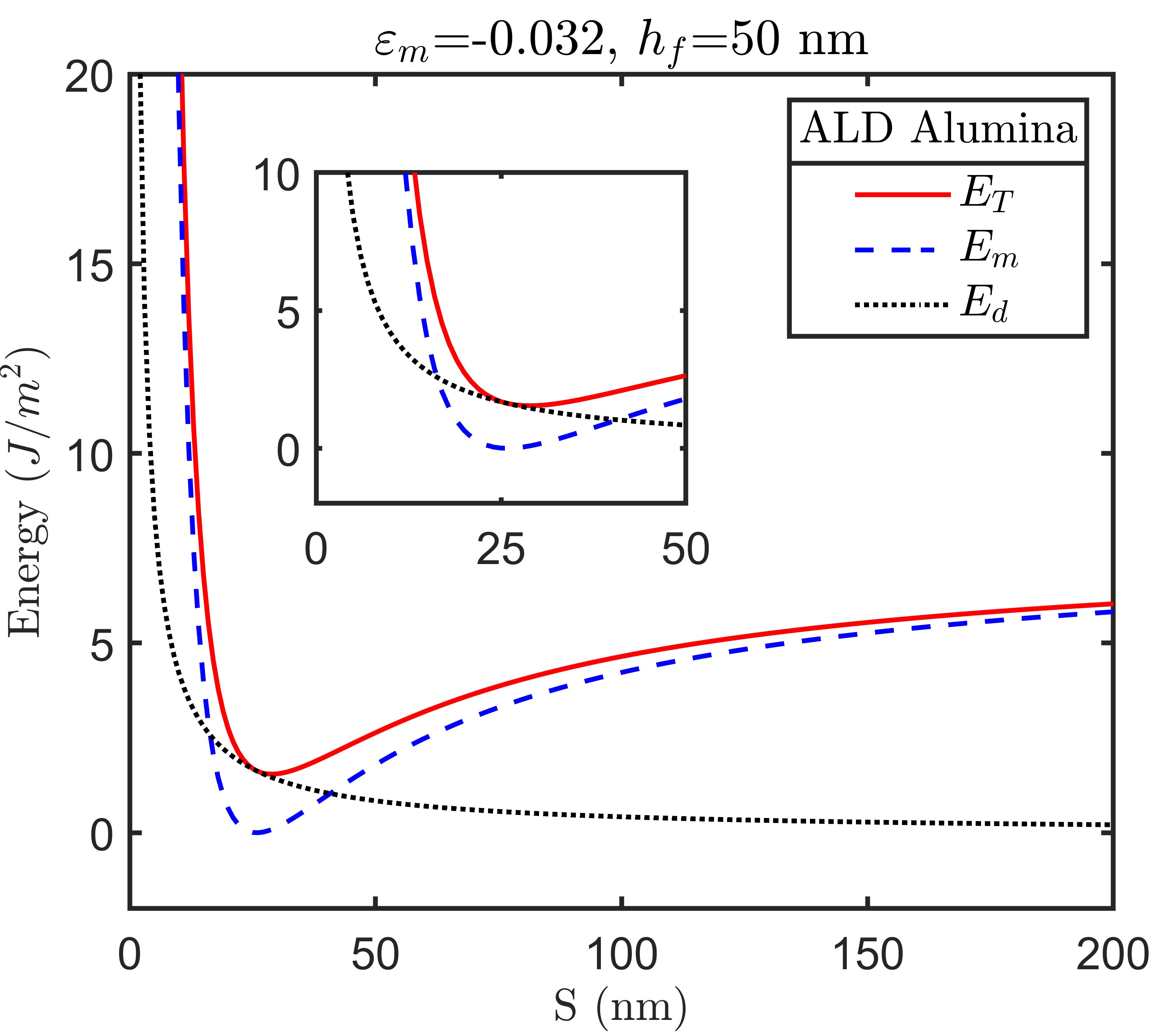}}
\qquad
\subfloat[]{
\includegraphics[width=0.40\linewidth]{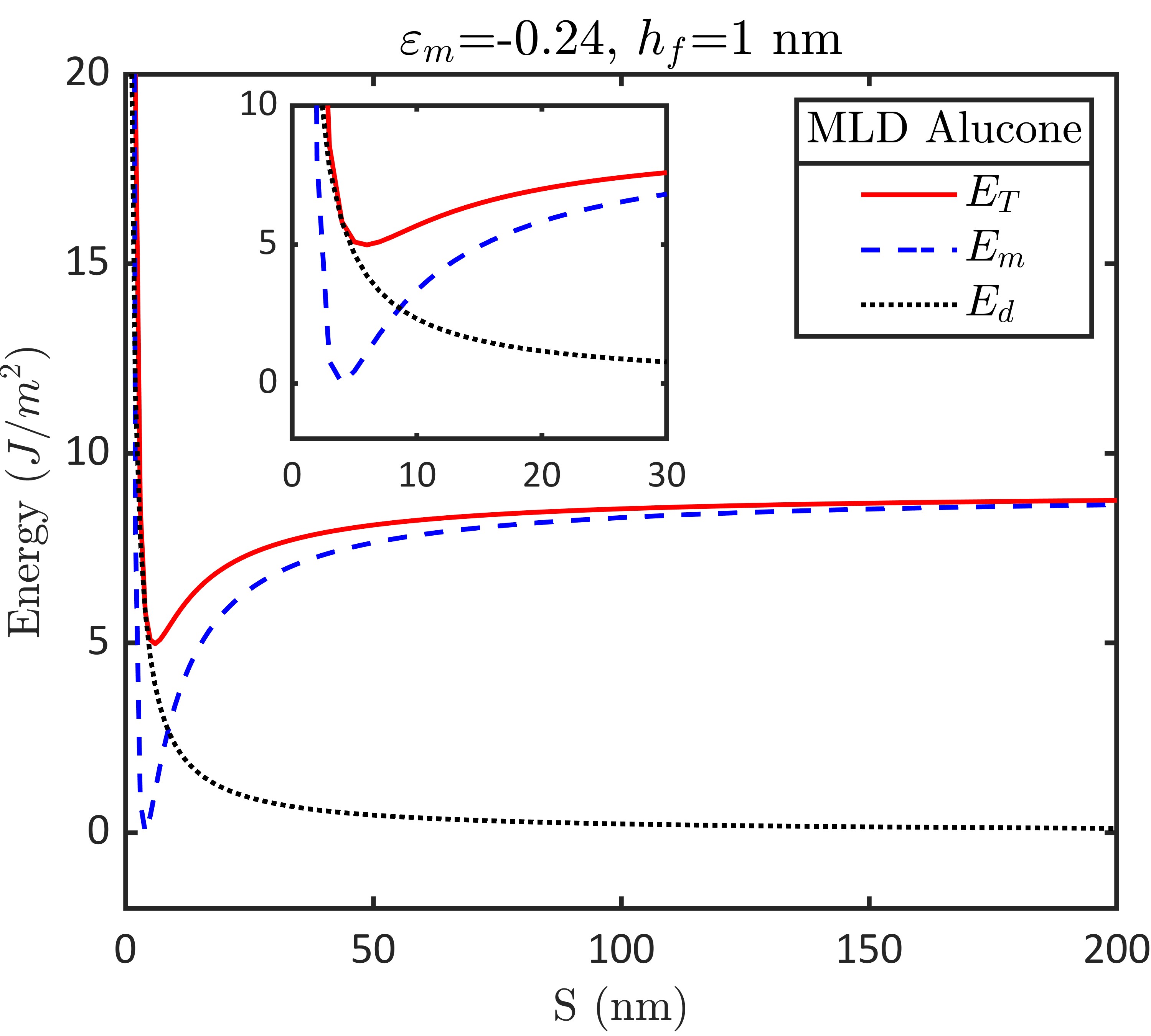}}
\qquad
\subfloat[]{
\label{}
\includegraphics[width=0.40\linewidth]{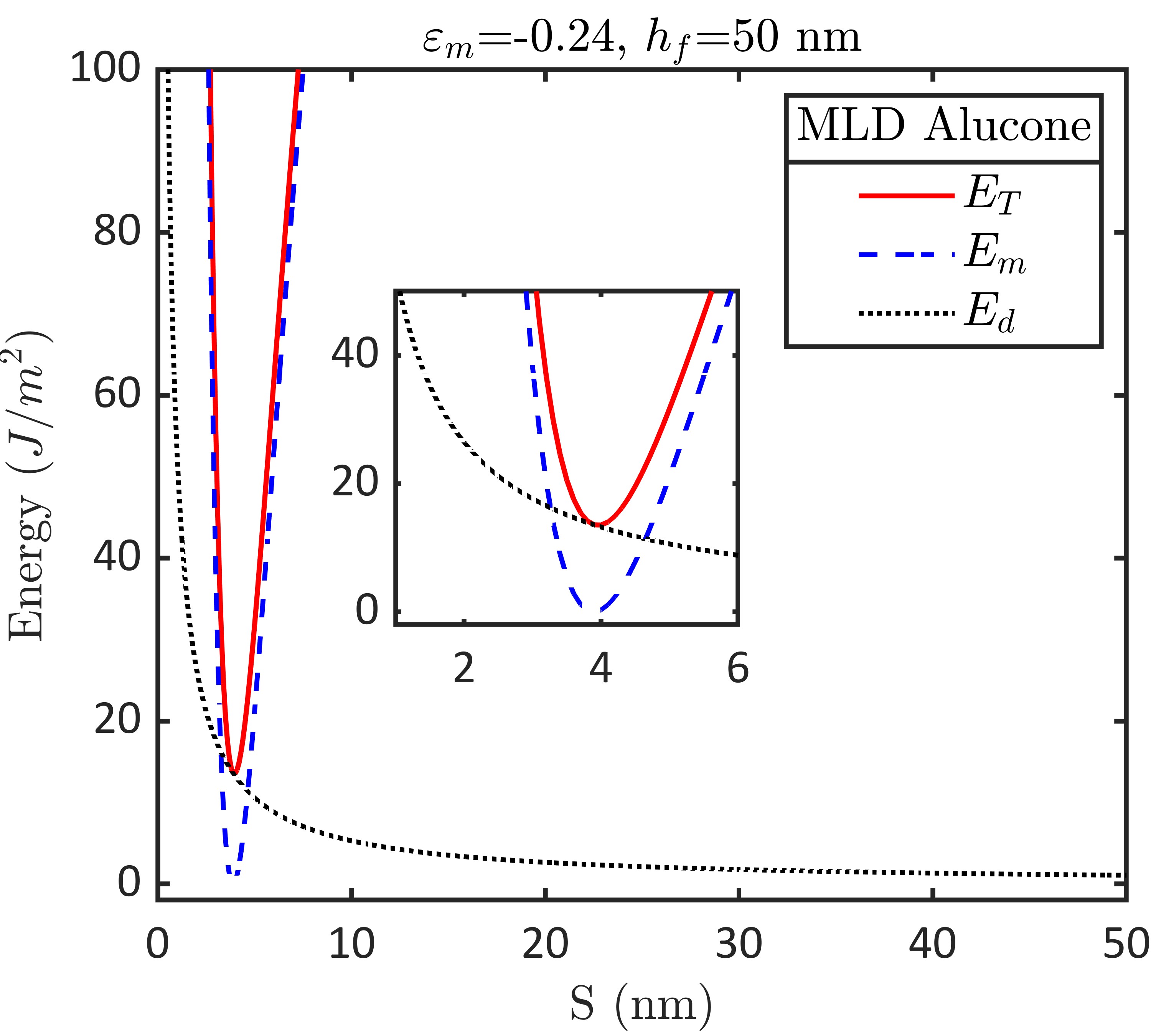}}
\caption{\label{fig:equilibrium_spacing}Elastic strain energy variation with dislocation spacing for ALD and MLD coatings-(a) $\varepsilon_\textrm{m}=-0.032$, $h_\textrm{f}$ = 1 nm, (b) $\varepsilon_\textrm{m}=-0.032$, $h_\textrm{f}$ = 50nm, (c) $\varepsilon_\textrm{m}=-0.24$, $h_\textrm{f}$ = 1 nm and (d) $\varepsilon_\textrm{m}=-0.24$, $h_\textrm{f}$ = 50 nm.}
\end{figure}
\fig{equilibrium_spacing} illustrates total elastic energy ($E_\textrm{T}$)  and its components ($E_\textrm{m},E_\textrm{d}$), which are depicted as a function of $S$ for lattice misfit strains: $\varepsilon_\textrm{m}=-0.24$ for MLD  alucone and $\varepsilon_\textrm{m}=-0.032$ for ALD alumina. To find the dislocation array's equilibrium configuration $S$ for a particular film thickness, the total energy $E_\textrm{T}$ in~\eq{total_energy} is minimized with respect to the separation distance among dislocations ($S$). It is clear from~\fig{equilibrium_spacing}(a) that for an atomically thin film of alumina ($h_\textrm{f}$ = 1 nm) with a low lattice misfit ($\varepsilon_\textrm{m}$), the total energy changes asymptotically with $S$. This shows that dislocation nucleation is unfavorable at such a thin film thickness. However, with a thicker film made of alumina ($h_\textrm{f}$ = 50 nm) shown in~\fig{equilibrium_spacing}(b)), dislocation nucleation is favorable, and the energy reached a minimum for a given value of $S\sim25$ nm. 

On the other hand, we noticed that for MLD alucone coating, since the misfit strain is significantly larger compared to the one produced by alumina, dislocation nucleation is favorable even at very small thicknesses of 1 nm, as illustrated in ~\fig{equilibrium_spacing}(c) and (d). The configuration at which energy reached a minimum was when dislocations were separated by $S\sim4$ nm for coating thicknesses between $h_f = 1 - 50$ nm. Point to be noted here, in~\fig{equilibrium_spacing}, we neglected the $E_{\textrm{core}}$ term for minimization of the energy as it requires an extremely large number for $E_{\textrm{core}}$ to have a significant impact on minimization of energy (for instance, for $E_{\textrm{core}}=7.5\times E_{\textrm{d}}$, $S$ changes from 4 to 5 nm). 

\begin{figure}[h]
\centering
\subfloat[]{
\label{}
\includegraphics[width=0.40\linewidth]{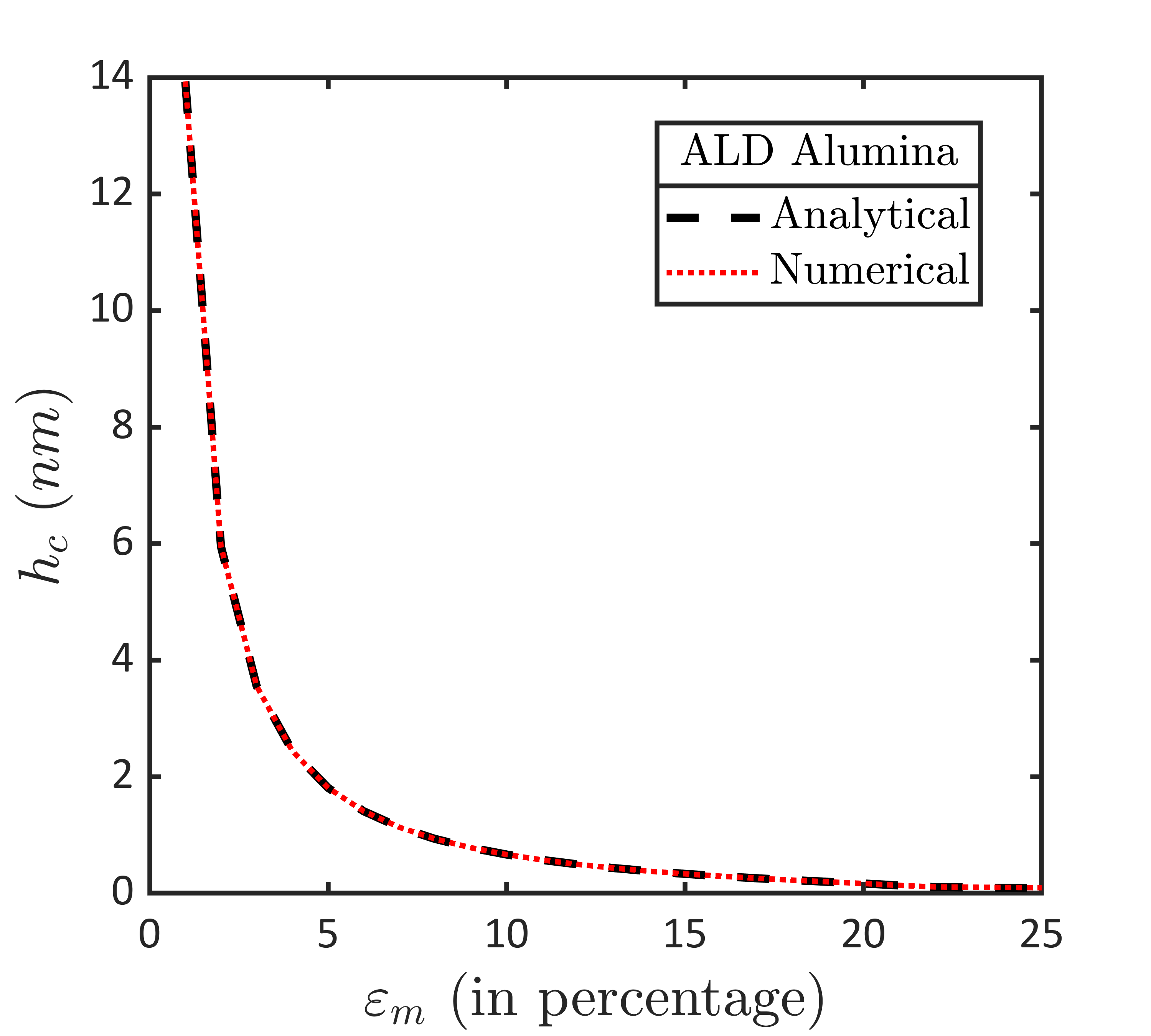}}
\qquad
\subfloat[]{
\label{}
\includegraphics[width=0.40\linewidth]{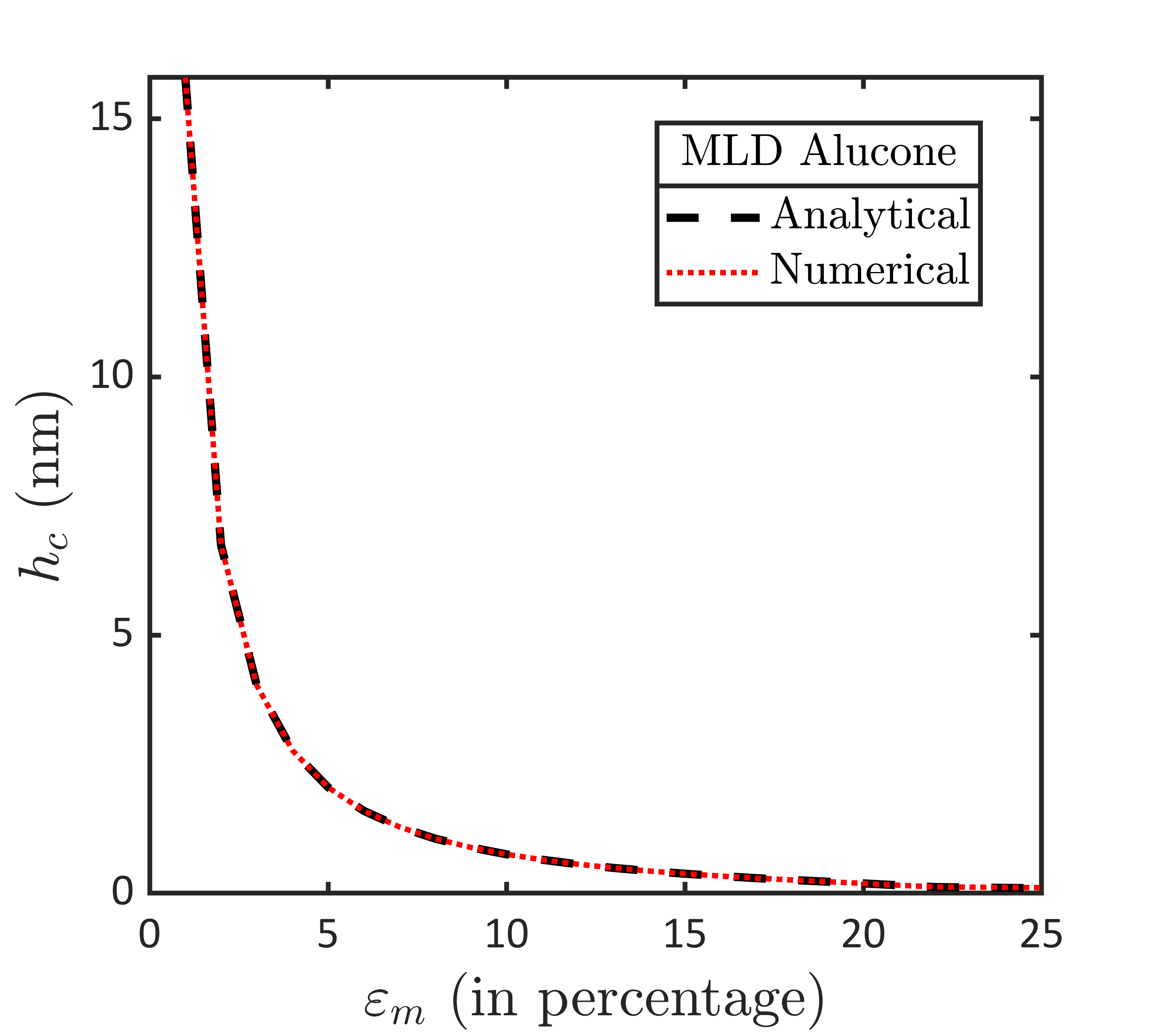}}
\caption{\label{fig:critical_thickness}Critical thickness $(h_\textrm{c})$ as a function of misfit
strain $\varepsilon_\textrm{m}$. The solid curve depicts the numerical solution, while the dashed curve shows the computed $h_\textrm{c}$ from an analytical Lambert $W-$function.}
\end{figure}
\par For nucleation of dislocations, the required critical film thickness ($h_\textrm{c}$) changes with lattice misfit. By generating misfit dislocations, strain/stress at the interface is relieved at the cost of reduced interface adhesion. This is caused by the strain energy accumulated during dislocation formation in the film when $h>h_\textrm{c}$. 
\par In order to identify the branch of $W(x)$ that accurately depicts the development of $h_\textrm{c}$, we need to consider a few factors. The critical thickness ($h_\textrm{c}$) to form dislocation should always be positive. As a result, both the pre-factor and the argument of the Lambert $W-$function consistently maintain the same sign, either positive or negative. Thus, we determine $W_\textrm{0}(x)$ branch accurately depicts the growth of $h_\textrm{c}$ varying the misfit strain ($\varepsilon_\textrm{m}$). This is because as the mismatch increases, $h_\textrm{c}$ must decrease, and for a misfit of $\varepsilon_{\textrm{mf}}=0$ at $x=0$ the critical thickness has to be infinite ($h_\textrm{c} = \infty$) which is satisfied as shown in~\fig{critical_thickness}. Additionally, a comparison between the analytical and numerical solutions of the computed analytical $h_\textrm{c}$ \emph{vs.} misfit strain is also presented in the figure. ~\fig{critical_thickness} (a) and (b) show relatively similar trends while solving~\eq{Critical_thickness} for ALD alumina and MLD alucone. This is because the anisotropic energy factor ($k$) of Zn and effective modulus ($M$) of thin films control the numerical values for $h_\textrm{c}$,  which are close in terms of numerical values for both the coating used in this study. Nevertheless, subtle differences appear in the behavior of $h_\textrm{c}$. Here, trust-region-reflective approaches based on nonlinear least-squares algorithms have been utilized to numerically solve the nonlinear algebraic ~\eq{Critical_thickness}~\cite{10.1137/0904038}. 
Therefore, a coherent interface is more likely to develop during the epitaxial growth of ALD alumina at lower lattice misfits  ($\varepsilon_\textrm{m}= -3.2\%$) of the film-substrate interface. Nucleation of dislocation is expected for high lattice misfit ($\varepsilon_\textrm{m}= -24 \%$) for MLD alucone as critical thickness decreases with increasing $\varepsilon_\textrm{m}$, and therefore, the effective misfit will be $0.8 \%$. 
\begin{figure}[!htb]
\centering
\subfloat[Thin film stresses.]{
\label{}
\includegraphics[width=0.40\linewidth]{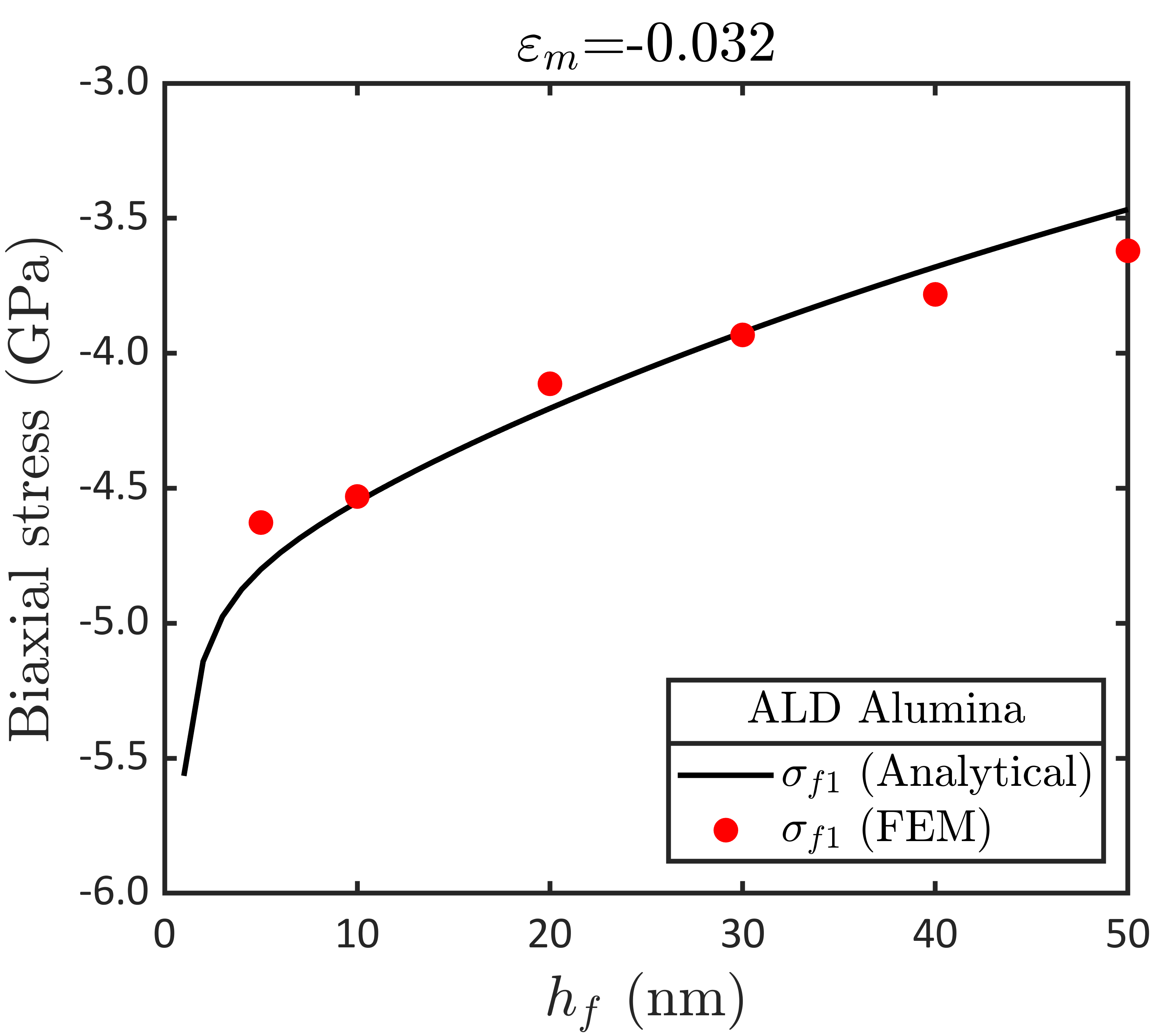}}
\qquad
\subfloat[Substrate stresses.]{
\label{}
\includegraphics[width=0.40\linewidth]{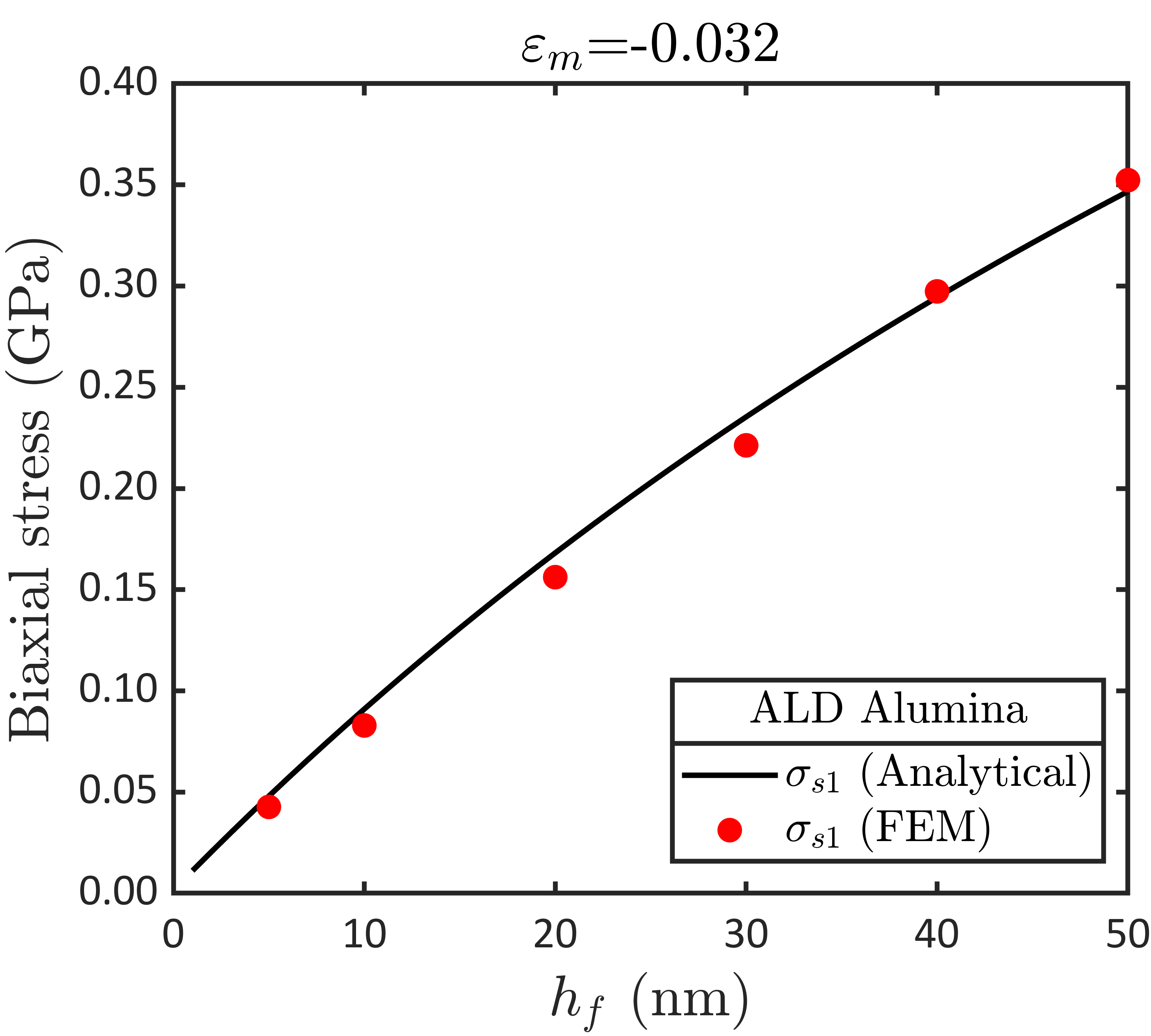}}
\qquad
\subfloat[Thin film stresses.]{
\label{}
\includegraphics[width=0.40\linewidth]{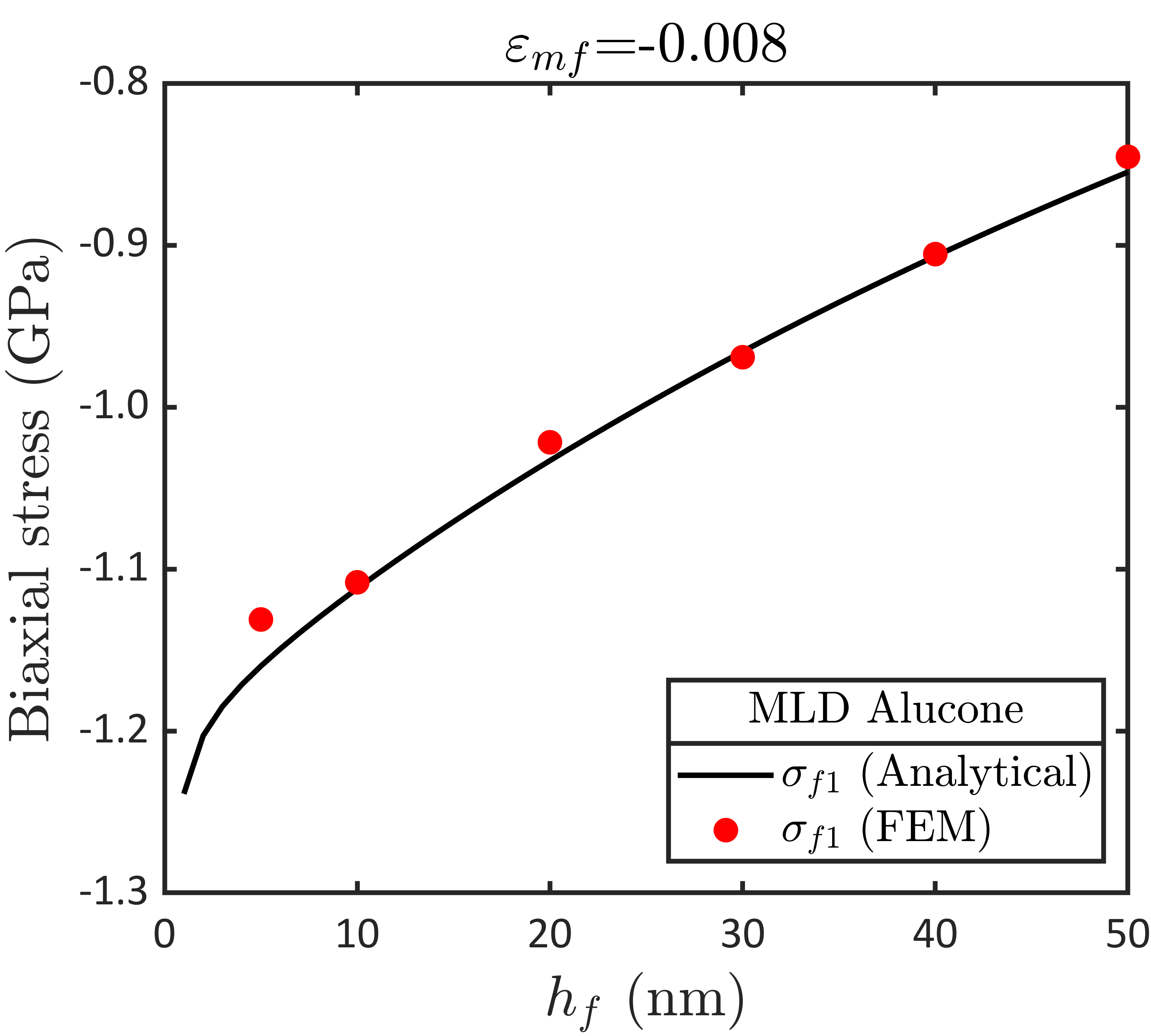}}
\qquad
\subfloat[Substrate stresses.]{
\label{}
\includegraphics[width=0.40\linewidth]{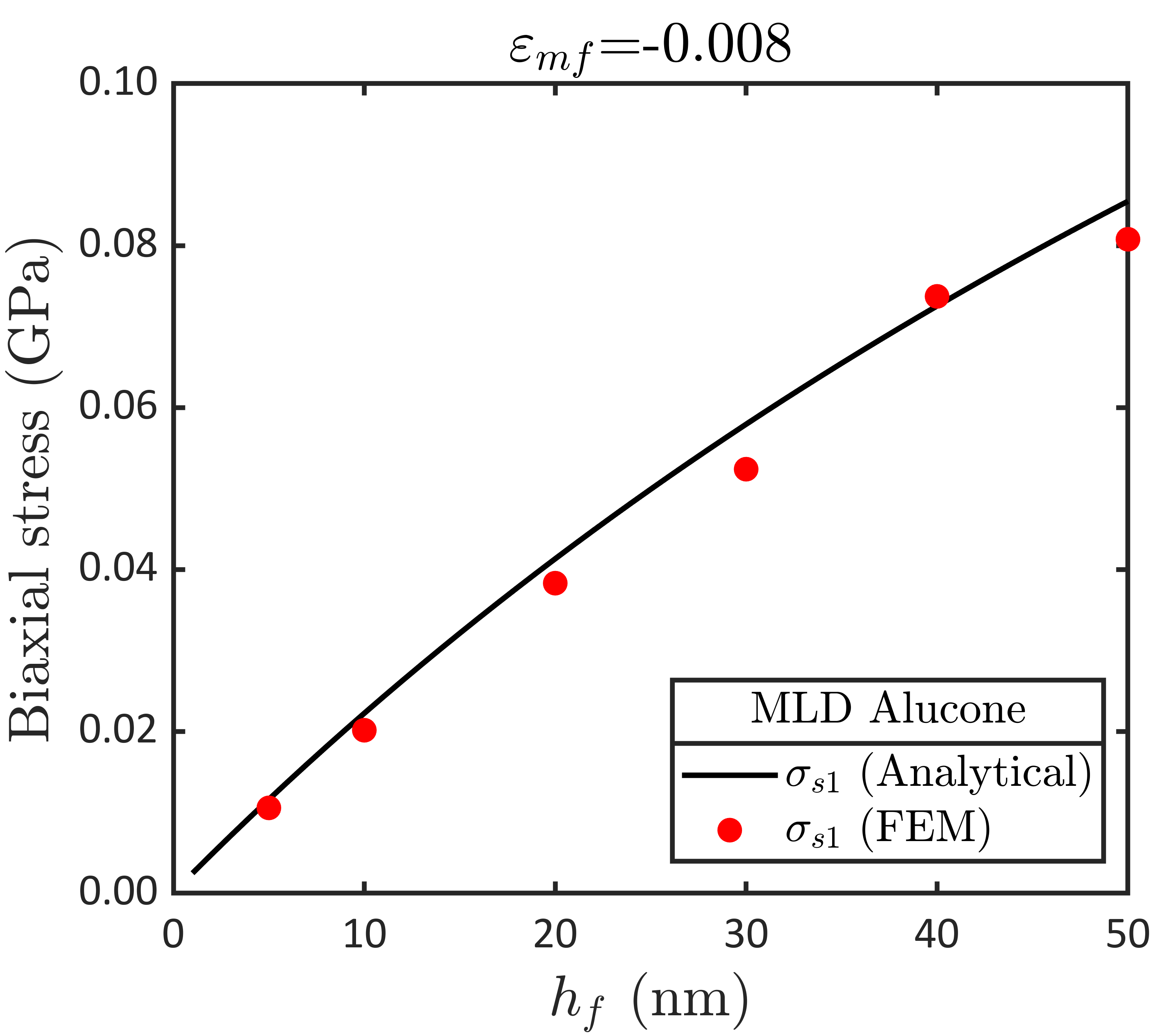}}
\caption{\label{fig:residual_stress_FEM} Evolution of biaxial stress profile due to lattice misfit strain - $\varepsilon_\textrm{m}=-0.032$ for ALD alumina (a) thin film, (b) substrate, and $\varepsilon_\textrm{mf}=-0.008$ for MLD alucone (c) thin film, (d) substrate. The comparison between the analytical model and FEA computed biaxial stress is plotted with varying thin film thickness. }
\end{figure}
\par To illustrate the impact of lattice misfit on the development of the interfacial stresses, two distinct coatings with distinct initial lattice misfits were considered. ~\fig{residual_stress_FEM} shows the evolution of the in-plane biaxial stress values in the film and substrate with the film's thickness ($h_f$) for alumina and alucone. From ~\fig{residual_stress_FEM}(a)-(d), it is evident that the lattice mismatch considerably adds to the stress amplitude for both coating systems. However, when thickness increases, stress decreases. One of the reasons for this phenomenon could be dislocation formation and a decrease in the interface's adhesion, thus lowering lattice misfit stress.  Underestimation of film stress is therefore possible if we ignore the influence of lattice misfit, as shown by~\fig{residual_stress_FEM}. 

\par Though the biaxial stress decreases with increasing thickness, high hetero-epitaxial (several GPa) stresses exist when the film thickness is small, corresponding to previous experimental studies. For example, for ALD alumina deposition on Si substrates, approximately 10.1 GPa in-plane residual stress was observed for 10 nm film~\cite{10.1016/j.sna.2010.09.018}. Notably, the biaxial stress values predicted here are estimated directly at the interface between film and substrate. While these values are extremely large, they only appear in a narrow portion of the system and tend to decay faster away from the interface. Nevertheless, the implications of these residual stresses could be extremely significant in the performance of Zn cathodes used in batteries.  
\par To examine the associated film thickness impact, FEA analysis of ALD/MLD thin films was conducted to display the relevance of the analytical equation. ~\fig{residual_stress_FEM} shows the in-plane stresses developed due to misfit for both alumina and alucone coating on zinc, respectively. ~\tab{material_properties} describes the material density, effective biaxial modulus, mean dislocation spacing, and lattice misfit of each component used in the FEA. In~\fig{residual_stress_FEM}, the analytical and average numerical thin film stresses with various film thicknesses of ALD/MLD films computed from FEA (\fig{finite_element_modeling}) are compared. The results show that the FEA and the analytically calculated stresses are comparable ($\approx\pm 5\%$). Noteworthy, the reported FEM stresses were calculated by taking average stress far from the edges. As depicted in ~\fig{residual_stress_FEM}, the effects of residual stresses decrease with the increase of thin film thicknesses.  
The edge effect close to the substrate's edge was ignored for the FEA findings while calculating the average stresses. The edge effect (resulting interfacial shearing stress, shown in~\fig{shear_stress_FEM}) causes the divergence between FEA and the estimated analytical stresses with the increase of film thickness. 
\begin{table}[h]
\caption{\label{tab:material_properties} List of material properties -  mass density ($\rho_m$) $g/cm^3$, effective biaxial modulus ($M$) (GPa), mean dislocation spacing ($S$) (nm), and misfit strain in percentage ($\varepsilon_{\textrm{m}}$).}
    \centering
    \begin{tabular}{c c  c c c c c}
\hline & $\rho_m$ &  $M_{11}$ & $S$ & $\varepsilon_{\textrm{m}}^{a}$  & $\varepsilon_{\textrm{m}}^{c}$ \\ 
 \hline \hline
Zinc & 7.14~\cite{10.1039/C4CP02878C}&   - & - & - & -\\
MLD alucone \\ deposited zinc & 2.35~\cite{10.1016/j.surfcoat.2013.06.098}&   150  & 4 & 0.80 & 0.20 \\
ALD alumina \\ deposited zinc & 3.00~\cite{10.1016/j.sna.2010.09.018} &  173 & 24 & 3.20 &  20.80 \\ \hline
    \end{tabular}
    \end{table}
\begin{figure}[h]
\centering
\subfloat[]{
\label{}
\includegraphics[width=0.42\linewidth]{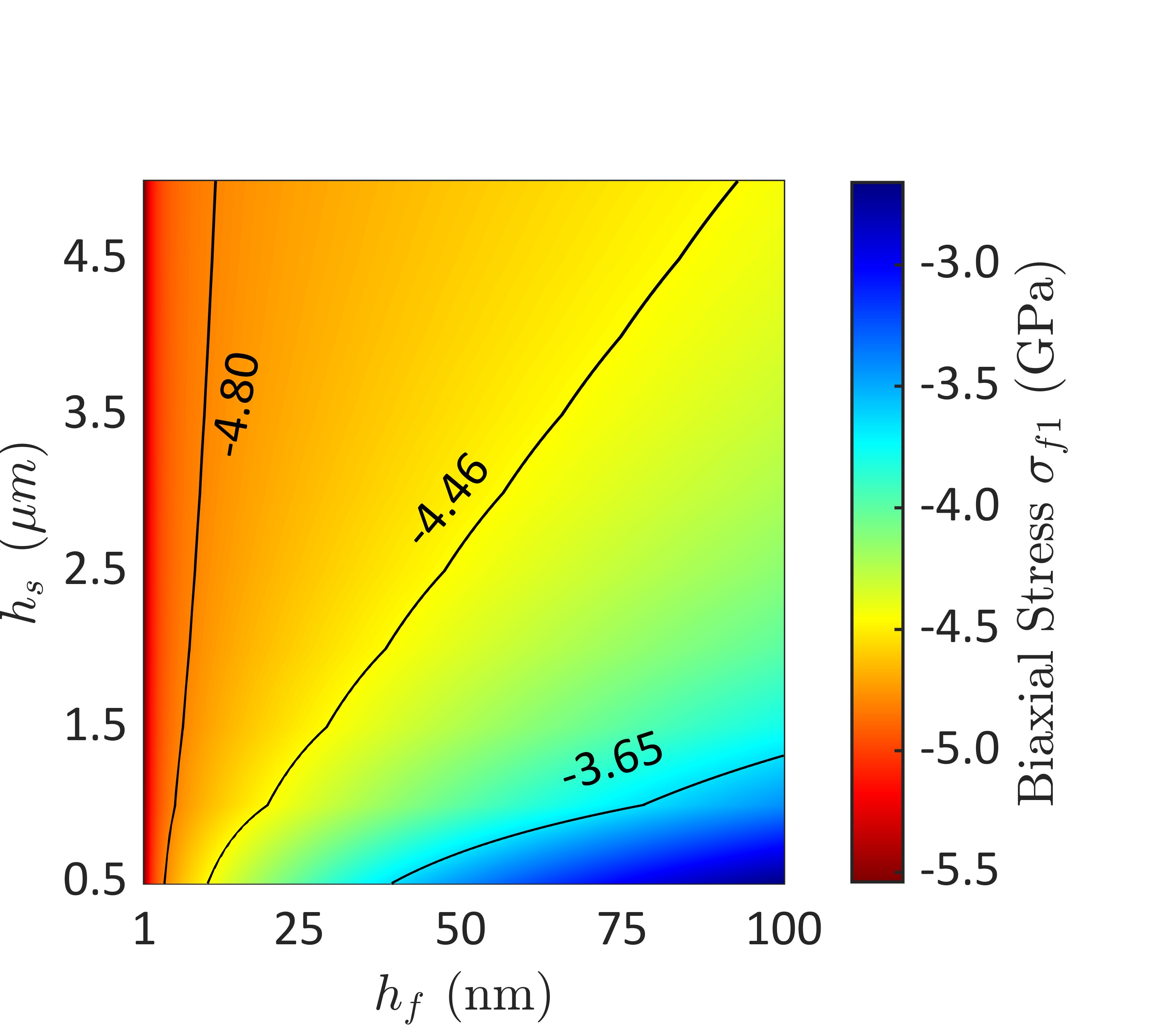}}
\qquad
\subfloat[]{
\label{}
\includegraphics[width=0.42\linewidth]{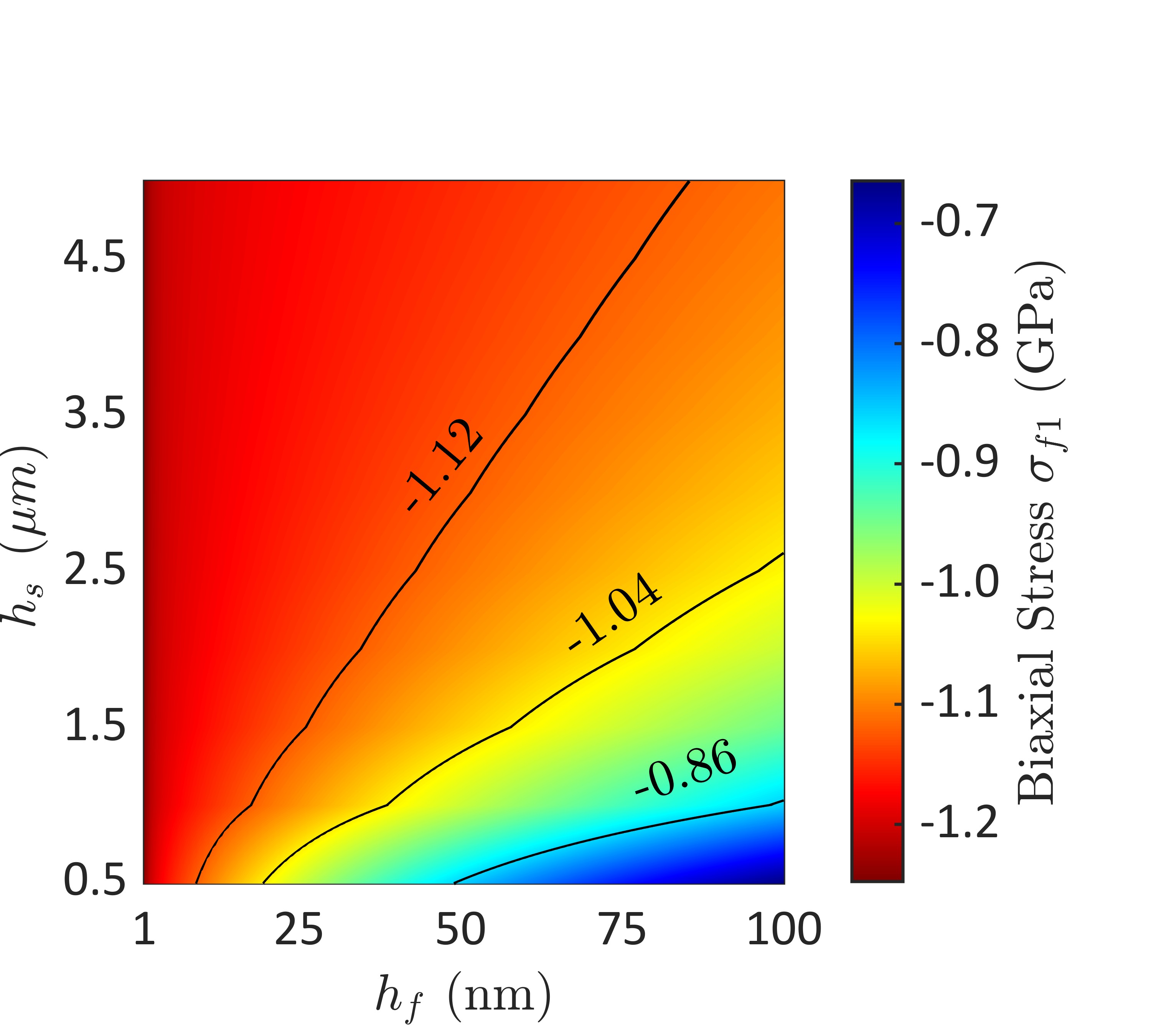}}
\caption{\label{fig:design_space} Two-dimensional mapping of the biaxial stress after hetero-epitaxial thin film position as a function of film thickness (${h_\textrm{f}}$) and substrate thickness (${h_\textrm{s}}$), showing the design space that leads to developed residual stress (a) ALD alumina, and (b) MLD alucone.}
\end{figure}
\par \fig{design_space} shows the design space of stresses according to different film and substrate thicknesses. The biaxial compressive stresses for ALD alumina vary from 3-5.5 GPa (\fig{design_space}(a)), whereas for MLD alucone, it varies from 0.7-1.2 GPa (\fig{design_space}(b)) with different film and substrate thicknesses. These significant compressive stresses can be accommodated by multilayer crystalline alumina thin films as nanometer-scale films show high yield strengths~\cite{10.1016/j.actamat.2009.08.073}. It can be predicted that as the film thickness of soft organic MLD alucone decreases, more dislocations would be entrapped in the thin film distant from the substrate-film interface, resulting in greater strain hardening rates~\cite{10.1177/1081286508092612}. Plastic deformation causes the stated array of interface dislocations to occur. Since each deposited dislocation reduces the equilibrium distance of dislocation ($S$) between interface dislocations (for instance, $S$=5 nm for 1 layer of MLD alucone, whereas $S$=4 nm for 5 layers of alucone coating), the strength increases during deformation. The plastic strain $\varepsilon_{\textrm{plastic}}$ can be estimated using $S$ during plastic deformation using $\varepsilon_{\textrm{plastic}} = \frac{b}{S}$~\cite{10.1016/S1359-6454(01)00170-7}.
\begin{figure}[h]
\centering
\subfloat[]{
\label{}
\includegraphics[width=0.42\linewidth]{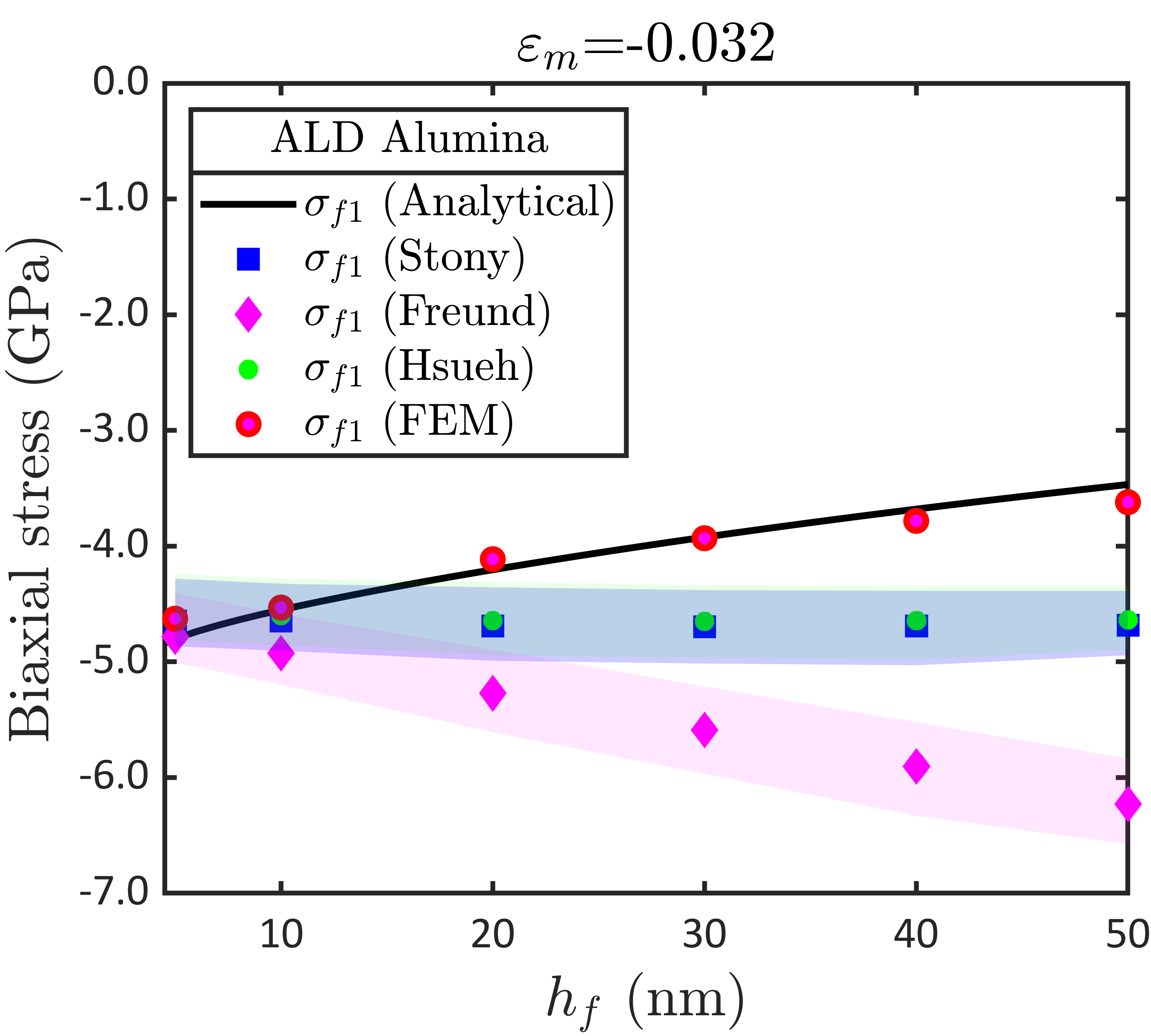}}
\qquad
\subfloat[]{
\label{}
\includegraphics[width=0.42\linewidth]{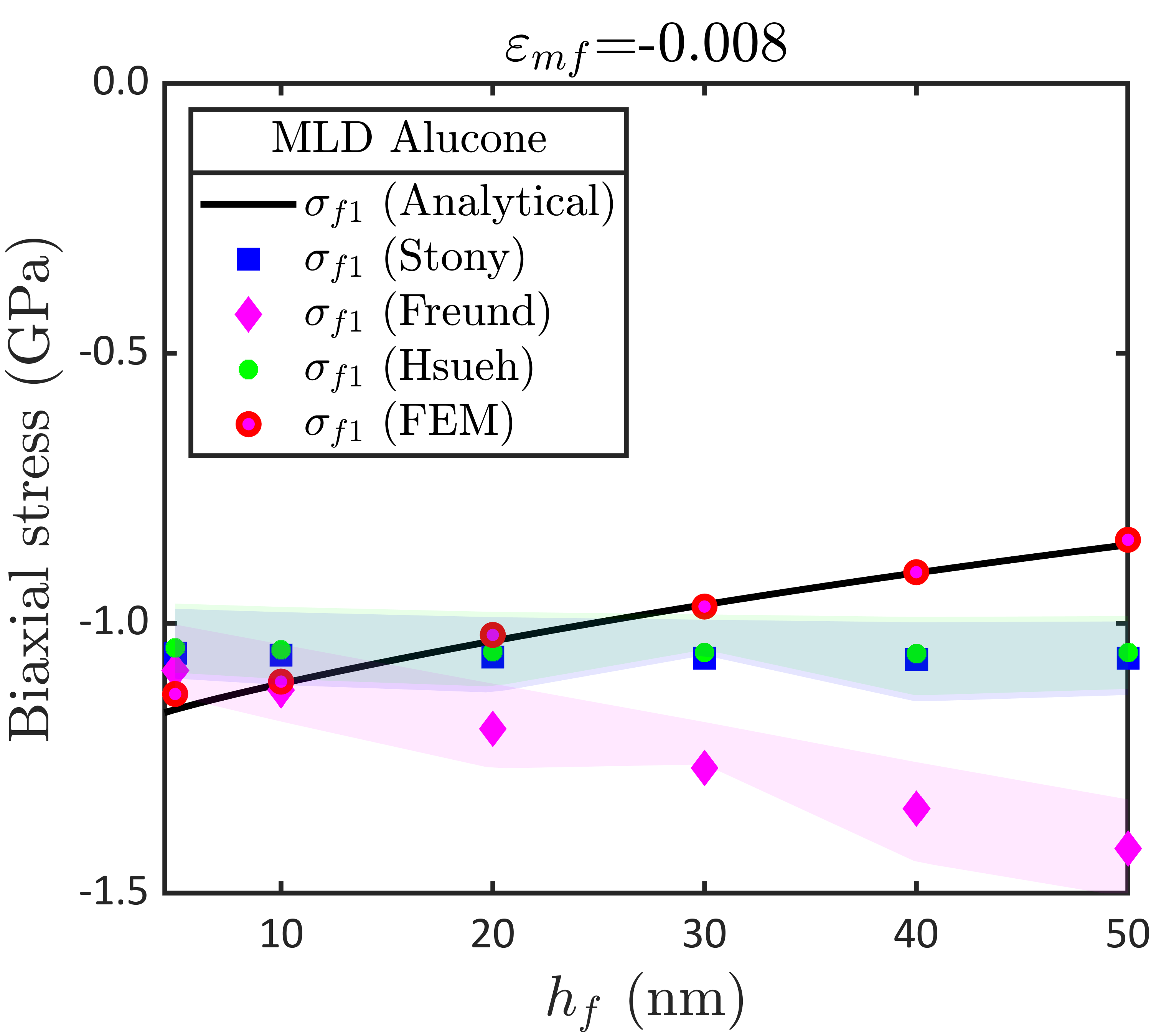}}
\caption{\label{fig:compare_formulas}The comparison among the analytical formula with the original Stoney's equation, Freund's equation, Hsueh's equation and FEA results for average biaxial stresses with film thickness - (a) ALD alumina, and (b) MLD alucone.}
\end{figure}
\par~\fig{compare_formulas} shows the comparison among the analytical formula with FEA, Stoney's ($\sigma_\mathrm{f}=\frac{\kappa E_\mathrm{s}^{*} h_\mathrm{s}^{2}}{6 \cdot h_\mathrm{f}}=\frac{E_\mathrm{s}^{*} h_\mathrm{s}^{2}}{6 \cdot h_\mathrm{f} \cdot \rho}$)~\cite{10.1098/rspa.1909.0021}, and modified Stoney's equations. Here, $\rho$ is the radius of curvature (RoC) $(\kappa=1 / \rho$ is the curvature). The modified Stoney's equation proposed by Hsueh has a better consistency with FEA, after the analytical model. Hsueh's model (${\sigma_\mathrm{f}}=\frac{E_\mathrm{s}^{*} h_\mathrm{s}^{2}}{6 h_\mathrm{f} \rho}\left[\frac{1+E_\mathrm{f}^{*} h_\mathrm{f}^{3} / E_\mathrm{s}^{*} h_\mathrm{s}^{3}}{1+h_\mathrm{f} / h_\mathrm{s}}\right]$)~\cite{10.1063/1.1478137} considered the film thickness and substrate bending effect by modifying Stoney's equation.\sloppy~Freund's modified Stoney's equation for thin substrate ($\sigma_\mathrm{f}=\frac{E_\mathrm{s}^{*} h_\mathrm{s}^{2}}{6 h_\mathrm{f} \rho \left(1+\frac{h_\mathrm{f}}{h_\mathrm{s}}\right)}\left[1+4 \frac{h_\mathrm{f}}{h_\mathrm{s}} \frac{E_\mathrm{f}^{*}}{E_\mathrm{s}^{*}}+6 \frac{h_\mathrm{f}^{2}}{h_\mathrm{s}^{2}} \frac{E_\mathrm{f}^{*}}{E_\mathrm{s}^{*}}+4 \frac{h_\mathrm{f}^{3}}{h_\mathrm{s}^{3}} \frac{E_\mathrm{f}^{*}}{E_\mathrm{s}^{*}}+\frac{h_\mathrm{f}^{4}}{h_\mathrm{s}^{4}} \frac{E_\mathrm{f}^{*}}{E_\mathrm{s}^{*}}\right]$)~\cite{10.1017/CBO9780511754715} also takes into account the film thickness effect by modifying Stoney's equation. However, computed results from Freund's equation diverge with increasing thickness. Biaxial stresses from all models get closer to FEA results with the decrease in thin film thickness. Here, the discrepancy in results between the Stoney and analytical equations depends on how we calculate the average $\mathrm{\rho}$ (e.g., over full width or partial width). The upper and lower bound of Stoney's formulas in ~\fig{compare_formulas} (a) and (b) are due to changes in areas used to calculate the average $\mathrm{\rho}$ (see SI~\sec{Radius_of_curvature} for further details).
\begin{figure}[h]
\centering
\subfloat[Radius of curvature for alumina.]{
\label{}
\includegraphics[width=0.40\linewidth]{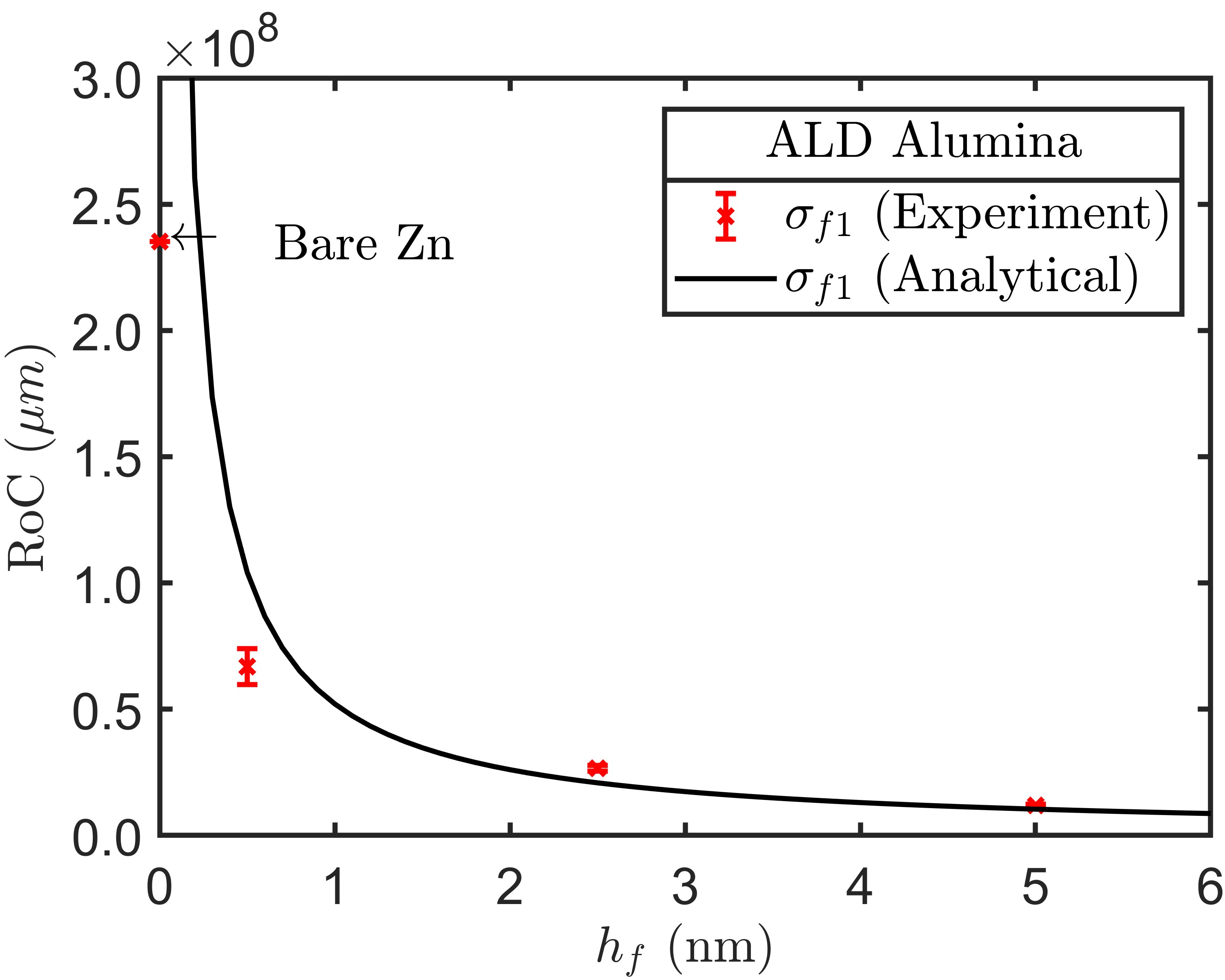}}
\qquad
\subfloat[Radius of curvature for alucone.]{
\label{}
\includegraphics[width=0.40\linewidth]{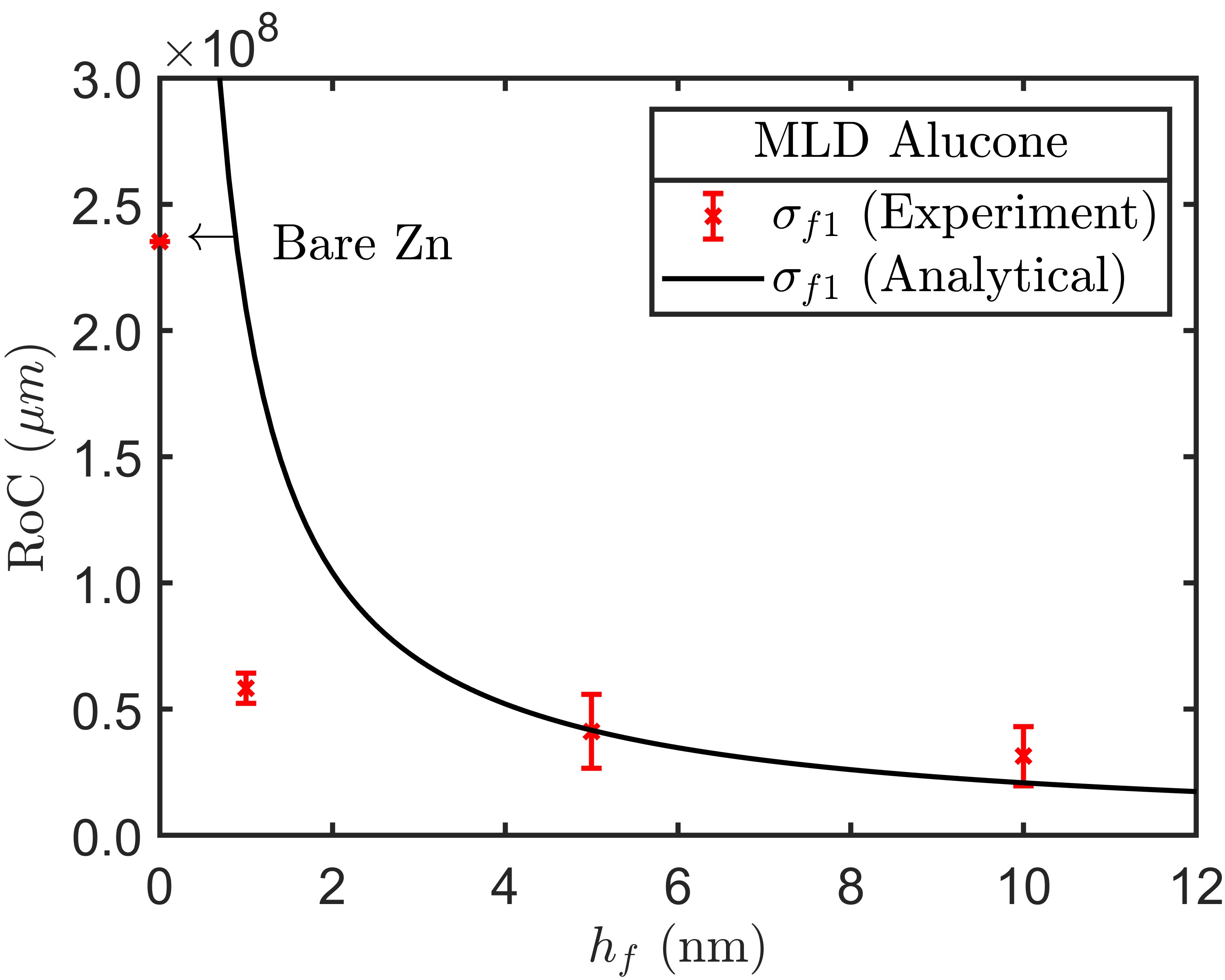}}
\qquad
\subfloat[Residual stress in film made by alumina.]{
\label{}
\includegraphics[width=0.40\linewidth]{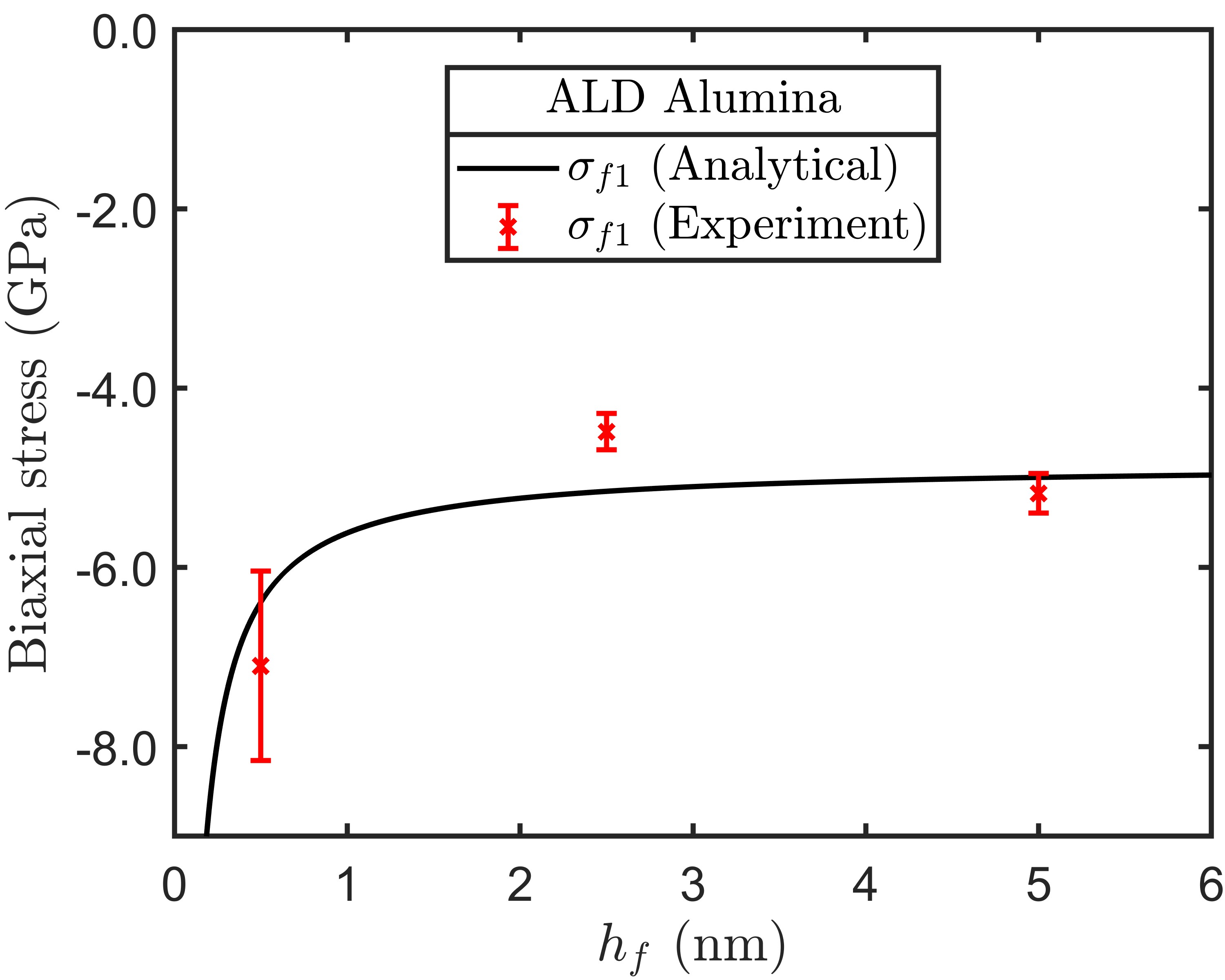}}
\qquad
\subfloat[Residual stress in film made by alucone.]{
\label{}
\includegraphics[width=0.40\linewidth]{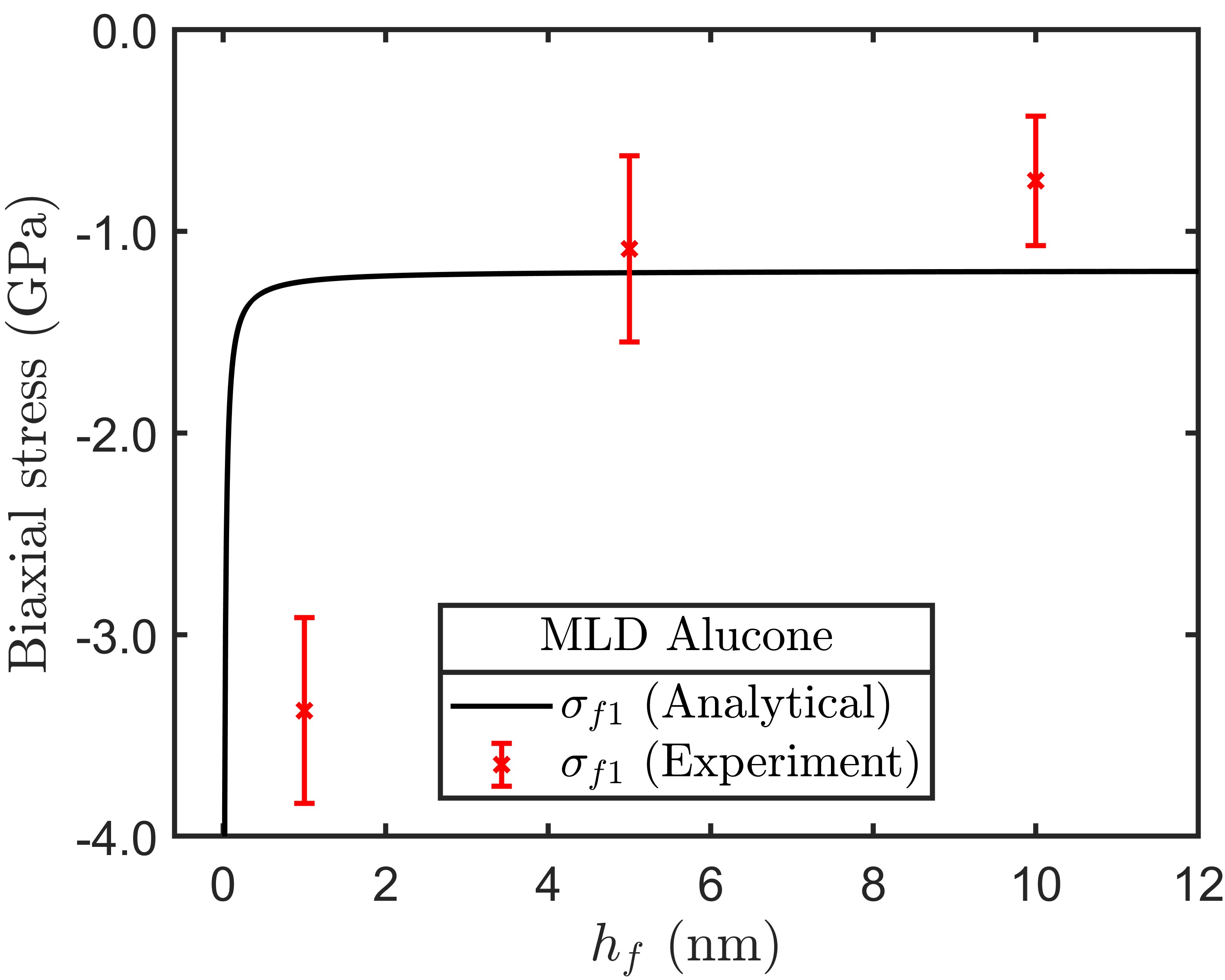}}
\caption{\label{fig:experimental_ROC_stress} Comparison of experimentally and theoretically measured radius of curvature (RoC, $\mathrm{\rho}$ in $\mu m$) - (a) ALD Alumina, and (b) MLD Alucone. Comparison of analytical residual stress with experimental residual stress derived from Stoney's formula- (c) ALD Alumina, and (d) MLD Alucone.}
\end{figure}
\subsection{Experimental evaluation of radius of curvature and residual stresses}
\par Radius of curvatures ($\mathrm{\rho}$) were measured experimentally for 1, 5, and 10 coated layers for both alumina and alucone coating on Zn surfaces as shown in~\fig{experimental_ROC_stress}. Here, the RoC for bare Zn is the average of three samples shown in~\fig{experiment_roc_bare_Zn}. Chu \textit{et. al.}~\cite{10.1063/1.334797} derived the radius of curvature for heteroepitaxial misfit-induced composite where the film is shorter than the substrate. Next, we modify the model to consider when the film is longer than the substrate for alumina and alucone coating over zinc. Because of the mutual constraints between the ALD/MLD film and metal substrate, elastic strains will develop to satisfy the need for strain compatibility in both the substrate and the film. Therefore, at the interface of the hetero-epitaxial system, the macroscopic boundary condition will be: ${a_\mathrm{f}[1-e_\mathrm{xx}^\mathrm{f}(F_\mathrm{f})+e_\mathrm{xx}^\mathrm{f}(M_\mathrm{f})] = a_\mathrm{s}[1+e_\mathrm{xx}^\mathrm{s}(F_\mathrm{s})-e_\mathrm{xx}^\mathrm{s}(M_\mathrm{s})]}$. Here, ${e_\mathrm{xx}^\mathrm{i}(F_\mathrm{i})}$ and ${e_\mathrm{xx}^\mathrm{i}(M_\mathrm{i})}$ are the effective strains due to mismatch force and mismatch moment for the film ($\mathrm{i=f}$) and substrate ($\mathrm{i=s}$), respectively. At equilibrium, the stress distribution inside the composite satisfies the force balance, moment balance, and boundary condition at the interface. Rewriting the formula for convex-up surface (film longer than bulk), we get ${\frac{1}{\rho} = \frac{E_\mathrm{f} h_\mathrm{f} h_\mathrm{s}(1+\frac{h_\mathrm{f}}{h_\mathrm{s}})\varepsilon_\mathrm{m}}{2A}}$ where, ${A = (1+\frac{a_\mathrm{s}}{a_\mathrm{f}}\frac{E_\mathrm{f}}{E_\mathrm{s}}\frac{h_\mathrm{f}}{h_\mathrm{s}})(\frac{E_\mathrm{s}I_\mathrm{s}}{L}+\frac{E_\mathrm{f}I_\mathrm{f}}{L}) - \frac{1}{4}E_\mathrm{f}h_\mathrm{f}^2h_\mathrm{s}(1+\frac{h_\mathrm{f}}{h_\mathrm{s}})(1+\frac{a_\mathrm{s}}{a_\mathrm{f}}\frac{h_\mathrm{s}}{h_\mathrm{f}})}$. Here, $L$ is the width of the interface.
\par~\fig{experimental_ROC_stress} shows the evolution of the RoC and residual stress for ALD/MLD coating as a function of the film thickness. The theoretical and experimental RoC profiles from~\fig{experimental_ROC_stress} (a) and (b) both show a nonlinear relation with increasing film thickness that the surface becomes more curved with the increase of thickness and RoC decreases. The experimental comparison between RoC for the experiment and the analytical expression for alumina coating is in reasonable agreement, whereas for alucone, more discrepancy is observed. This discrepancy could be due to several effects neglected, including the development of residual deformation during the handling/deposition process, thermal effects, and other residual stresses that might develop during the manufacturing of the films not considered in our model. 
\par ~\fig{experimental_ROC_stress} (c) and (d) show that the calculated residual stresses are in the GPa range, similar to the analytical model, and follow a similar trend that stress decreases with increasing thickness. However, the standard deviation is high for a few coating cases, which resembles a high standard deviation in calculating $\mathrm{\rho}$. Also, at this small length scale, it is well known that linear elasticity might not be valid to predict stresses accurately. The stresses for 1C MLD alucone are far from the analytical model, which is proportional to the mismatch of radius or curvature for 1C MLD alucone from the theoretical calculation. Sources of uncertainties could include the profilometer tips' limitations in capturing the surface roughness for 1 nm thin films and the uncertainties of experimental environments. These considerations point out the difficulty of experimentally measuring these residual stress values and, conversely, the need for models that can predict this behavior. 
\section{\label{sec:Conclusion}Conclusion}
%


We presented a novel bottom-up approach to investigate residual stresses due to the lattice misfit of ALD/MLD coatings. Leveraging \textit{ab initio} informed continuum models, we studied the impact of residual stresses in Zn films typically used in Zn-ion batteries. Notably, we have found that even though the thicknesses of these coatings are a few nanometers, they can greatly impact the residual stresses of Zn films. In particular, at the interface between Zn/coating, large compressive stresses develop (e.g., between $\sim3-5$ GPa for ALD alumina and $\sim0.85-1.2$ GPa for MLD alucone), which diminish with coating thickness. Our continuum model was validated through finite element analysis and indirect experimental measurements, demonstrating reasonable agreement with these methodologies. Our study suggests that residual stresses due to ALD/MLD coatings -largely considered negligible effect without any rigorous evidence- can have significant importance in the chemo-mechanical performance of Zn thin films used in batteries. Thus, our study suggests that the lattice misfit-induced residual stress made by ALD/MLD coatings must be considered when predicting dendrite formation of the anodic operation of batteries to understand better and evaluate the morphology of battery interfaces.

\section{\label{sec:Acknowledgements}Acknowledgements}
We acknowledge the support of the New Frontiers in Research Fund (NFRFE-2019-01095) and the Natural Sciences and Engineering Research Council of Canada (NSERC) through the Discovery Grant under Award Application Number 2016-06114. M.G. gratefully acknowledges the financial support from the Department of Mechanical Engineering at UBC through the Four Years Fellowship,  the Collaborative Research Mobility Award (UBC CRMA), the UBC Eminence program (Battery Innovation Cluster), and the Institute for Computing, Information and Cognitive Systems (ICICS). This research was supported through high-performance computational resources and services provided by Advanced Research Computing at the University of British Columbia and the Digital Research Alliance of Canada.
\bibliographystyle{ieeetr}
\bibliography{manuscript.bib}

\clearpage
\onecolumngrid
\raggedbottom
\setcounter{section}{0}
\setcounter{equation}{0}
\setcounter{figure}{0}
\setcounter{table}{0}
\setcounter{page}{1}
\makeatletter
\renewcommand{\thesection}{S\arabic{section}}
\renewcommand{\theequation}{S\arabic{equation}}
\renewcommand{\thefigure}{S\arabic{figure}}
\renewcommand{\thetable}{S\arabic{table}}
\renewcommand{\thepage}{SM\arabic{page}}

\renewcommand{\bibnumfmt}[1]{[S#1]}
\renewcommand{\citenumfont}[1]{S#1}

\begin{center}
{\Large \bf Supporting Information}
\end{center}

\section{\label{sec:ALD_MLD}Atomic layer depostion}
Atomic layer deposition (ALD)~\cite{suntola1977method} is an advantageous thin film deposition approaches that allows lower deposition temperatures and precise reactions between surface functional groups and chemical precursors. However, the concern is the stability of the thin film, as most nanoparticle thin films (NTFs) are subject to experience mechanical failure under moderate abrasion, even with a small load. This instability of thin film is likely due to the weak adhesion of NTFs to the substrate and the weak cohesive strength between substrate-thin film nanoparticles. By analyzing the Zn surface with ALD coating, we highlight the importance of a stable coating to improve the performance of AZIB.
ALD achieves the remarkable ability to precisely manage material deposition with atomic scale by following a sequential exposure of gas-phase precursors to the substrate, which leads to saturated and self-limiting surface reactions ~\cite{10.1002/adma.200700079, 10.1039/D0CS00156B}. 
Structural and interfacial problems have been addressed using these thin surface coating films on the electrodes.
Furthermore, the ALD technique can control nucleation sites, such as hydroxyl groups, heteroatom-doping sites, and defect sites, to deposit nanoparticles on substrates~\cite{10.1016/j.mattod.2014.04.026}.

Huibing~\emph{et al.}~\cite{10.1039/D0TA00748J} used the ALD technique to develop an alumina ($\mathrm{Al_{2}O_{3}}$) coating, which boosts the Zn anodes' rechargeability for AZIBs addressing the aforesaid problems.
$\mathrm{Al_{2}O_{3}}$ coating on Zn anodes suppressed the growth of Zn dendrites, enhanced the wettability of Zn, and inhibited corrosion which significantly improved the lifetime of the system.
ALD growth process has benefits over other coating approaches in several parameters, such as uniformity of deposition and conformity of films with tunable thickness~\cite{10.1080/14686996.2019.1599694}. ALD growth of $\mathrm{Al_{2}O_{3}}$ has been studied broadly throughout the last couple of years, and the growth mechanism of ALD $\mathrm{Al_{2}O_{3}}$ has been established on the chemical vapor deposition (CVD) reaction~\cite{10.1016/S0040-6090(02)00438-8}:
\begin{equation}
\label{eq:CVD_Alumina}
\mathrm{2Al(CH_{3})_{3}+H_{2}O} = \mathrm{Al_{2}O_{3}+6CH_{4}}
\end{equation} 
\begin{figure}[h]
\centering
\includegraphics[width=0.80\linewidth]{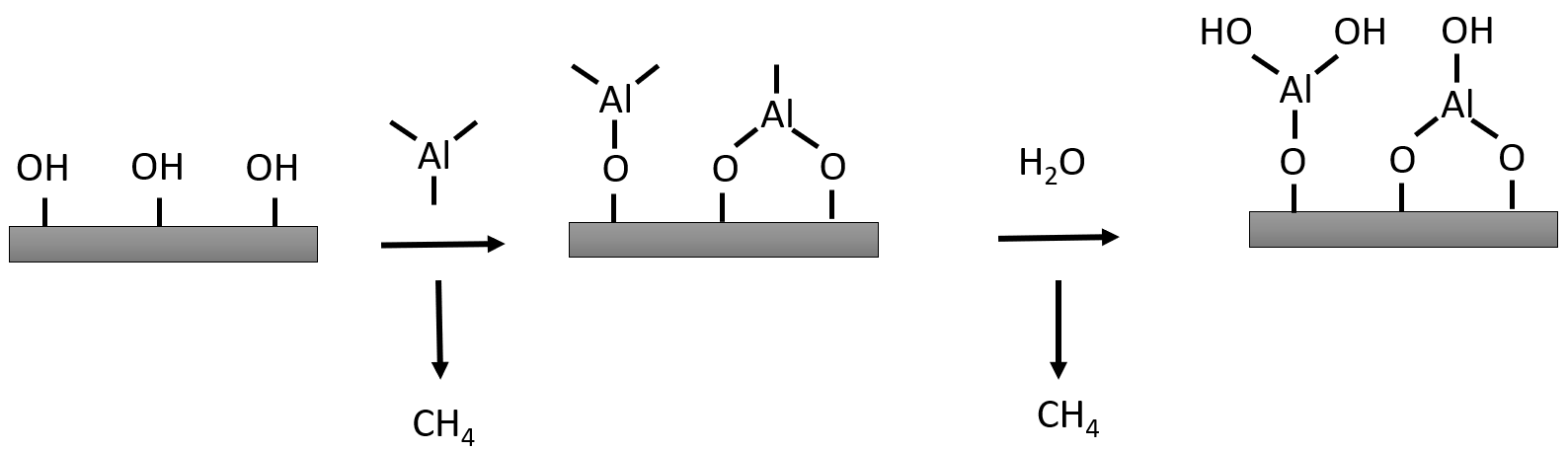}
\caption{Schematic representation of atomic layer deposition process of alumina on Zn substrate.}
\label{fig:ALD_Alumina_schematic}
\end{figure}
For ALD $\mathrm{Al_{2}O_{3}}$ growth, two precursors trimethylaluminum (TMA) and $\mathrm{H_{2}O}$ are exposed to the Zn surface sequentially for self-saturated surface reactions as shown in~\fig{ALD_Alumina_schematic}. Firstly, TMA is exposed to react with the surface hydroxyl groups. This reaction continues until the hydroxyl group in the surface and TMA reaction reach completion. Secondly, excess TMA and by-products ($\mathrm{CH_{4}}$) are purged away. Thirdly, $\mathrm{H_{2}O}$ is exposed to the surface, and the same self-saturation reactions are carried on. The $\mathrm{H_{2}O}$ precursor reacts with surface methyl groups from the previous reaction. Fourthly, the excess $\mathrm{H_{2}O}$ and by-products ($\mathrm{CH_{4}}$) are purged away. The exposure of TMA (A) and $\mathrm{H_{2}O}$ (B) compose one AB cycle. These AB growth cycles are repeated until the desired coating thickness is achieved. The TMA and $\mathrm{H_{2}O}$ form ALD $\mathrm{Al_{2}O_{3}}$ as shown here~\cite{10.1021/cm0304546, 10.1021/cm050704d, 10.1063/1.3567912, 10.1016/S0040-6090(96)08934-1}:
\begin{equation}
\begin{split}
\label{eq:ALD_Alumina}
\mathrm{(A)\: Al-OH^{*} + Al(CH_{3})_{3}} &= \mathrm{Al-O-(CH_{3})^{*}_{2}+CH_{4}} \\
\mathrm{(B)\: Al-CH^{*}_{3} +H_{2}O} &= \mathrm{ Al-OH^{*}+CH_{4}}.\end{split}
\end{equation} 
Here, the asterisks indicate the surface components. When executed in an ABAB... cycles arrangement, these self-saturating reactions provide atomically controlled $\mathrm{Al_{2}O_{3}}$ thin film. Here, the growth of $\mathrm{Al_{2}O_{3}}$ is linear with the recurrence number of AB cycles. Previous studies have shown the conformal and linear growth process of $\mathrm{Al_{2}O_{3}}$ using TMA and $\mathrm{H_{2}O}$ \cite{10.1021/cm050704d}.
\subsection{\label{sec:relaxted_structure}Relaxed structure}
The bond lengths for the relaxed structure are shown in Table \ref{tab:bond_length}.
\begin{table}[H]
    \caption{\label{tab:bond_length}Bond lengths of relaxed ALD alumina model.}
    \centering
      \begin{tabular}{cc}
    	Atom & Bond Length (Å)\\ \hline
        $\textrm{Al - O}$ & $1.87-1.94$\\
        $\textrm{O - H}$ & $0.97-1.00$\\
        $\textrm{Zn - O}$ & $2.00-2.26$\\
      \end{tabular}
\end{table}  
\subsection{\label{sec:vdW_calculation}Van der Waals interactions}
\begin{figure}[h]
\centering
\includegraphics[width=0.45\linewidth]{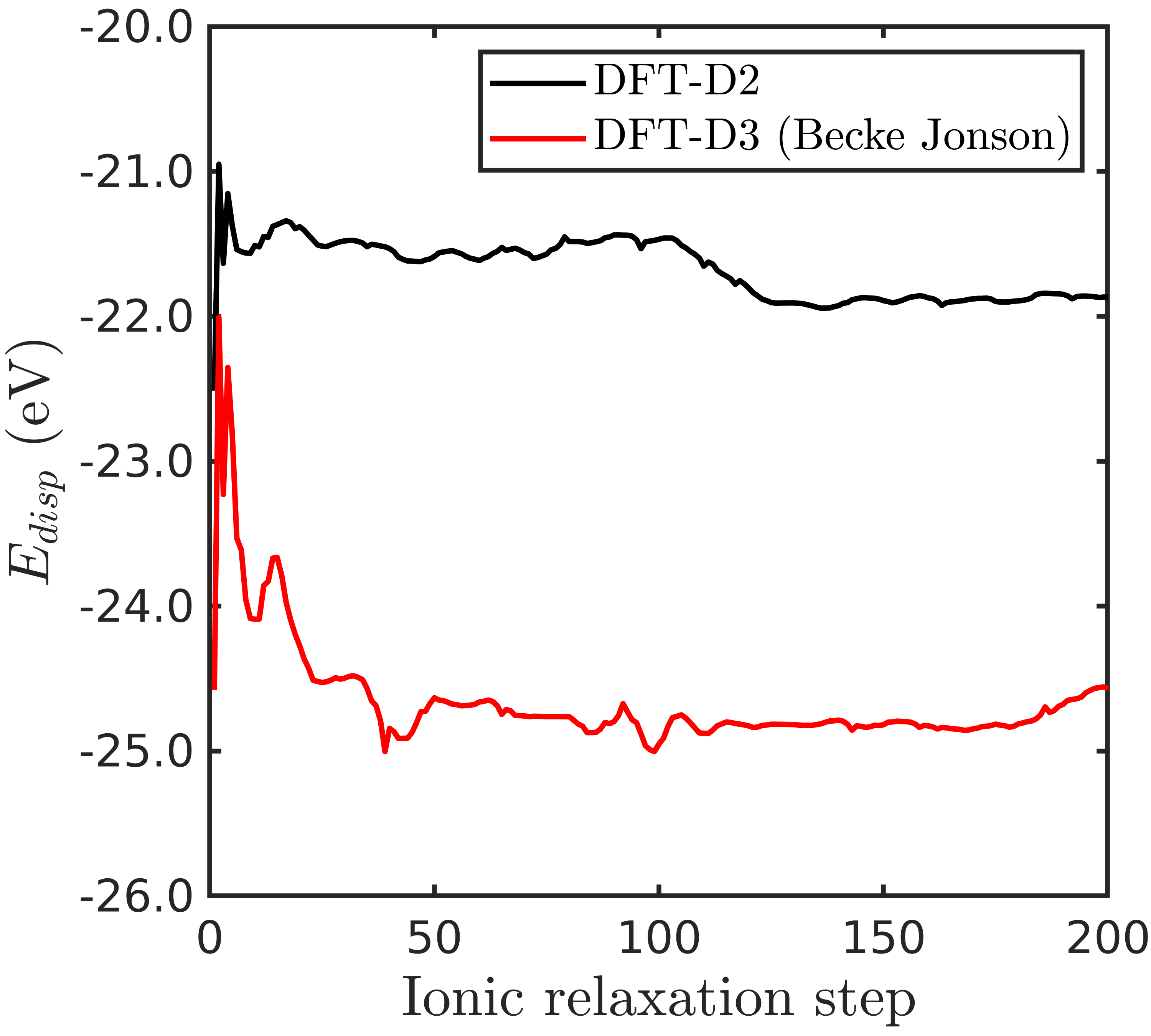}
\caption{Contribution of van der Waals interactions with ionic relaxation steps. Here, the correction term $E_\mathrm{{disp}}$ contains the dispersion coefficient and damping function to evaluate interactions over a suitably chosen cutoff radius.}
\label{fig:vdW}
\end{figure}

In this study, we neglected the effects of van der Waals (vdW) forces in the simulation model for understanding the surface properties~\cite{10.1038/sdata.2016.80}. On average, the effect of dispersion force is -21.5 and -24.6 eV for DFT-D2\cite{10.1002/jcc.20495} and DFT-D3\cite{10.1063/1.3382344} (Becke-Jonson), respectively, as displayed in~\fig{vdW}. The vdW contribution is approximately 7.5 and 8.5 percent compared to the total ground state energy, which sets an error bar of $7.5-8.5\%$ for calculating surface properties. Thus, having measured the error bound, we proceeded to neglect the vdW contributions due to their relatively small effect. 
\subsection{\label{Adsorption_energy}Adsorption energy}
The adsorption process of $\mathrm{Al_{2}O_{3}}$ on hydroxylated Zn surface using the ALD technique can be represented using the following equation, the \emph{adsorption equation},
\begin{equation}
\begin{split}
\label{eq:adsorption_eqn}
\mathrm{6Zn-OH + 6Al(CH_{3})_{3}+9H_{2}O} = \mathrm{6 Zn - O - Al -(OH)_{2}+15 CH_{4}}.
\end{split}
\end{equation} 

The adsorption energy of ALD alumina on a hydroxylated Zn surface is calculated as follows:
\begin{equation}
\begin{split}
\label{eq:adsorption_energy_eqn}
E_{ads} = (6E_{\mathrm{Zn - O - Al -(OH)_{2}}} + 15E_{\mathrm{{CH_{4}}}}- 6E_{\mathrm{Zn-OH}}- 6E_{\mathrm{Al(CH_{3})_{3}}}- 9E_{\mathrm{H_{2}O}})/6,
\end{split}
\end{equation} 
where, $E_{\mathrm{Zn - O - Al -(OH)_{2}}}$ is the total energy of ALD alumina adsorbed hydroxylated Zn surface, $E_{\mathrm{Zn-OH}}$ is the total energy of hydroxylated Zn surface, $E_{\mathrm{Al(CH_{3})_{3}}}$ is the total energy of trimethyl aluminum, $E_{\mathrm{H_{2}O}}$ is the total energy of $\mathrm{{H_{2}O}}$ and $E_{\mathrm{CH_{4}}}$ is the total energy of $\mathrm{{CH_{4}}}$. By definition from equation~\eq{adsorption_energy_eqn}, a negative value for $\mathrm{E_{ads}}$ states an exothermic reaction with favorable attractive adsorption of precursors on the surface.

The simulated results of the adsorption energy were:
\begin{center}\fbox{\parbox{0.5\linewidth}{favourable
 	$E_{\mathrm{ads}} = -1.94\text{\:eV}$ (without vdW)\\
 	$E_{\mathrm{ads,DFT-D2}} = -5.37\text{\:eV}$\\
 	$E_{\mathrm{ads, DFT-D3}} = -5.78\text{\:eV}$}}
\end{center}

As expected, from the analysis shown in~\fig{vdW}, the inclusion of the van der Waals correction term ($\mathrm{E_{disp}}$) raises the overall adsorption energy compared to the pure functional. The variation in contribution in adsorption energy is in the order of $-3.43$ to $-3.84$ eV per absorbate molecule.

\subsection{Interpolating the Strain-Dependent Surface Energy}
The values of the first and second derivatives of the surface energy as a function of the components of the strain tensor are shown in Table \ref{tab:FDM_HOLMES_LME}. 

\begin{table}[H]
    \caption{\label{tab:FDM_HOLMES_LME} Comparison among Finite Difference Method (FDM), Local Maximum Entropy (LMS) and Higher Order LME Scheme (HOLMES) method for slope ($\mathrm{d}$) and curvature ($\mathrm{d^{'}}$) to evaluate surface elastic constants $(\mathrm{{\it{C}}_{ijkl}^{s}})$ for $\mathrm{Al_{2}O_{3}}$ deposited hydroxylated zinc thin film.}
    \centering
    \begin{tabular}{c c c c c }
Biaxial strain \\($\varepsilon_{11},\varepsilon_{22}$) & FDM (4th order) & LME & HOLMES \\ 
 \hline
$d_{11}\:({d{\gamma}/d{\varepsilon}_{11}}$) & -0.11 & -0.11 & -0.11 \\
$d_{22}\:({d{\gamma}/d{\varepsilon}_{22}}$) & -0.12 & -0.12 & -0.12 \\ 
$d^{'}_{11}\:({d^{2}{\gamma}/d{\varepsilon}_{11}}^{2}$) & 9.38 & 8.89 & 9.12 \\ 
$d^{'}_{22}\:({d^{2}{\gamma}/d{\varepsilon}_{22}}^{2}$) & 5.75 & 5.89 & 5.82 \\ 
$d^{'}_{12}\:({d^{2}{\gamma}/d{\varepsilon}_{11}}d{\varepsilon}_{22}$) & 1.78 & 1.84 & 1.63 \\
\hline
Biaxial strain \\($\varepsilon_{12},\varepsilon_{22}$) & FDM (4th order) & LME & HOLMES \\ 
\hline
$d_{12}\:({d{\gamma}/d{\varepsilon}_{11}}$) & -0.12 & -0.12 & -0.12 \\
$d_{22}\:({d{\gamma}/d{\varepsilon}_{22}}$) & 0.00 & 0.00 & 0.00 \\ 
$d^{'}_{12}\:({d^{2}{\gamma}/d{\varepsilon}_{12}}^{2}$) & 5.75 & 5.89 & 5.83 \\ 
$d^{'}_{22}\:({d^{2}{\gamma}/d{\varepsilon}_{22}}^{2}$) & 0.00 & 0.00 & 0.00 \\ 
$d^{'}_{12}\:({d^{2}{\gamma}/d{\varepsilon}_{12}}d{\varepsilon}_{22}$) & 0.68 & 0.40 & 0.99 \\
\hline
Biaxial strain \\($\varepsilon_{11},\varepsilon_{12}$) & FDM (4th order) & LME & HOLMES \\ 
\hline
$d_{11}\:({d{\gamma}/d{\varepsilon}_{11}}$) & -0.11 & -0.11 & -0.11 \\
$d_{12}\:({d{\gamma}/d{\varepsilon}_{12}}$) & 0.00 & 0.00 & 0.00 \\ 
$d^{'}_{11}\:({d^{2}{\gamma}/d{\varepsilon}_{11}}^{2}$) & 9.38 & 8.89 & 9.12 \\ 
$d^{'}_{12}\:({d^{2}{\gamma}/d{\varepsilon}_{12}}^{2}$) & 0.00 & 0.00 & 0.00 \\ 
$d^{'}_{12}\:({d^{2}{\gamma}/d{\varepsilon}_{11}}d{\varepsilon}_{12}$) & 0.06 & 0.01 & 0.10 \\
\end{tabular}
\end{table}
\begin{figure}[H]  
\centering
\label{}
\includegraphics[width=0.40\linewidth]{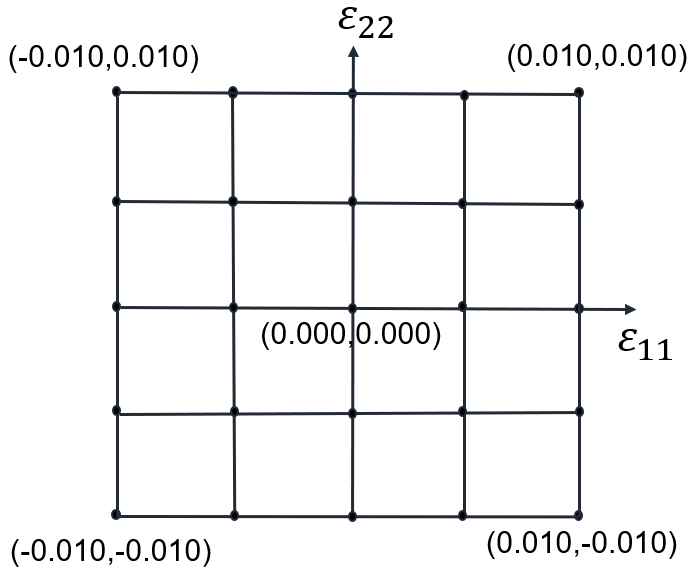}
\qquad
\caption{
\label{fig:schematic_surface_energy_with_strain} Schematic diagram of infinitesimal strain grid applied on the thin film.}
\end{figure}
\section{\label{sec:Implementation_of_multi_layer_misfit}Implementation of multi-layer misfit}
\par We modified the Vasp program and added option ISIF =8, which allowed a principal degree of freedom for cell volume and atom positions but not for cell shape (c/a ratio and angle between a,b,c axis will be fixed). Main.F file was modified in the following way to incorporate only ionic positions and cell volume as degrees-of-freedom by ISIF=8. 
\begin{verbatim} 
IF ((DYN%ISIF<5).OR.(DYN%ISIF==8))
IF ((DYN%ISIF==7).OR.(DYN%ISIF==8))
\end{verbatim}
\section{\label{sec:Radius_of_curvature}Radius of curvature calculation}
\begin{figure}[H]
\centering
\subfloat[]{
\label{}
\includegraphics[width=0.44\linewidth]{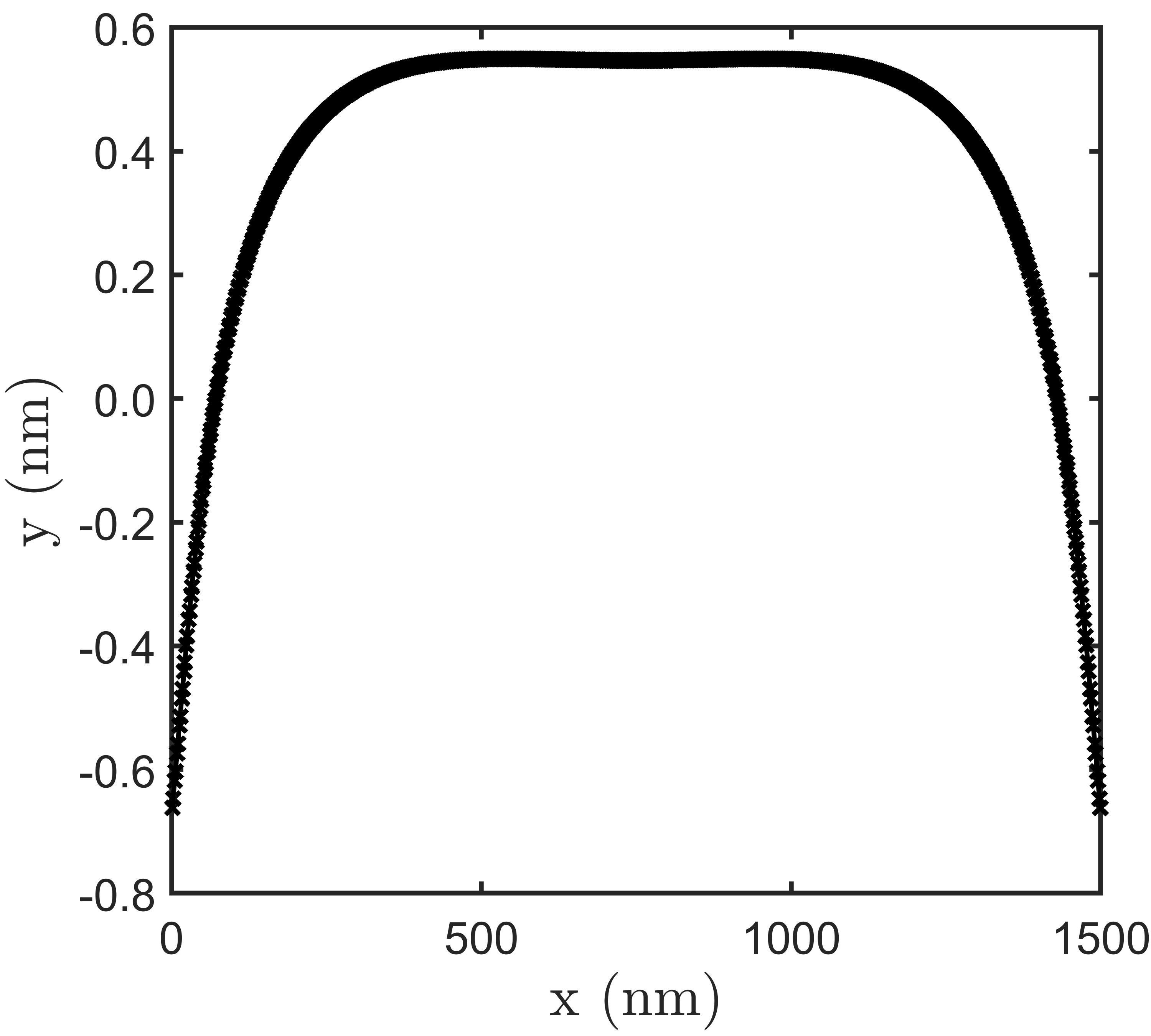}}
\qquad
\subfloat[]{
\label{}
\includegraphics[width=0.44\linewidth]{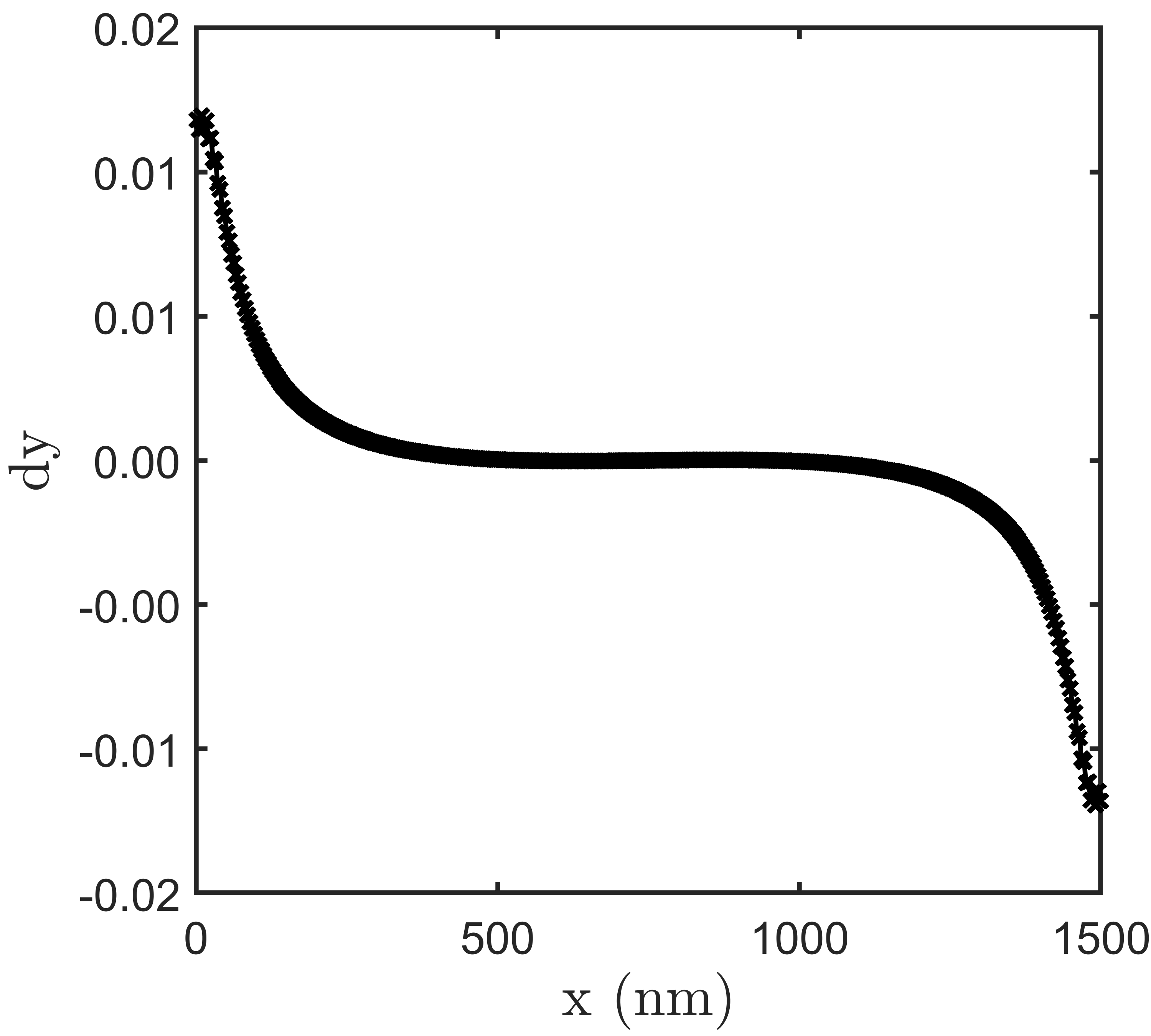}}
\qquad
\subfloat[]{
\label{}
\includegraphics[width=0.44\linewidth]{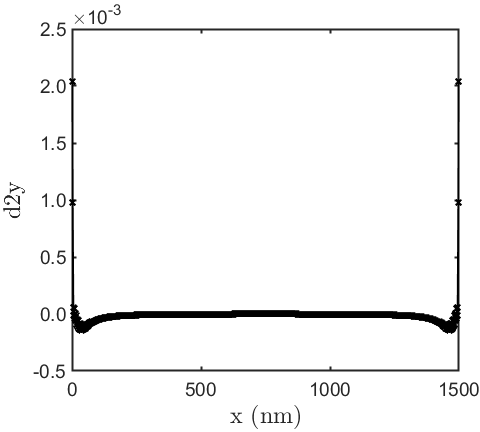}}
\qquad
\subfloat[]{
\label{}
\includegraphics[width=0.45\linewidth]{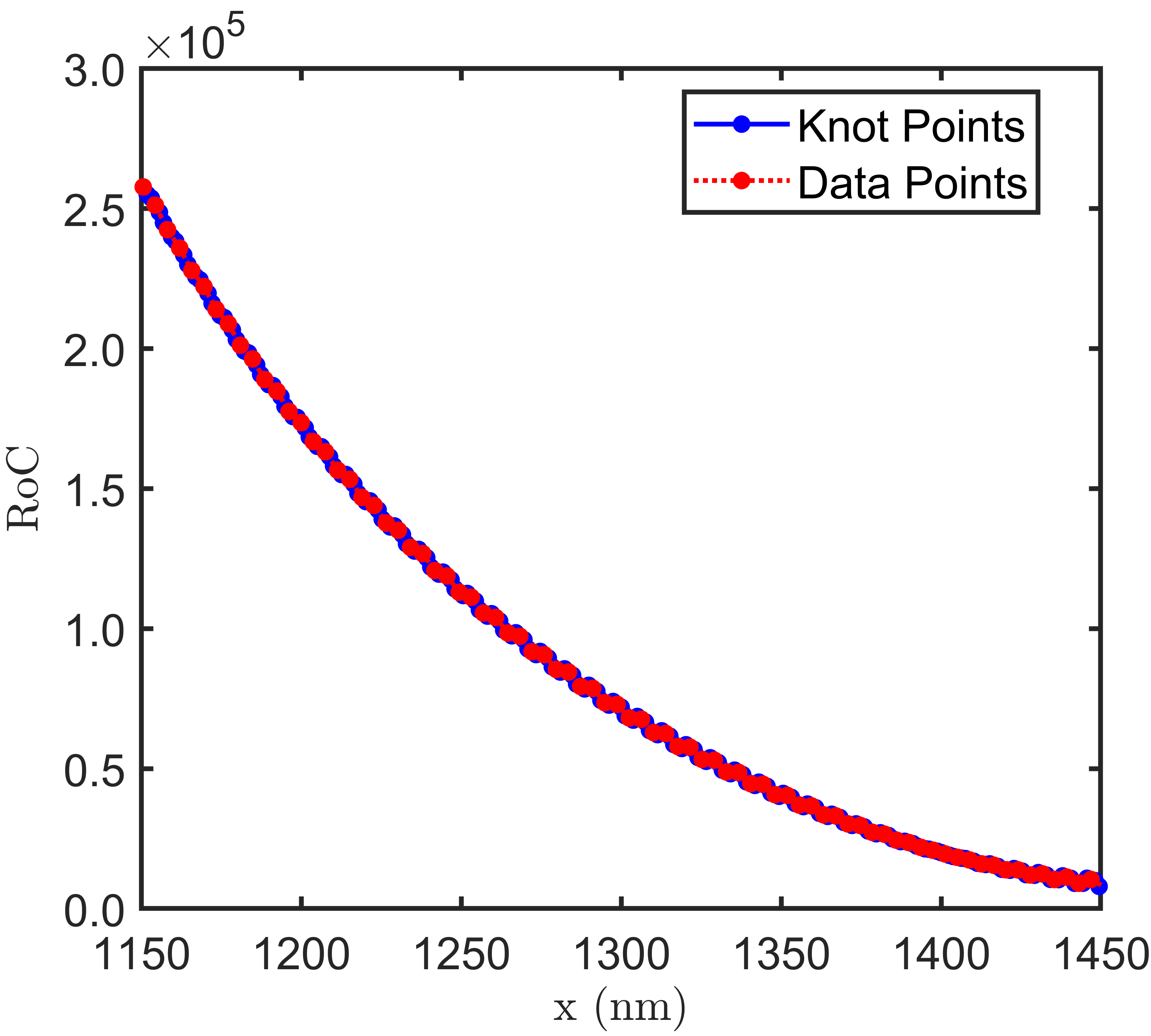}}
\caption {
\label{fig:ROC} The radius of curvature (RoC $ = \frac{(1+(\frac{dy}{dx})^2)^{3/2}}{\lvert \frac{d^2y}{dx^2} \rvert}$) is calculated by extracting (a) the displacements y from FEA model and by calculating the derivative (b) $\frac{dy}{dx}$ and (c) $\frac{d^2y}{dx^2}$. The RoC along the x direction for 50 nm MLD alucone film is shown in (d).  }
\end{figure}
We measured the surface displacement after alumina and alucone deposition from FEA and calculated the average radius of curvature (RoC) to use in Stoney's equations as shown in~\fig{ROC}. The upper and lower bound using the shaded area in~\fig{compare_formulas}(a) and (b) were due to the change of $\mathrm{\rho}$. For example, for a 50 nm MLD alucone film, the RoC was $1.099\times 10^{5}$ $\mathrm{nm}$ for the maximum range ($x$= 1100 - 1500 nm), which represents biaxial stresses at the lower bound of the shaded area in~\fig{compare_formulas}(b). Before $x$=1100 nm, the curvature was infinite as the middle section of the surface was flat. If we change the range ($x$ = 1150 - 1350 nm), the average RoC was $1.25\times 10^{5}$ $\mathrm{nm}$, and the measured stresses using Stoney's equations were closer to the analytical formula as shown by the upper bound of the shaded area. The average curvature was calculated in the range of $x$ = 1200 - 1300 nm for biaxial stresses represented by different points inside the shaded area. 

\begin{figure}[H]
\centering
\label{}
\includegraphics[width=0.45\linewidth]{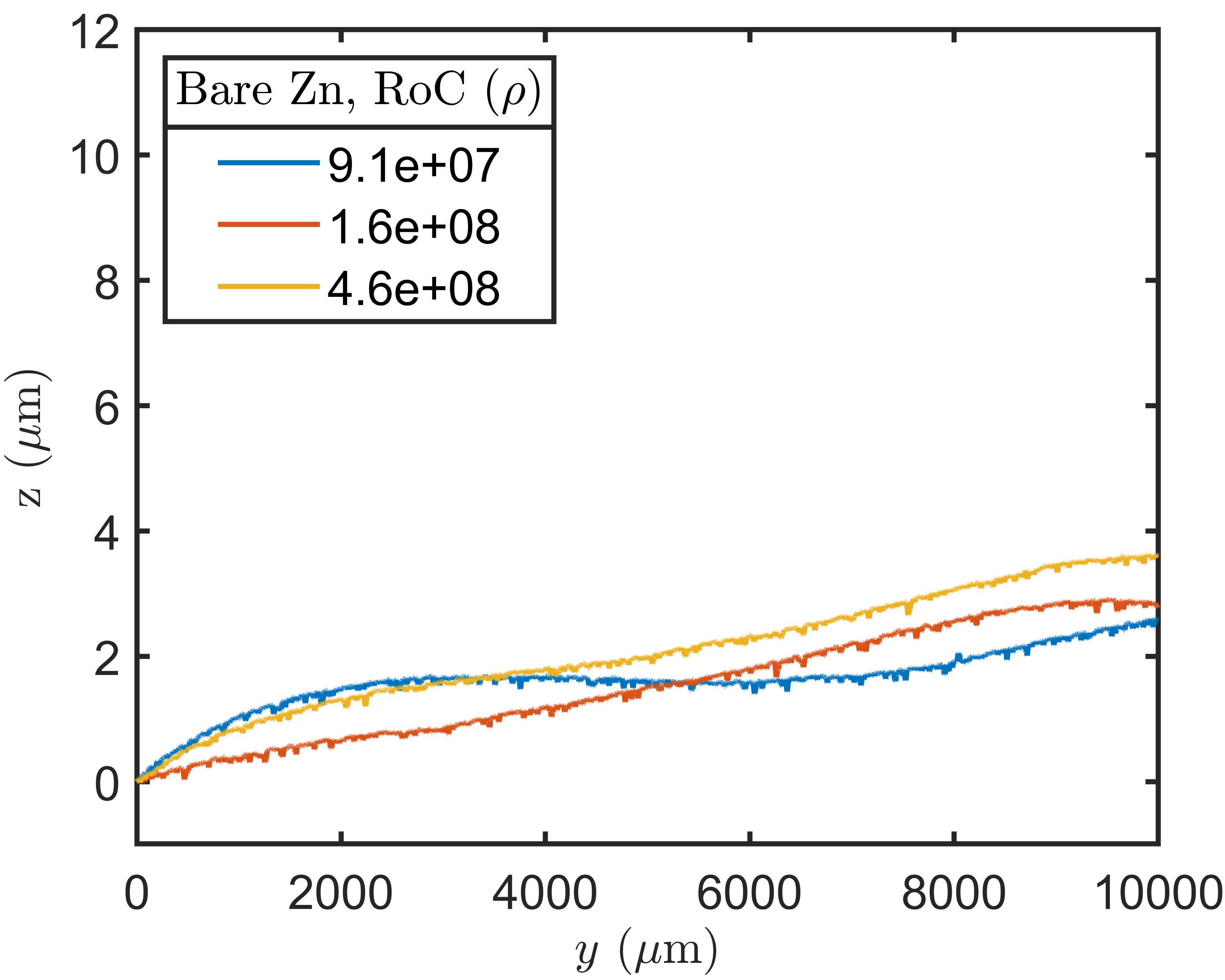}
\caption {
\label{fig:experiment_roc_bare_Zn} Radius of curvature (RoC) of bare Zn from scanning profilometer experiment for scanning of 10 mm. }
\end{figure}
\section{Orientational mismatch between film and bulk material}
\par Consider a thin film epitaxially grown on the flat surface of a moderately thick substrate. In this configuration, let us introduce a rectangular coordinate system with its origin located on the substrate-film interface. The interface of this coordinate frame, which we refer to as the \emph{global coordinate frame}, denotes $x_{1}x_{2}$-plane as the interface, and the $x_{3}$ -axis expands into the film. If the thin film properties are specified with respect to some ‘natural’ rectangular $x^{\star}_{1}x^{\star}_{2}x^{\star}_{3}$ -coordinate frame directed to the film material itself (where ‘$\star$’ is employed to denote that the physical properties in terms of the material coordinate system), the relation between local and global coordinate system will be $x_{k}^{\star} = Q_{ki}x_{i}$. Here, the rotation matrix ($Q$) can be determined from the unit vectors ($e$) as $Q_{ki} = e_{k}^{\star}.e_{i}$. Therefore, the strain tensor and mismatch stress from the material to the global coordinate system can be presented as~\cite{10.1017/CBO9780511754715}:
  \begin{equation} \label{eq:eqns_for_coordiate_transformation}
  \begin{split} 
   \varepsilon_{ij}^{\star(m)}&=Q_{ik}\varepsilon_{kl}^{m}Q_{jl}\\
   \sigma_{i}^{\star(m)}&=c_{ij}^{\star}\varepsilon_{j}^{\star(m)}\\
   \sigma_{ij}^{m}&=Q_{ki}\sigma_{kl}^{\star (m)}Q_{lj}\\
   c_{ijkl} &= Q_{mi}Q_{nj}Q_{pk}Q_{ql}C^{\star}_{mnpq}\\
   \sigma_{i}^{m} &= c_{ij}\varepsilon_{j}^{m}
   \end{split}
   \end{equation} 
Point to be noted here, before converting a component defined by a tensor of any degree of~\eq{eqns_for_coordiate_transformation} from one coordinate system to another, the array must have been converted into its \emph{bona-fide} tensor form~\cite{10.1017/CBO9780511754715}. For example, column matrices ($\mathrm{\sigma_{i}}$, $\mathrm{\varepsilon_{j}}$) have to be converted to their original form ($\mathrm{\sigma_{ij}}$, $\mathrm{\varepsilon_{ij}}$).
\section{\label{sec:Comparing_Analytical_Model_with_Finite_Element_Analysis} Finite element analysis}
\begin{figure}[H]
\centering
\subfloat[In-plane stresses developed by ALD alumina coating of $h_f=30$ nm, and $h_s=500$ nm. ]{
\label{}
\includegraphics[width=0.90\linewidth]{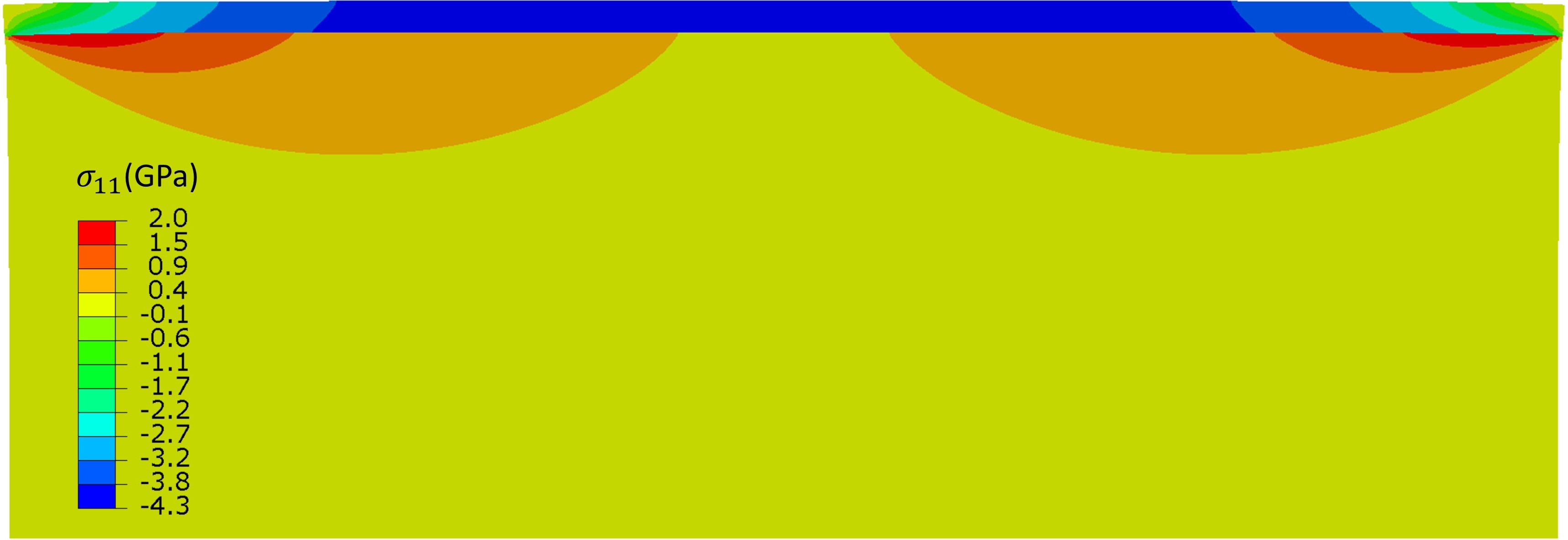}}
\qquad
\subfloat[In-plane stresses developed by MLD alucone coating of $h_f=30$ nm, and $h_s=500$ nm.]{
\label{}
\includegraphics[width=0.90\linewidth]{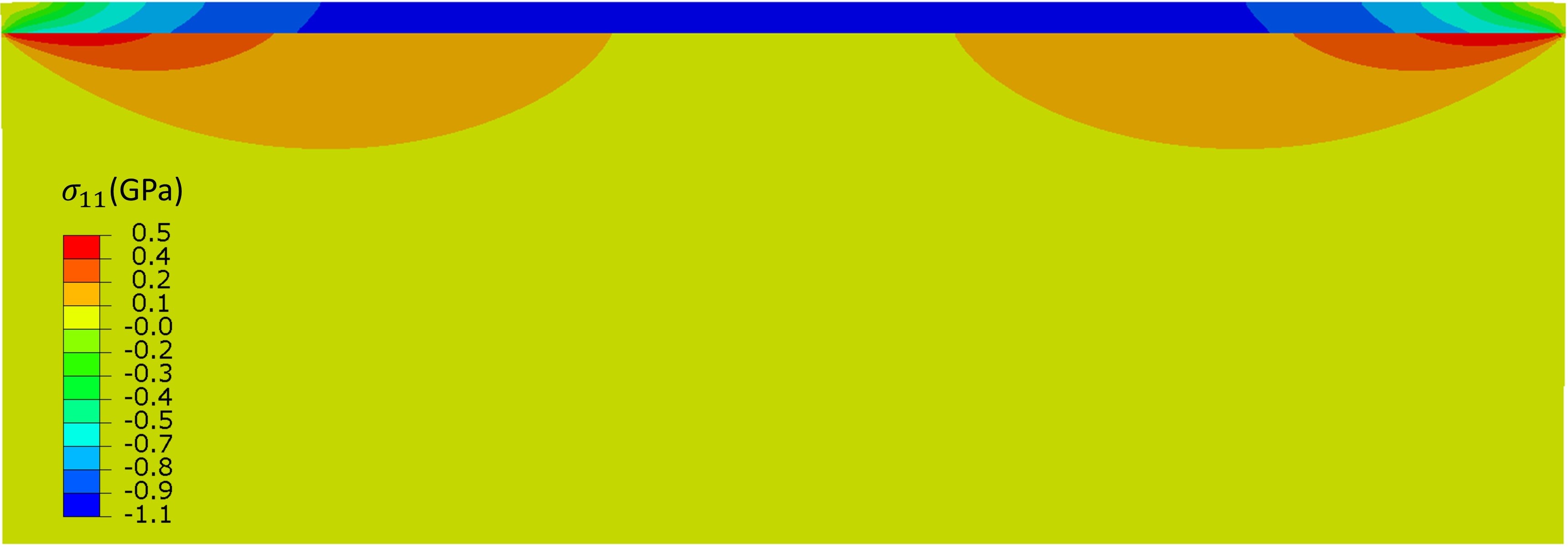}}
\caption {
30~nm thin film on 500 nm thick substrate (a) ALD alumina and (B) MLD alucone. Here, the width (L) of the film-substrate system is 1500 nm.
\label{fig:finite_element_modeling}}
\end{figure}
\section{Interfacial shear stress from FEA}
\begin{figure}[h]
\centering
\subfloat[]{
\label{}
\includegraphics[width=0.43\linewidth]{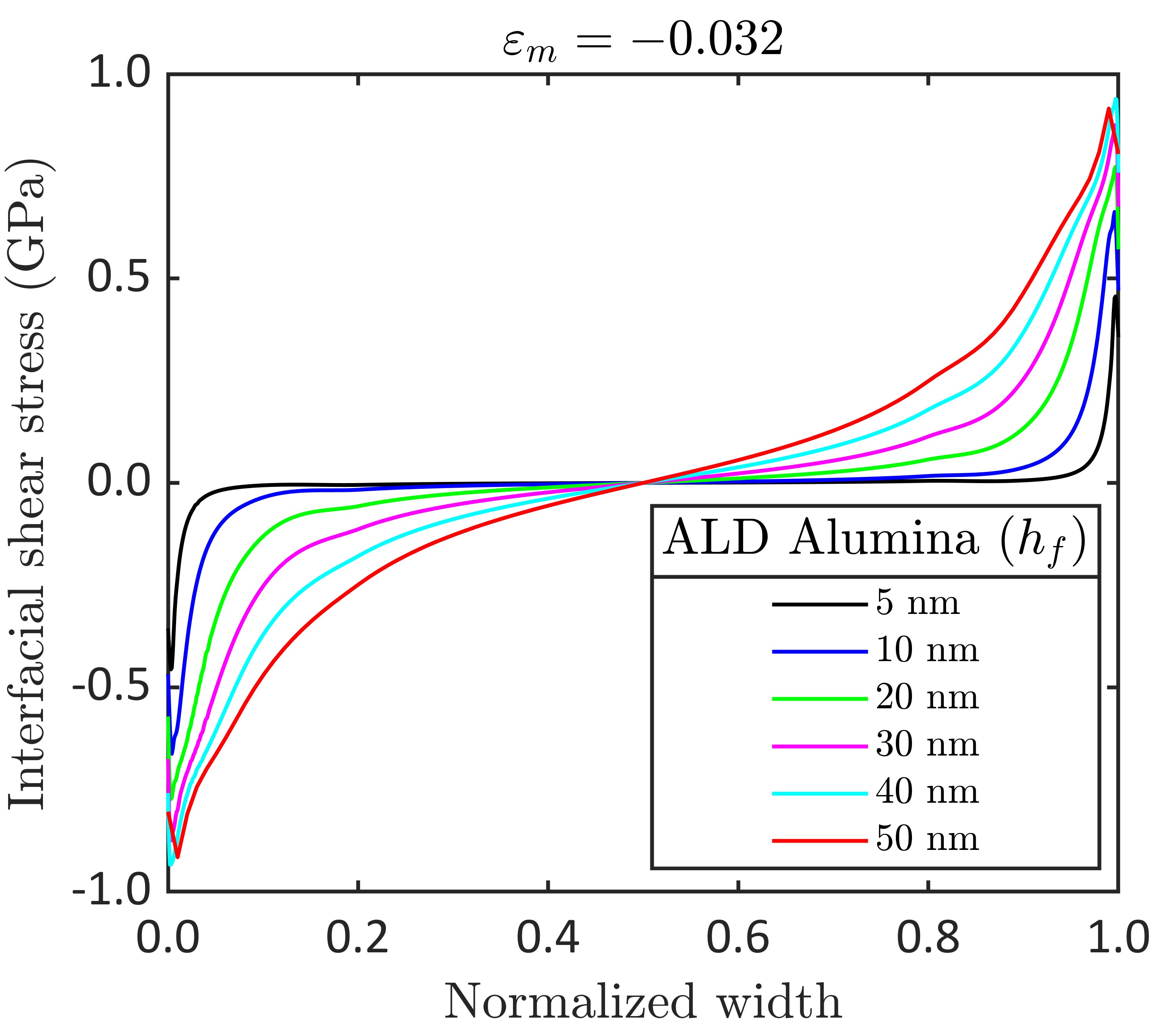}}
\qquad
\subfloat[]{
\label{}
\includegraphics[width=0.43\linewidth]{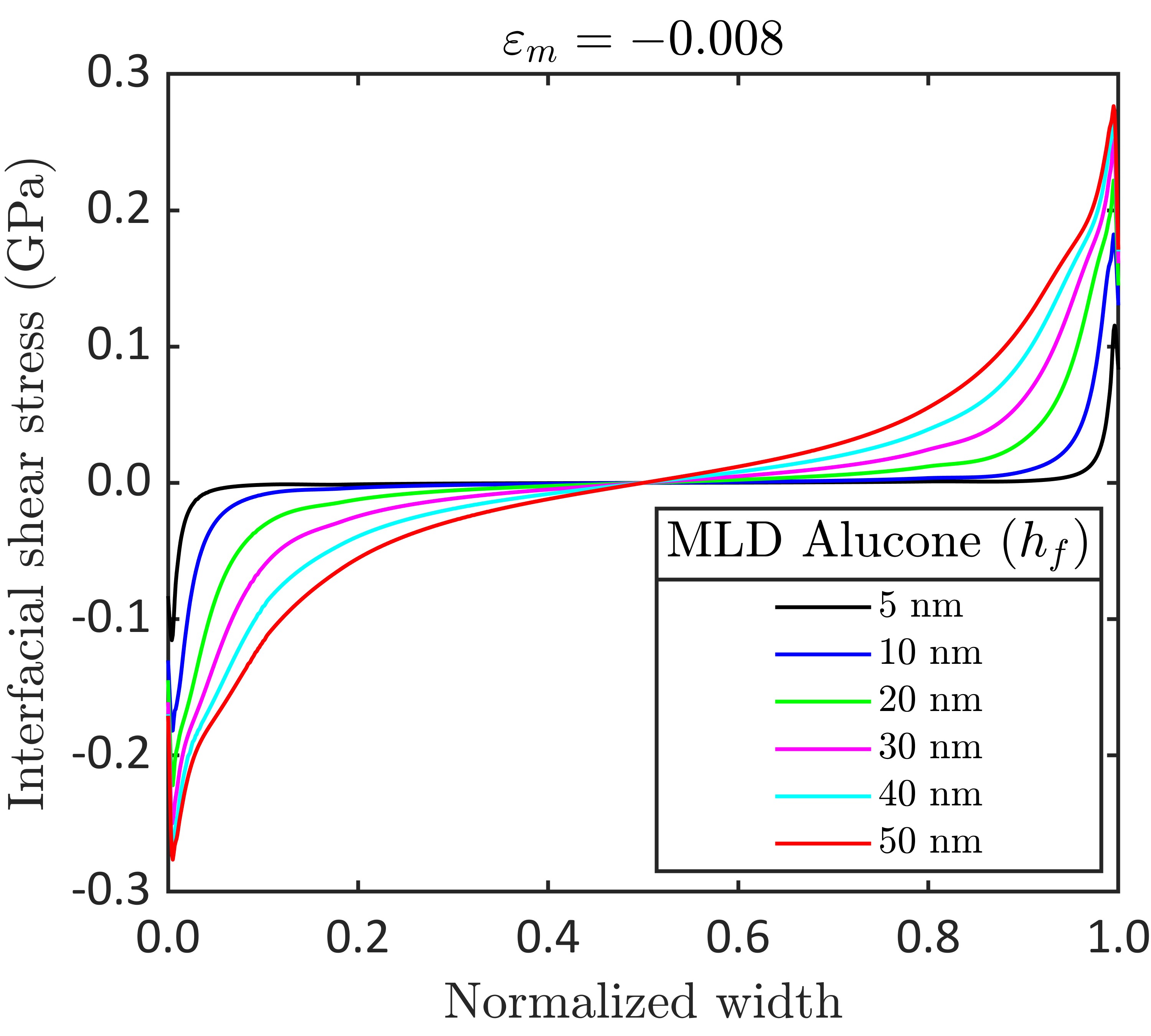}}
\caption{\label{fig:shear_stress_FEM}Effects of interfacial shearing stress with varying thin film thickness (a) ALD alumina, and (b) MLD alucone.}
\end{figure}
\par According to the FEA results, when the interface has complete bonding, the interfacial shearing stresses were at their lowest in the inner area and reached their highest near the edge (see Fig. \ref{fig:finite_element_modeling}). The computed results of shearing stress produced by FEA with various film thicknesses are shown in~\fig{shear_stress_FEM} where the thicker film layer has a higher shearing stress at the interface. The findings show that the inner position's stress distribution is relatively homogeneous and magnitude close to zero. However, the distribution of shearing stresses changes quickly near the edge because of the so-called ``edge effect", reaching its maximum values there. The interfacial shear stress increases significantly toward either side.
%

\end{document}